\documentclass[12pt,letterpaper,a4paper]{article}
\usepackage[includeheadfoot,
            marginratio={1:1,2:3},
            width=450pt,
            height=700pt,]{geometry}
\usepackage{amsmath}
\usepackage{amsfonts}
\usepackage{amssymb}
\usepackage{graphicx}
\usepackage{cancel}

\usepackage{empheq}
\usepackage{color}
\usepackage{hyperref}

\usepackage{float}
\restylefloat{table}
\restylefloat{figure}

\newcommand{\nc}{\newcommand}
\nc{\beq}{\begin{equation}}
\nc{\eeq}{\end{equation}}
\nc{\bea}{\begin{eqnarray}}
\nc{\eea}{\end{eqnarray}}

\def\ov{\overline}

\newcommand{\eq}[1]{\begin{equation}
                     \begin{split} #1 \end{split}
                     \end{equation}}

\allowdisplaybreaks[2]
\begin{document}

\vspace{1.5cm}
\begin{center}
{\LARGE
A symplectic rearrangement of the four dimensional non-geometric scalar potential}
\vspace{0.4cm}
\end{center}

\vspace{0.35cm}
\begin{center}
Pramod Shukla\footnote{Email: pkshukla@to.infn.it}
\end{center}

\vspace{0.1cm}
\begin{center}
{Universit\'a di Torino, Dipartimento di Fisica and I.N.F.N.-sezione di Torino
\vskip0.02cm
Via P. Giuria 1, I-10125 Torino, Italy \footnote{From October 1, 2015, the address has been changed to {\it ICTP, Strada Costiera 11, Trieste 34014, Italy}, Email: shukla.pramod@ictp.it.}}
\end{center}

\vspace{1cm}


\begin{abstract}
We present a symplectic rearrangement of the effective four-dimensional non-geometric scalar potential resulting from type IIB superstring  compactification on Calabi Yau  orientifolds. The strategy has two main steps. In the first step, we rewrite the four dimensional scalar potential utilizing some interesting flux combinations which we call {\it new generalized flux orbits}. After invoking a couple of non-trivial symplectic relations, in the second step, we further rearrange all the pieces of scalar potential into a completely `symplectic-formulation' which involves only the symplectic ingredients (such as period matrix etc.) without the need of knowing Calabi Yau metric. Moreover, the scalar potential under consideration is induced by a generic tree level K\"{a}hler potential and (non-geometric) flux superpotential for arbitrary numbers of complex structure moduli, K\"ahler moduli and odd-axions. Finally, we exemplify our symplectic formulation for the two well known toroidal examples based on type IIB superstring compactification on ${\mathbb T}^6/{({\mathbb Z}_2 \times {\mathbb Z}_2)}$-orientifold and ${\mathbb T}^6/{{\mathbb Z}_4}$-orientifold.
\end{abstract}

\clearpage

\tableofcontents



\section{Introduction}
For more than a decade, moduli stabilization has been among the most challenging goals of realistic model building attempts in superstring compactifications. In this regard, a lot of attractive progress has been made in type II orientifold compactification in recent years \cite{Kachru:2003aw, Balasubramanian:2005zx, Grana:2005jc, Blumenhagen:2006ci, Douglas:2006es, Denef:2005mm, Blumenhagen:2007sm}. Turing on various possible fluxes on the internal background induces effective potentials for the moduli and hence create the possibility of fluxes being utilized for moduli stabilization and in search of string vacua \cite{deCarlos:2009qm, Danielsson:2012by, Blaback:2013ht, Damian:2013dq, Damian:2013dwa, Hassler:2014mla, Blumenhagen:2015qda,Blumenhagen:2015kja}. Moreover, interesting connections between the toolkits of superstring flux-compactifications and the gauged supergravities have given the platform for approaching phenomenology based goals from two directions  \cite{ Derendinger:2004jn, Derendinger:2005ph, Shelton:2005cf, Aldazabal:2006up, Dall'Agata:2009gv, Aldazabal:2011yz, Aldazabal:2011nj,Geissbuhler:2011mx,Grana:2012rr,Dibitetto:2012rk, Villadoro:2005cu}.  A consistent incorporation of various kinds of possible fluxes makes the compactification background richer and more flexible for model building. For example, inclusion of non-geometric flux breaks the no-scale structure of low energy 4D type IIB supergravity, and opens the possibility of stabilizing {\it all} moduli at tree level. However, the task does not remain as simple as many technical challenges are inevitable, and the same have led to enormous amount of progress in recent years \cite{Derendinger:2004jn, Derendinger:2005ph, Shelton:2005cf, Dall'Agata:2009gv, Aldazabal:2011yz, Aldazabal:2011nj,Geissbuhler:2011mx,Grana:2012rr,Dibitetto:2012rk, Kachru:2002sk,Hellerman:2002ax,Dabholkar:2002sy,Hull:2004in, Andriot:2012wx, Andriot:2012an, Andriot:2011uh,Blumenhagen:2013hva,Andriot:2013xca,Andriot:2014qla,Blair:2014zba}. For example the resulting 4D scalar potential are very often so huge in concrete examples (say in Type IIB on ${\mathbb T}^6/({\mathbb Z}_2\times{\mathbb Z}_2)$ orientifold) that even it gets hard to analytically solve the extremization conditions, and one has to look either for simplified ansatz by switching off certain flux components at a time, or else one has to opt for an involved numerical analysis.  

There have been close connections between the symplectic geometry and effective potentials of type II supergravity theories \cite{Ceresole:1995ca, D'Auria:2007ay}, and the role of symplectic geometry gets crucially important while dealing with Calabi Yau orientifolds. The reason for the same being the fact that unlike toroidal orientifold examples, one does not know the explicit analytic representation of Calabi Yau metric needed to express the effective potential. However, in the context of type IIB orietifolds with the presence of standard NS-NS three-form flux ($H_3$) and RR three-form flux ($F_3$), it has been shown that the complete four dimensional scalar potential (derived from $F/D$-term contributions) could be expressed via merely using the period matrices and without the need of CY metric \cite{Taylor:1999ii, Blumenhagen:2003vr}.  A chain of  successive $T$-duality operations on $H$-flux of type II orientifold theories lead to various geometric and non-geometric fluxes, namely $\omega, \, Q$ and $R$-fluxes. Moreover, S-duality invariance of type IIB superstring compactification demands for including additional $P$-fluxes S-dual to non-geometric $Q$-flux \cite{Aldazabal:2008zza,Font:2008vd,Guarino:2008ik, Hull:2004in, Aldazabal:2008zza, Kumar:1996zx, Hull:2003kr}. Now the question arises if it could be possible to take the next step to include generalized (non-geometric) fluxes in symplectic formalism of \cite{Taylor:1999ii, Blumenhagen:2003vr}.

Moreover, in the context of non-geometric flux compactifications, there have been great amount of studies via considering the 4D effective potential merely derived by knowing the K\"ahler and super-potentials \cite{Danielsson:2012by, Blaback:2013ht, Damian:2013dq, Damian:2013dwa, Blumenhagen:2013hva, Villadoro:2005cu, Robbins:2007yv, Ihl:2007ah, Gao:2015nra}, and without having a good understanding of their ten dimensional origin. Some significant steps have been taken towards exploring the form of non-geometric 10D action via Double Field Theory (DFT) \footnote{A much better understanding of the ten dimensional origin of the 4D non-geometric scalar potential has been proposed very recently in a nice work \cite{Blumenhagen:2015lta} via considering dimensional reduction of DFT.} \cite{Andriot:2013xca, Andriot:2011uh} as well as supergravity \cite{Villadoro:2005cu, Blumenhagen:2013hva, Gao:2015nra, Shukla:2015rua, Shukla:2015bca}. In this regard, toroidal orientifolds have been always in the center of attraction because of their relatively simpler structure to perform explicit computations, and so toroidal setups have served as promising toolkits. For example the knowledge of metric has helped in anticipating the ten-dimensional origin of the geometric flux dependent \cite{Villadoro:2005cu} as well as the non-geometric flux dependent potentials \cite{Blumenhagen:2013hva} via a {\it dimensional oxidation process} in the ${\mathbb T}^6/{\left({\mathbb Z}_2 \times {\mathbb Z}_2\right)}$ toroidal orientifolds of type IIA and its T-dual type IIB model. Later on, this {\it dimensional oxidation process} has been further extended with the inclusion of P-flux, the S-dual to non-geometric Q-flux in \cite{Gao:2015nra} as well as with the inclusion of odd axions $B_2/C_2$ and geometric flux ($\omega$) as well as non-geometric flux ($R$) in \cite{Shukla:2015rua,Shukla:2015bca} leading to the appearance of peculiar flux combinations which are called as {\it new generalized flux orbits}. 

In this article, we aim to extend the symplectic formalism of \cite{Taylor:1999ii, Blumenhagen:2003vr} by rewriting the four dimensional non-geometric scalar potential in terms of symplectic ingredients.  To be specific, in the context of type IIB non-geometric Calabi Yau orientifold compactification, we will consider generic tree level K\"ahler and (non-geometric) flux super-potentials for arbitrary number of moduli/axions, and rearrange the $F/D$ term contributions using new generalized flux orbits and some symplectic relations. 

The article is organized as follows: Section \ref{sec_Basics} provides a very brief review of type IIB non-geometric flux compactification relevant for the present work. In Section \ref{sec_using-orbits}, we compute the four dimensional non-geometric scalar potential resulting from the generic tree level expressions of K\"ahler - and super-potentials valid for arbitrary numbers of complex structure moduli, K\"ahler moduli and odd-axions. Subsequently as a first step, we rewrite the scalar potential into a compact manner via using interesting flux combinations what we call {\it new generalize flux orbits}.  In section \ref{sec_using-symplecticIdentities}, we invoke a couple of symplectic relations to further rewrite the first rearrangement of section \ref{sec_using-orbits} into an entirely symplectic and very compact fashion. In section \ref{sec_2examples}, we illustrate the utility of the same for rewriting the 4D scalar potentials  of two concrete well-known examples of Type IIB superstring compactifications ${\mathbb T}^6/({\mathbb Z}_2 \times {\mathbb Z}_2)$ and ${\mathbb T}^6/{\mathbb Z}_4$ orientfolds. Finally, in section \ref{sec_conclusion}, we provide an overall conclusion followed by an appendix of additional useful symplectic relations.

\section{Preliminaries}
\label{sec_Basics}
Let us consider Type IIB superstring theory compactified on an orientifold of a
Calabi-Yau threefold $X$. 

\subsection{Splitting of various cohomologies under orientifold action}
The admissible orientifold projections can be classified by their action on the
K\"ahler form $J$ and the holomorphic three-form $\Omega_3$ of
the Calabi-Yau, given as under \cite{Grimm:2004uq}:
\begin{eqnarray}
\label{eq:orientifold}
 {\cal O}= \begin{cases}
                       \Omega_p\, \sigma  & : \, 
                       \sigma^*(J)=J\,,  \, \, \sigma^*(\Omega_3)=\Omega_3 \, ,\\[0.1cm]
                       (-)^{F_L}\,\Omega_p\, \sigma & :\, 
        \sigma^*(J)=J\,, \, \, \sigma^*(\Omega_3)=-\Omega_3\, ,
\end{cases}
\end{eqnarray}
where $\Omega_p$ is the world-sheet parity, $F_L$ is the left-moving space-time fermion number, and $\sigma$ is a holomorphic, isometric
 involution. The first choice leads to orientifold with $O5/O9$-planes
whereas the second choice to $O3/O7$-planes.
The massless states in the four dimensional effective theory are in one-to-one correspondence
with harmonic forms which are either  even or odd
under the action of $\sigma$, and these do generate the equivariant  cohomology groups $H^{p,q}_\pm (X)$. Let us fix our conventions as those of \cite{Robbins:2007yv}, and denote the bases  of even/odd two-forms as $(\mu_\alpha, \, \nu_a)$ while four-forms as $(\tilde{\mu}_\alpha, \, \tilde{\nu}_a)$ where $\alpha\in h^{1,1}_+(X), \, a\in h^{1,1}_-(X)$ \footnote{For explicit construction of some type-IIB toroidal/CY orientifold examples with odd-axions, see \cite{Lust:2006zg,Lust:2006zh,Blumenhagen:2008zz,Cicoli:2012vw,Gao:2013rra,Gao:2013pra}.}. Also, we denote the zero- and six- even forms as ${\bf 1}$ and $\Phi_6$ respectively. The definitions of integration over the intersection of various cohomology bases are,
\bea
\label{eq:intersection}
& & \hskip-0.7cm \int_X \Phi_6 = f, \, \, \int_X \, \mu_\alpha \wedge \tilde{\mu}^\beta = \hat{d}_\alpha^{\, \, \, \beta} , \, \, \int_X \, \nu_a \wedge \tilde{\nu}^b = {d}_a^{\, \, \,b} \\
& & \hskip-0.7cm \int_X \, \mu_\alpha \wedge \mu_\beta \wedge \mu_\gamma = k_{\alpha \beta \gamma}, \, \, \, \, \int_X \, \mu_\alpha \wedge \nu_a \wedge \nu_b = \hat{k}_{\alpha a b} \nonumber
\eea
Note that if four-form basis is appropriately chosen to be dual of the two-form basis, one will of course have $\hat{d}_\alpha^{\, \, \, \beta} = \hat{\delta}_\alpha^{\, \, \, \beta}$ and ${d}_a^{\, \, \,b} = {\delta}_a^{\, \, \,b}$. However for the present work, we follow the conventions of \cite{Robbins:2007yv}, and take the generic case. Considering the bases for the even/odd cohomologies $H^3_\pm(X)$ of three-forms as symplectic pairs $(a_K, b^J)$ and $({\cal A}_\Lambda, {\cal B}^\Delta)$ respectively, we fix the normalization as under,
\bea
\int_X a_K \wedge b^J = \delta_K{}^J, \, \, \, \, \, \int_X {\cal A}_\Lambda \wedge {\cal B}^\Delta = \delta_\Lambda{}^\Delta
\eea
Here, for the orientifold choice with $O3/O7$-planes, $K\in \{1, ..., h^{2,1}_+\}$ and $\Lambda\in \{0, ..., h^{2,1}_-\}$ while for $O5/O9$-planes, one has $K\in \{0, ..., h^{2,1}_+\}$ and $\Lambda\in \{1, ..., h^{2,1}_-\}$.

Now, the various field ingredients can be expanded in appropriate bases of the equivariant cohomologies. For example, the K\"{a}hler form $J$, the
two-forms $B_2$,  $C_2$ and the R-R four-form $C_4$ can be expanded as \cite{Grimm:2004uq}
\bea
\label{eq:fieldExpansions}
& & J = t^\alpha\, \mu_\alpha,  \,\,\,\quad  B_2= b^a\, \nu_a , \,\,\, \quad C_2 =c^a\, \nu_a \, \\
& & C_4 = D_2^{\alpha}\wedge \mu_\alpha + V^{K}\wedge a_K + U_{K}\wedge b^K + {\rho}_{\alpha} \, \tilde\mu^\alpha \nonumber
\eea
where $t^\alpha$ is string-frame two-cycle volume moduli, while $b^a, \, c^a$ and $\rho_\alpha$ are various axions. Further, ($V^K$, $U_K$) forms a dual pair of space-time one-forms
and $D_2^{\alpha}$ is a space-time two-form dual to the scalar field $\rho_\alpha$.
Also, since $\sigma^*$ reflects the holomorphic three-form $\Omega_3$, we have $h^{2,1}_-(X)$ complex structure moduli $z^{\tilde a}$ appearing as complex scalars. 
Now, we consider a complex multi-form of even degree $\Phi_c^{even}$ defined as \cite{Benmachiche:2006df},
\bea
& & \hskip-1cm \Phi_c^{even} = e^{B_2} \wedge C_{RR} + i \, e^{-\phi} Re(e^{B_2+i\, J})\\
& & \hskip-0.0cm \equiv \tau + G^a \, \,\nu_a + T_\alpha \, \, \tilde{\mu}^\alpha \, ,\nonumber
\eea
which suggests the following forms for the Einstein-frame chiral variables appearing in ${N}=1$ 4D-effective theory,
\bea
\label{eq:N=1_coords}
& & \tau = C_0 + \, i \, e^{-\phi} \, \, , \, \, \, \, \, G^a= c^a + \tau \, b^a \, ,\\
& & \hskip-0.05cm T_\alpha= \left({\rho}_\alpha +  \hat{\kappa}_{\alpha a b} c^a b^b + \frac{1}{2} \, \tau \, \hat{\kappa}_{\alpha a b} b^a \, b^b \right)  -\frac{i}{2} \, \kappa_{\alpha\beta\gamma} t^\beta t^\gamma\, ,\nonumber
\eea
where $\kappa_{\alpha\beta\gamma}=(\hat{d^{-1}})_\alpha^{ \, \,\delta} \, k_{\delta\beta\gamma}$ and $\hat{\kappa}_{\alpha a b} = (\hat{d^{-1}})_\alpha^{ \, \,\delta} \, \hat{k}_{\delta a b}$. It is worth to mention that as compared to chiral variables defined in \cite{Grimm:2004uq}, we have rescaled our $T_\alpha$ by a factor of $2/(3 i)$  along with a sign flip in NS-NS axion $b^a$.

\subsection{Four dimensional effective scalar potential}
The dynamics of low energy effective supergravity action is encoded in three building blocks; namely a K\"{a}hler potential ($K$), a holomorphic superpotential ($W$) and a holomorphic gauge kinetic function ($\hat{\cal G}$) written in terms of appropriate chiral variables. Subsequently, the total ${N}=1$ scalar potential can be computed from various $F/D$-term contributions via
\bea
\label{eq:Vtot}
& & V=e^{K}\Big(K^{i\bar\jmath}D_i W\, D_{\bar\jmath} \ov W-3\, |W|^2\Big) + \frac{1}{2} (Re \, \, \, \hat{\cal G})^{{-1}{JK}} \, D_J D_K \, . \nonumber
\eea
Let us provide some details on the basic ingredients needed to generate the scalar potential $V$.
\subsubsection*{K\"ahler potential ($K$) and moduli space metrices}
Using appropriate chiral variables, a generic form of the tree level K\"{a}hler potential can be written as a sum of two pieces motivated from their underlying ${N}=2$ special K\"ahler and quaternionic structure, and the same is give as under,
\bea
\label{eq:K}
& & \hskip1.5cm K := K_{cs} + K_q, \quad {\rm where}\\
& & K_{cs}= -\ln\left(i\int_{X}\Omega_3\wedge{\bar\Omega_3}\right), \quad K_q= - \ln\left(-i(\tau-\ov\tau)\right) -2\ln{\cal V}_E \, .\nonumber
\eea
Here, the involutively-odd holomorphic three-form $\Omega_3$ generically depends on the complex structure moduli ($z^k$) and can be written out in terms of period vectors, 
\bea
& &  \Omega_3\, \equiv  {\cal X}^\Lambda \, {\cal A}_\Lambda - \, {\cal F}_{\Lambda} \, {\cal B}^\Lambda \,  
\eea
via using a genetic tree level pre-potential  as under,
\bea
\label{eq:prepotential}
& & {\cal F} = (X^0)^2 \, \, f({z^i}) \,, \quad \quad  f({z^i}) = \frac{1}{6} \,{l_{ijk} \, z^i\, z^j \, z^k} +  \frac{1}{2} \,{l_{ij} \, z^i\, z^j} +  \,{l_{i} \, z^i} +  \frac{1}{2} \,{l_{0}}
\eea 
where special coordinates $z^i =\frac{\delta^i_\Lambda \, X^\Lambda}{X^0}$ are used, and $ l_{ijk}$ are triple intersection numbers on the Mirror Calabi Yau. Further,  the quantities $l_{ij}, l_i$ are real numbers while $l_0$ is a pure imaginary number \cite{Hosono:1994av,Arends:2014qca}.  In general, $f({z^i})$ will have an infinite series of non-perturbative contributions (say ${\cal F}_{\rm inst.}(z^i)$), however  for the current purpose, we are assuming the large complex structure limit to suppress the same. Now, the overall internal volume ${\cal V}_E$ in the Einstein frame can be generically written in terms of two-cycle volume moduli as below,
\begin{eqnarray}
& & {\cal V}_E = \frac{1}{6} \,{k_{\alpha \beta \gamma} \, t^\alpha\, t^\beta \, t^{\gamma}} 
\end{eqnarray}
The well known fact which can be seen from the structure of the K\"ahler potential (\ref{eq:K}) is that the total moduli space metric is block diagonal with one block corresponding to the complex structure moduli while the other one involving K\"ahler moduli, odd-axions and axion-dilaton. So the scalar potential computations via utilizing K\"ahler derivatives and metrices gets much simplified.

One should represent ${\cal V}_E$ in terms of chiral variables ($\tau, G^a, T_\alpha$) as defiend in eqn. (\ref{eq:N=1_coords}) for computing the moduli space K\"ahler metrices, and for doing this, one needs to invert the expression of $T_\alpha$. Though it is not possible to do it for a general CY orientifold compactification, nevertheless one can still represent the K\"ahler matrix components into another suitable form involving two cycle volume moduli ($t^\alpha$), the dilaton ($s$), NS-NS $B_2$ axion ($b^a$) and triple intersection numbers ($\kappa_{\alpha\beta\gamma}$ and $\hat{\kappa}_{\alpha a b}$) \cite{Grimm:2004uq}. For the present work, let us just recollect the relevant K\"ahler derivatives along with the inverse K\"ahler metric components as under,
\begin{eqnarray}
\label{eq:derK}
& & K_\tau = \frac{i}{2 \,s }\left(1 + 2 \, s \, {\cal G}_{ab} \, b^a \, b^b\right) = - K_{\ov \tau} \\
& & \hskip-1cm K_{G^a} = -2 \, i \, {\cal G}_{ab} \, b^b = - K_{\ov{G}^a}, \quad K_{T_\alpha} = -\frac{ 3\, i \, \hat{d}_\beta{}^\alpha \, t^\beta}{k_0} = - K_{\ov{T}_\alpha}, \nonumber
\end{eqnarray}
and
\begin{eqnarray}
\label{eq:InvK}
& & K^{\tau \ov{\tau}} =  4 \, s^2, \quad K^{G^a \, \ov{\tau}} =  4 \, s^2 \, b^a , \quad K^{T_\alpha \, \ov{\tau}} = 2 \, s^2 \, \hat{\kappa}_{\alpha a b} b^a b^b, \\
& & K^{G^a \, \ov{G}^b} = s\, {\cal G}^{ab} + 4 \, s^2\, b^a b^b , \quad K^{T_\alpha \, \ov{G}^a} = \,s \, {\cal G}^{ab} \, \hat{\kappa}_{\alpha b c} b^c + 2 \, s^2 \hat{\kappa}_{\alpha b c} b^b b^c \, b^a , \nonumber\\
& & K^{T_\alpha \, \ov{T}_\beta} = \frac{4}{9}\, k_0^2\, \tilde{\cal G}_{\alpha \beta}  + s \, {\cal G}^{ab}\, {\hat{\kappa}_{\alpha a c} b^c} \, {\hat{\kappa}_{\beta b d} b^d} + s^2\, {\hat{\kappa}_{\alpha a b} b^a b^b} \, {\hat{\kappa}_{\beta c d} b^c b^d}, \nonumber
\end{eqnarray}
where $\tilde{\cal G}_{\alpha \beta}=\left(({\hat{d}^{-1}})_{\alpha}{}^{\alpha'}\, {\cal G}_{\alpha' \beta'}\, ({\hat{d}^{-1}})_{\beta}{}^{\beta'}\right)$. Also, for writing K\"ahler metric we have used $\hat{\kappa}_{\alpha a b} = (\hat{d}^{-1})_\alpha{}^{\beta}\hat{k}_{\beta a b}$ along with the following short hand notations for ${\cal G}$ and ${\cal G}^{-1}$ components,
\begin{eqnarray}
\label{eq:genMetrices}
& & {\cal G}_{\alpha \beta} = -\frac{3}{2} \, \left( \frac{k_{\alpha \beta}}{k_0} - \frac{3}{2} \frac{k_\alpha \,k_\beta}{k_0^2}\right), \quad \quad \quad {\cal G}^{ab} = -\frac{2}{3} \, k_0\, \hat{k}^{ab} \, \\
& & {\cal G}^{\alpha \beta} = -\frac{2}{3} \, k_0 \, k^{\alpha \beta} + 2 \, t^\alpha \, t^\beta , \quad \quad \quad \quad \quad {\cal G}_{ab} = -\frac{3}{2} \frac{\hat{k}_{ab}}{k_0} \nonumber
\end{eqnarray} 
Moreover we have introduced $k_0 = 6 \, {\cal V}_E = k_\alpha\, t^\alpha$, $k_\alpha =k_{\alpha \beta} \, t^\beta$, $k_{\alpha\beta} = k_{\alpha\beta\gamma} \, t^\gamma$ and $\hat{k}_{a b} = \hat{k}_{\alpha a b} \, t^\alpha$. Let us mention that apart from a slight difference in the definition of chiral variables (\ref{eq:N=1_coords}), due to the presence of $\hat{d}^{\alpha}_\beta$ and $d^a_b$ matrices, there is a further slight change in the expressions of various components of inverse K\"ahler metric as compared to the ones given in \cite{Grimm:2004uq}. One can show that the dependence on these $d$-matrices can be picked up via considering the fact that $\frac{\partial {\cal V}_E}{\partial T_\alpha}$ scales with $\hat{d}^\alpha_{\beta}\, \frac{\partial {\cal V}_E}{\partial T_\beta}$ as can be anticipated directly from $6\, {\cal V}_E := {k}_{\alpha\beta\gamma} t^\alpha \, \, t^\beta \, t^{\gamma}= (t^\delta \, \hat{d}_\delta{}^\alpha) ({\kappa}_{\alpha\beta\gamma} \, t^\beta \, t^{\gamma})$. Taking these into account, one finds that,
\bea
& & \frac{4}{9}\, k_0^2 \tilde{\cal G}_{\alpha \beta} = ({\hat{d}^{-1}})_{\alpha}{}^{\alpha'} \left(k_{\alpha'}\, k_{\beta'} - \frac{2}{3} \, k_0 \, k_{\alpha'\beta'}\, \right)\, ({\hat{d}^{-1}})_{\beta}{}^{\beta'}\\
& & \hskip3cm = \left(\kappa_\alpha\, \kappa_\beta \, - 4\, {\cal V}_E\, ({\hat{d}^{-1}})_{\alpha}{}^{\alpha'}\, k_{\alpha' \beta'}\,  ({\hat{d}^{-1}})_{\beta}{}^{\beta'} \right) \nonumber
\eea
Note that one often uses orientifold constructions  such that $\hat{d}_\alpha{}^\beta={\delta}_\alpha{}^\beta$ and ${d}_a{}^b={\delta}_a{}^b$, and so one will not need to take care of these extra normalizations, however in cases otherwise, e.g. in the second example `Example B', the same is important as we will see later.

\subsubsection*{Non-geometric flux superpotential ($W$)}
Turning on various fluxes on the internal background induces a non-trivial flux superpotential \cite{Taylor:1999ii}. To construct a generic form of the superpotential, one has to understand the splitting of various geometric as well as non-geometric fluxes into the suitable orientifold even/odd bases. Moreover, it is important to note that in a given setup, all flux-components will not be generically allowed under the full orietifold action ${\cal O} = \Omega_p (-)^{F_L} \sigma$. For example, only geometric flux $\omega$ and non-geometric flux $R$ remain invariant under  $(\Omega_p (-)^{F_L})$, while the standard fluxes $(F, H)$ and non-geometric flux $(Q)$ are anti-invariant \cite{Blumenhagen:2015kja, Robbins:2007yv}. Therefore, under the full orientifold action, we can only have the following flux-components 
\bea
& & \hskip-0.10cm F\equiv \left(F_\Lambda, F^\Lambda\right),  H\equiv \left(H_\Lambda, H^\Lambda\right), \omega\equiv \left({\omega}_a{}^\Lambda, {\omega}_{a \Lambda} , \hat{\omega}_\alpha{}^K, \hat{\omega}_{\alpha K}\right),\nonumber\\
& &  R\equiv \left(R_K, R^K \right), \, \, \, Q\equiv \left({Q}^{a{}K}, \, {Q}^{a}{}_{K}, \, \hat{Q}^{\alpha{}\Lambda} , \, \hat{Q}^{\alpha}{}_{\Lambda}\right), 
\eea
For writing a general flux-superpotential, one needs to define a twisted differential operator, ${\cal D}$ involving the actions from all the NS-NS (non-)geometric fluxes as \cite{Robbins:2007yv}, 
\bea
\label{eq:twistedD}
& & {\cal D} = d + H \wedge.  + \omega \triangleleft . + Q \triangleright. + R \bullet \, \, .
\eea
The action of operator $\triangleleft, \triangleright$ and $\bullet$ on a $p$-form changes it into a $(p+1)$, $(p-1)$ and $(p-3)$-form respectively. Considering various flux-actions on the different  even/odd bases to result in even/odd three-forms, we have \cite{Robbins:2007yv},
\bea
\label{eq:action1}
& & H = {H}^\Lambda {\cal A}_\Lambda + H_\Lambda \, \,{\cal B}^\Lambda, \, \, \, \, \, F = {F}^\Lambda {\cal A}_\Lambda + F_\Lambda \, \,{\cal B}^\Delta,\nonumber\\
& & \omega_a \equiv (\omega \triangleleft \nu_a) = {\omega}_a{}^\Lambda \, {\cal A}_\Lambda + \omega_{a{}\Lambda} {\cal B}^\Lambda, \nonumber\\
& & \hat{Q}^{\alpha}\equiv (Q \triangleright {\tilde\mu}^\alpha) = \hat{Q}^{\alpha{}\Lambda} {\cal A}_\Lambda + \hat{Q}^{\alpha}{}_{\Lambda} {\cal B}^\Lambda \\
& & \hat{\omega}_\alpha \equiv (\omega \triangleleft \mu_\alpha) = \hat{\omega}_\alpha{}^K a_K + \hat{\omega}_{\alpha{}K} b^K, \nonumber\\
& & {Q}^{a}\equiv (Q \triangleright \tilde{\nu}^a) = {Q}^{a{}K} \, a_K + Q^{a}{}_{K} b^K, \nonumber\\
& & R\bullet \Phi = R^K a_K + R_K b^K \, . \nonumber
\eea
The first three lines involve flux components counted via `odd-index' $\Lambda\in{h^{2,1}_-(X)}$ while the later three have `even-index' $K\in {h^{2,1}_+(X)}$. Using definitions in (\ref{eq:action1}), we have the following additional useful non-trivial actions of fluxes on various 3-form even/odd basis elements \cite{Robbins:2007yv},
\bea
\label{eq:action2}
& & \hskip-0.4cm H \wedge {\cal A}_\Lambda = - f^{-1} H_\Lambda \,\, \Phi_6, \, \, \quad \quad \quad \quad H \wedge {\cal B}^\Lambda = f^{-1} H^\Lambda \, \,\Phi_6 \\
& & \hskip-0.4cm \omega\triangleleft {\cal A}_\Lambda=-\left({d}^{-1}\right)_a{}^b \,{\omega}_{b \Lambda} \, \tilde{\nu}^a, \, \quad  \quad \quad \omega\triangleleft {\cal B}^\Lambda= \left({d}^{-1}\right)_a{}^b \, {\omega}_{b}{}^{\Lambda} \, \tilde{\nu}^a\nonumber\\
& & \hskip-0.4cm Q\triangleright {\cal A}_\Lambda=-\left(\hat{d}^{-1}\right)_\alpha{}^\beta \,\hat{Q}^\alpha_{\Lambda} \, {\mu}_\beta, \, \quad \quad  \, Q\triangleright {\cal B}^\Lambda=\left(\hat{d}^{-1}\right)_\alpha{}^\beta \, \hat{Q}^{\alpha \Lambda} \,{\mu}_\beta ,\nonumber
\eea
and
\bea
& & \hskip-0.4cm R\bullet a_K = -\, f^{-1} \, R_K \, {\bf 1}, \quad  \quad \quad \quad \,  \quad R \bullet b^K = f^{-1} \, R^K \,{\bf 1}\nonumber\\
& & \hskip-0.4cm \omega\triangleleft a_K=- \, \left(\hat{d}^{-1}\right)_\alpha{}^\beta \hat{\omega}_{\beta K} \, \tilde{\mu}^\alpha, \quad  \quad \omega\triangleleft b^K=\left(\hat{d}^{-1}\right)_\alpha{}^\beta \hat{\omega}_{\beta}{}^{K} \, \tilde{\mu}^\alpha\nonumber\\
& & \hskip-0.4cm Q\triangleright a_K=-\left({d}^{-1}\right)_a{}^b {Q}^a_{K} \,{\nu}_b,\quad  \quad \quad Q\triangleright b^K=\left({d}^{-1}\right)_a{}^b \,{Q}^{a K} \, {\nu}_b \, .\nonumber
\eea
With these ingredients in hand, a generic form of flux superpotential is as under,
\bea
\label{eq:W1}
& & \hskip-0.5cm W = -\int_X \biggl[F+ {\cal D} \Phi_c^{even}\biggr]_3 \wedge \Omega_3  =- \int_{X} \biggl[{F} +\tau \, {H} + \, \omega_a {G}^a + \, {\hat Q}^{\alpha} \,{T}_\alpha \biggr]_3 \wedge \Omega_3. 
\eea
This generic flux superpotential $W$ can be equivalently written as,
\begin{eqnarray}
\label{eq:W_gen}
& & W = e_\Lambda \, {\cal X}^\Lambda + m^\Lambda \, {\cal F}_\Lambda,
\end{eqnarray} 
where
\begin{eqnarray}
\label{eq:eANDm}
& &  e_\Lambda = {F}_\Lambda + \tau \, H_\Lambda + \omega_{a\, \Lambda}\, G^a+ \hat{Q}^\alpha{}_{\Lambda} \, T_\alpha, \, \\
& & \quad \quad \quad \quad \quad m^\Lambda = {F}^\Lambda + \tau \, {H}^\Lambda + {\omega}_a{}^{\Lambda}\, G^a + \hat{Q}^{\alpha \,\Lambda} \, T_\alpha \,.\nonumber
\end{eqnarray} 
Using the superpotential (\ref{eq:W_gen}), one can compute the various derivatives with respect to chiral variables, $\tau, G^a$ and $T_\alpha$ as followings,
\begin{eqnarray}
\label{eq:DerW}
&  W_\tau = H_\Lambda \, {\cal X}^\Lambda + {H}^\Lambda\, {\cal F}_\Lambda,  \quad & \ov{W}_{\ov \tau} = H_\Delta \, \ov{{\cal X}}^\Delta + {H}^\Delta\, {\ov {\cal F}}_\Delta\nonumber\\
&  W_{G^a} = \omega_{a \Lambda}\, {\cal X}^\Lambda + {\omega}_a{}^\Lambda\, {\cal F}_\Lambda, \quad & \ov{W}_{\ov{G}^a} = \omega_{a \Delta}\, \ov{\cal X}^\Delta + {\omega}_a{}^\Delta\, {\ov {\cal F}}_\Delta\nonumber\\
& W_{T_\alpha} = \hat{Q}^\alpha{}_\Lambda \, {\cal X}^\Lambda + \hat{Q}^{\alpha \Lambda}\, {\cal F}_\Lambda, \quad & \ov{W}_{\ov{T}_\alpha} = \hat{Q}^\alpha{}_\Delta \, \ov{\cal X}^\Delta + \hat{Q}^{\alpha \Delta}\, \ov{\cal F}_\Delta\, .
\end{eqnarray}
Note that, only $\omega_a$ and $\hat{Q}^\alpha$ components are allowed by the choice of involution to contribute into the superpotential, and in order to turn-on the non-geometric $R$-fluxes, one has to induce D-terms via implementing a non-trivial even sector of $H^{2,1}(X)$-cohomology \cite{Robbins:2007yv, Shukla:2015bca, Shukla:2015rua, Blumenhagen:2015lta}. However, for the cases with homomorphic involutions with $h^{2,1}_-(CY/{\cal O}) = 0$, which one often adopts in moduli stabilization and subsequent phenomenological purposes, no such D-terms involving non-geometric $R$-flux will be induced. 

\subsubsection*{The D-terms ($D_K, D^K$)}
In the presence of a non-trivial sector of even (2,1)-cohomology, i.e. for $h^{2,1}_+(X)\ne 0$, there are additional D-term contributions to the four dimensional scalar potential. Following the strategy of \cite{Robbins:2007yv}, the same can be determined via considering the following gauge transformations of RR potentials $C_{RR} = C_0 + C_2 + C_4$,
\bea
& & \hskip-1.0cm C_{RR} \longrightarrow  C_{RR} + {\cal D} (\lambda^K a_K + \lambda_K b^K )\\
& & \hskip0cm \supset \left(C_0 - f^{-1} R_K \lambda^K + f^{-1} R^K \lambda_K\right) + \left(c^b -(d^{-1})_a{}^b Q^a{}_K \lambda^K + (d^{-1})_a{}^b Q^{a K} \lambda_K\right) \nu_b \nonumber\\
& & \hskip0cm + \left(\rho_\alpha -({\hat{d}}^{-1})_\alpha{}^\beta \, \hat{\omega}_{\beta K} \lambda^K + ({\hat{d}}^{-1})_\alpha{}^\beta \, \hat{\omega}_{\beta}{}^{K} \lambda_K\right) \tilde{\mu}^\alpha\nonumber
\eea
Recollection of various pieces suggests the following two D-terms being generated by the gauge transformations,
\bea
& & \hskip-1cm D_K = -i \, \biggl[ f^{-1} R_K \, (\partial_\tau K) + (d^{-1})_b{}^a Q^b{}_K \, (\partial_a K) + ({\hat{d}}^{-1})_\alpha{}^\beta \, \hat{\omega}_{\beta K}\, (\partial^\alpha K) \biggr] \\
& & \hskip-1cm D^K = i \, \biggl[ f^{-1} R^K \, (\partial_\tau K) + (d^{-1})_b{}^a Q^{b K} \, (\partial_a K) + ({\hat{d}}^{-1})_\alpha{}^\beta \, \hat{\omega}_{\beta}{}^{K}\, (\partial^\alpha K) \biggr] \nonumber
\eea
Now using the expressions for tree level K\"ahler derivatives (\ref{eq:derK}), one finds 
\bea
\label{eq:DtermOld}
& & \hskip-1.2cm \quad D_K = \frac{1}{2\, s\,{\cal V}_E}\, \biggl[ \frac{R_K}{f} \, \left({\cal V}_E -\frac{s}{2}\hat{k}_{\alpha a b} t^\alpha b^a b^b\right) + s\, (d^{-1})_b{}^a Q^b{}_K \, \hat{k}_{\alpha a c} t^\alpha b^c - s\, t^\alpha \, \,\hat{\omega}_{\alpha K}\, \biggr]\\
& & \hskip-1.2cm \quad D^K = -\frac{1}{2\, s\,{\cal V}_E}\, \biggl[ \frac{R^K}{f} \, \left({\cal V}_E -\frac{s}{2}\hat{k}_{\alpha a b} t^\alpha b^a b^b\right) + s\, (d^{-1})_b{}^a Q^{b K} \, \hat{k}_{\alpha a c} t^\alpha b^c - s\, t^\alpha \, \,\hat{\omega}_{\alpha}{}^{K}\, \biggr]\nonumber
\eea

\subsection{New generalized flux orbits}
A closer investigation of the symplectic vectors $(e_\Lambda, m^\Lambda)$ and ($D_K, D^K$), which are responsible for generating $F$-term and $D$-term contributions to the scalar potential, suggests for defining peculiar flux combination as  {\it new generalized flux orbits} \cite{Shukla:2015bca,Shukla:2015rua}. The flux orbits in NS-NS sector with orientifold odd-indices $k \in h^{2,1}_-(X)$ are given as,
\bea
\label{eq:OddOrbitA}
& &  {\mathbb H}_\Lambda = H_\Lambda + \omega_{a\Lambda} \, {b}^a + \hat{Q}^\alpha{}_\Lambda \, \left(\frac{1}{2}\, \hat{\kappa}_{\alpha a b}\, b^a b^b\right) \nonumber\\
& &  {\mathbb H}^\Lambda = H^\Lambda + \omega_{a}{}^{\Lambda} \, {b}^a + \hat{Q}^{\alpha \Lambda} \, \left(\frac{1}{2}\, \hat{\kappa}_{\alpha a b}\, b^a b^b\right)\\
& & \hskip-1cm {\mathbb\mho}_{a\Lambda} = \omega_{a\Lambda} + \hat{Q}^\alpha{}_\Lambda \, \left(\hat{\kappa}_{\alpha a b}\, b^b\right), \quad {\mathbb\mho}_{a}{}^{\Lambda} = \omega_{a}{}^{\Lambda} + \hat{Q}^{\alpha \Lambda} \, \left(\hat{\kappa}_{\alpha a b}\, b^b\right) \nonumber\\
& & \hskip1cm \hat{\mathbb Q}^\alpha{}_\Lambda = \hat{Q}^\alpha{}_\Lambda , \quad \hat{\mathbb Q}^{\alpha \Lambda} = \hat{Q}^{\alpha \Lambda} \nonumber
\eea
while the flux components of even-index $K\in h^{2,1}_+(X)$ are given as, 
\bea
\label{eq:OddOrbitB}
& & \hat{\mho}_{\alpha K} = \hat{\omega}_{\alpha K}\, - (d^{-1})_b{}^a  \, Q^{b}{}_{K} \, \, \left(\hat{k}_{\alpha a c} \, b^c \right) + f^{-1} \, \, R_K \, \left(\frac{1}{2}\hat{k}_{\alpha a b} \, b^a \,b^b\right)\nonumber\\
& & \hat{\mho}_{\alpha}{}^{K} =\hat{\omega}_{\alpha}{}^{K}\, - (d^{-1})_b{}^a \, \, Q^{b K} \, \left(\hat{k}_{\alpha a c} \, b^c\right)+ f^{-1} \, \, R^K \, \left(\frac{1}{2}\hat{k}_{\alpha a b} \, b^a \,b^b\right)\\
& & \hskip-1.5cm {\mathbb Q}^{a}{}_{K} =  {Q}^{a}{}_{K} + f^{-1} \,\,d_b{}^a \, (R_K \bullet b^b), \quad {\mathbb Q}^{a{}K} = {Q}^{a{}K} + f^{-1} \,\, d_b{}^a  \, (R^K \bullet \, b^b), \nonumber\\
& & \hskip3cm {\mathbb R}_K = \, R_K, \quad {\mathbb R}^K = \, R^K \, .\nonumber
\eea
In the first set of orbits ({\ref{eq:OddOrbitA}), we have used $\hat{\kappa}_{\alpha a b} =({\hat{d}^{-1}})_\alpha^{ \, \,\delta} \, \hat{k}_{\delta a b}$. Now, the RR three-form flux orbits are generalized in the following form, 
\bea
\label{eq:OddOrbitC}
& &  \hskip-0.9cm {\mathbb F}_\Lambda = F_\Lambda + \omega_{a\Lambda} \, {c}^a + \hat{Q}^\alpha{}_\Lambda \, \left({\rho}_\alpha + \hat{\kappa}_{\alpha a b} c^a b^b\right)   + c_0 {\mathbb H}_\Lambda, \\
& & \hskip2cm {\mathbb F}^\Lambda = F^\Lambda + \omega_a{}^\Lambda \, {c}^a + \hat{Q}^{\alpha \Lambda} \, \left({\rho}_\alpha + \hat{\kappa}_{\alpha a b} c^a b^b\right)\, + c_0 \, {\mathbb H}^\Lambda . \, \nonumber
\eea
Using these flux orbits along with the definitions of chiral variables in eqn. (\ref{eq:N=1_coords}), the symplectic vectors $(e_\Lambda, m^\Lambda)$ and ($D_K, D^K$)  are compactly written as under,
\bea
& &  \hskip-1.5cm e_\Lambda = \,{\mathbb F}_\Lambda  + i \, \left(s \, {\mathbb H}_\Lambda \right)- i\, \left(\hat{\mathbb Q}^{\alpha}{}_{\Lambda}\, \,\sigma_\alpha \right), \,  \quad \nonumber\\
& & \hskip1.0cm \, m^\Lambda =\,{\mathbb F}^\Lambda + i \, \left(s \, {\mathbb H}^\Lambda \right)- i\, \left(\hat{\mathbb{Q}}^{\alpha \Lambda}\, \, \sigma_\alpha \right), 
\eea
and
\bea
\label{eq:D-termCompact}
& & \hskip-1.5cm  D_K  = \frac{1}{2\, s\,{\cal V}_E}\, \biggl[ f^{-1} R_K \, {\cal V}_E - s\, t^\alpha \, \hat{\mho}_{\alpha K} \biggr]\, , \quad \nonumber\\
& & \hskip1.0cm  D^K  = -\frac{1}{2\, s\,{\cal V}_E}\, \biggl[ f^{-1} R^K \, {\cal V}_E - s\, t^\alpha \,\hat{\mho}_{\alpha}{}^{K} \biggr] \, , 
\eea
where the symbol $\sigma_\alpha$ represents Einstein-frame four-cycle volume given as: $\sigma_\alpha = \frac{1}{2}\,{\kappa}_{\alpha \beta \gamma} t^\beta t^\gamma$.

\section{Rearrangement of scalar potential: Step 1}
\label{sec_using-orbits}
Here we provide a detailed computation of the ${N}=1$ four dimensional effective scalar potential for the type IIB superstring theory compactified on a Calabi Yau orientifold. {\it Our aim is to perform the most genetric tree level analysis with arbitrary number of moduli and axions, i.e. for $h^{1,1}_+(CY/{\cal O})$ number of complexified K\"ahler moduli $T_\alpha$, $h^{1,1}_-(CY/{\cal O})$ number of complexified odd-axions $G^a$ as well as $h^{2,1}_-(CY/{\cal O})$ number of complex structure moduli $z^i$.} In the process of doing the taxonomy of various pieces of F-term scalar potential, we will utilize new generalized flux orbits (\ref{eq:OddOrbitA})-(\ref{eq:OddOrbitC}) which has been proposed in \cite{Blumenhagen:2013hva,Gao:2015nra,Shukla:2015bca,Shukla:2015rua} in a series of iterative attempts. 
\subsection{Using first set of symplectic relations}
Let us start with the scalar potential analysis via considering the following splitting of pieces coming from generic ${N} = 1$ F-term contribution,
\begin{eqnarray}
\label{eq:V_gen}
& & \hskip-2cm e^{- K} \, V_F = K^{{\cal A} \ov {\cal B}} \, (D_{\cal A} W) \, (\ov D_{\ov {\cal B}} \ov{W}) -3 |W|^2 \equiv V_{cs} + V_{k}\, .
\end{eqnarray}
Here we have separated the complex structure and rest of the moduli dependent piece as facilitated by the block diagonal form of K\"ahler metric in these two sectors, and 
\begin{eqnarray}
\label{eq:VcsVk}
& & \hskip-1.5cm V_{cs} =  K^{{i} \ov {j}} \, (D_{i} W) \, (\ov D_{\ov {j}} \ov{W}), \quad V_{k} =  K^{{A} \ov {B}} \, (D_{A} W) \, (\ov D_{\ov {B}} \ov{W}) -3 |W|^2 
\end{eqnarray}
The indices $(i,j)$ corresponds to complex structure moduli $z^i$'s while the other indices $(A,B)$ are counted in rest of the chiral variables $\{\tau, G^a, T_\alpha\}$. {\it Now the plan is to rewrite the F-term scalar potential into various pieces which could be expressed in terms of components of period-matrix}, and we will do it in three parts. 

\subsubsection*{Part 1: } In the first step of simplification, we use the following symplectic identity \cite{Ceresole:1995ca},
\begin{eqnarray}
\label{eq:Identity1}
&& K^{i \ov j} \, (D_i {\cal X}^{\Lambda}) \, (\ov D_{\ov j} \ov{{\cal X}^{\Delta}}) = - \ov{{\cal X}^{\Lambda}} \,  {\cal X}^{\Delta} - \frac{1}{2} \, e^{-K_{cs}} \, {\rm Im{\cal N}}^{\Lambda \Delta}
\end{eqnarray}
where the period matrix ${\cal N}$ for the involutively odd (2,1)-cohomology sector is defined as under,
\bea
\label{eq:periodN}
& & {\cal N}_{\Lambda\Delta} = \ov{\cal F}_{\Lambda\Delta} + 2 \, i \, \frac{Im({\cal F}_{\Lambda\Gamma}) \, {\cal X}^\Gamma X^\Sigma \, (Im{\cal F}_{\Sigma \Delta}) }{Im({\cal F}_{\Gamma\Sigma}) {\cal X}^\Gamma X^\Sigma}
\eea
Using period matrix components, one can introduce the following definitions of the new-matrices (${\cal M}$) for computing the hodge star of various odd-three forms \cite{Ceresole:1995ca}, 
\begin{eqnarray}
\label{stardef}
&& \star \, {\cal A}_\Lambda =  {\cal M}_{\Lambda}^{\, \, \, \, \, \Sigma} \, \, {\cal A}_\Sigma + {\cal M}_{\Lambda \Sigma}\, \, {\cal B}^\Sigma, \, \, \, \, \, {\rm and} \, \\
& & \hskip2cm \, \, \, \star\, {\cal B}^\Lambda = {\cal M}^{\Lambda \Sigma} \, \, {\cal A}_\Sigma + {\cal M}^\Lambda_{\, \, \, \, \Sigma} \, \, {\cal B}^\Sigma \, \nonumber
\end{eqnarray}
where we also define the following useful components to be heavily utilized in the present work,
\begin{eqnarray}
\label{coff}
&& {\cal M}^{\Lambda \Delta} = {\rm Im{\cal N}}^{\Lambda \Delta} \nonumber\\
&& {\cal M}_{\Lambda}^{\, \, \, \, \, \Delta}  = {\rm Re{\cal N}}_{\Lambda \Gamma} \, \, {\rm Im{\cal N}}^{\Gamma \Delta} \\
&& {\cal M}^\Lambda_{\, \, \, \, \Delta} =- \left({\cal M}_{\Lambda}^{\, \, \, \, \, \Delta}\right)^{T} \nonumber\\
&& {\cal M}_{\Lambda \Delta}\, =  -{\rm Im{\cal N}}_{\Lambda \Delta} -{\rm Re{\cal N}}_{\Lambda \Sigma} \, \, {\rm Im{\cal N}}^{\Sigma \Gamma}\, \, {\rm Re{\cal N}}_{\Gamma \Delta} \nonumber
\end{eqnarray}
Now, we can split the piece $V_{cs}$ into two parts as follows
\begin{eqnarray}
& & \hskip-0.5cm V_{cs} = K^{i \ov j} \, (D_i W) \, (\ov D_{\ov j} \ov{W}) = V_{cs1} + V_{cs2} \, ,
\eea
where the two pieces are further simplified  as,
\bea
\label{eq:Vcs1aAndb}
& & \hskip-0.6cm V_{cs1} = -\frac{1}{2} \, e^{-K_{cs}} \, \left(e_\Lambda + m^\Sigma \ov {\cal N}_{\Sigma \Lambda} \right) \, {\rm Im{\cal N}}^{\Lambda \Delta} \,   \left(\ov e_\Delta + \ov m^\Gamma {\cal N}_{\Gamma \Delta} \right)\\
& & \hskip0.2cm =-\frac{1}{2} \, e^{-K_{cs}} \, \left(e_\Lambda \, {\cal M}^{\Lambda \Delta} \, \ov e_\Delta - e_\Lambda \, {\cal M}^{\Lambda}_{\, \, \, \Delta} \, \ov m^\Delta+ \ov e^\Lambda \, {\cal M}_{\Lambda}^{\, \, \, \Delta} \, m_\Delta- m^\Lambda \, {\cal M}_{\Lambda \Delta} \, \ov m^\Delta\right)\nonumber\\
& & \hskip2cm +\frac{i}{2} \, e^{-K_{cs}} \,\left(\ov e_\Lambda  m^\Lambda - e_\Lambda  \, \ov m^\Lambda \right) \nonumber\\
& & := V_{cs1}^{(1)} + V_{cs1}^{(2)} \, .\nonumber
\eea
and
\bea
& & \hskip-0.6cm V_{cs2} = - \left(e_\Lambda + m^\Sigma \ov {\cal N}_{\Sigma \Lambda} \right) \, \left(\ov {\cal X}^\Lambda {\cal X}^\Delta \right) \,   \left(\ov e_\Delta + \ov m^\Gamma {\cal N}_{\Gamma \Delta} \right) \\
&& \hskip0.2cm =  - e_\Lambda \, (\ov {\cal X}^\Lambda {\cal X}^\Delta) \, \ov e_\Delta - m^\Lambda \, (\ov {\cal F}_\Lambda {\cal X}^\Delta) \, \ov e_\Delta - e_\Lambda \, (\ov {\cal X}^\Lambda {\cal F}_\Delta) \, \ov m^\Delta - m^\Lambda \, (\ov {\cal F}_\Lambda {\cal F}_\Delta) \, \ov m^\Delta  \nonumber\\
& & \hskip0.2cm = - \left(e_\Delta \, \ov {\cal X}^\Delta + m^\Delta \,\ov {\cal F}_\Delta \right) \left(\ov e_\Lambda \, {\cal X}^\Lambda + \ov m^\Lambda \, {\cal F}_\Lambda \right)   \nonumber
\end{eqnarray} 
where in the last line of $V_{cs2}$, an exchange of indices $\Lambda \leftrightarrow \Delta$ has been utilized. 
The reason for this splitting is the fact that the second piece of $V_{cs1}$ is nullified via using a set of tadpole conditions and NS-NS Binachi identities. This point will be detailed later on when we will see the explicit expressions of $e_\Lambda$ and $m^\Lambda$ written in terms of NS-NS and RR generalized flux orbits. 

\subsubsection*{Part 2:}
Now, we take the piece $V_k$ of the scalar potential (\ref{eq:VcsVk}), and consider a taxonomy of pieces recollected  as under
\begin{eqnarray}
& & \hskip-3cm V_k = \left(K^{{A} \ov {B}} \, K_A \, K_{\ov {B}} |W|^2 -\,3 |W|^2\right) + \left( K^{{A} \ov {B}} \, W_A \, \ov{W}_{\ov B} \right) \nonumber\\
& & + K^{{A} \ov {B}} \, \left( (K_A \, W) \, \ov{W}_{\ov B} \, + W_A\, (K_{\ov {B}} \ov W) \right) 
\end{eqnarray}
Using the derivatives of K\"ahler potential (\ref{eq:derK}) and inverse K\"ahler metric (\ref{eq:InvK}), one finds the following useful relations,
\begin{eqnarray}
& & {K}_A\, {K}^{{A} \ov {\tau}} \,  = (\tau -\ov \tau) = - {K}^{{\tau} \ov {B}} \, {K}_{\ov B} \nonumber\\
& & {K}_A\, {K}^{{A} \ov {G^a}} \,  = (G^a -\ov G^a) = - {K}^{{G^a} \ov {B}} \, {K}_{\ov B} \\
& & {K}_A\, {K}^{{A} \ov {T_\alpha}} \, = (T_\alpha -\ov T_\alpha) = - {K}^{{T_\alpha} \ov {B}} \,  {K}_{\ov B}\nonumber
\end{eqnarray}
using which, one gets the well known no-scale relation,
\begin{eqnarray}
 K^{{A} \ov {B}} \, K_A \, K_{\ov {B}} =4 \,.
\end{eqnarray}
This simplifies the piece $V_k$ as under
\begin{eqnarray}
\label{eq:Vk2}
& & \hskip-3cm V_k =|W|^2  + K^{{A} \ov {B}} \, \left( (K_A \, W) \, \ov{W}_{\ov B} \, + W_A\, (K_{\ov {B}} \ov W) \right)  + \left( K^{{A} \ov {B}} \, W_A \, \ov{W}_{\ov B} \right)
\end{eqnarray}
A closer investigation of the second pieces shows that,
\begin{eqnarray}
\label{eq:cross2}
& & \hskip-1.5cm K^{{A} \ov {B}} \, \left( (K_A \, W) \, \ov{W}_{\ov B} \, + W_A\, (K_{\ov {B}} \ov W) \right)= (\tau -\ov \tau) \, \left(W \ov{W}_{\ov \tau} - \ov{W} \, {W}_{\tau}\right)  \nonumber\\
& & \hskip1cm +  (G^a -\ov{G}^a) \, \left(W \ov{W}_{\ov{G}^a}- \ov{W}\,  {W}_{{G}^a}\right) + (T_\alpha -\ov{T}_\alpha) \, \left(W \ov{W}_{\ov{T}_\alpha} - \ov{W}\, {W}_{{T}_\alpha}\right)\nonumber\\
& & \hskip-1cm = - 2 \, |W|^2 + W \, \left(e_\Delta \, \ov {\cal X}^\Delta + m^\Delta \,\ov {\cal F}_\Delta \right) + \ov W\, \left(\ov e_\Lambda \, {\cal X}^\Lambda + \ov m^\Lambda \, {\cal F}_\Lambda \right)
\end{eqnarray}
Here in simplifying the last step, we have used the fact that superpotential (\ref{eq:W_gen}) is a linear function in chiral variables $\tau, G^a$ and $T_\alpha$ which results in following relations,
\bea
& & \tau \, W_\tau + G^a \,W_{G^a} + T_\alpha \, W_{T_\alpha} = W - \left(F_\Lambda {\cal X}^\Lambda+ F^\Lambda {\cal F}_\Lambda \right) \\
& & \ov \tau \, \ov W_{\ov \tau} + \ov G^a \,\ov W_{\ov G^a} + \ov T_\alpha \, \ov W_{\ov T_\alpha} = \ov W - \left(F_\Delta \ov {\cal X}^\Delta+ F^\Delta \ov{\cal F}_\Delta \right) \nonumber
\eea
This follows directly from derivatives of superpotential given in eqn. (\ref{eq:DerW}).

\subsubsection*{Part: 3}
After observing eqn. (\ref{eq:cross2}), we find that the first two (of the three) pieces of $V_k$ given in eqn. (\ref{eq:Vk2}) can be recombined with $V_{cs2}$. Now we can have a new rearrangement of the total $F$-term scalar potential into three pieces as under,
\begin{eqnarray}
& & e^{-K}\, V_F = V_1 + V_2 + V_3
\end{eqnarray}
where we consider a new collection of pieces given as under,
\begin{eqnarray}
& & \hskip-1cm V_1 := V_{cs1} \\
& & \hskip-1cm V_2:= V_{cs2} + \left(K^{{A} \ov {B}} \, K_A \, K_{\ov {B}} |W|^2 -\,3 |W|^2\right) + K^{{A} \ov {B}} \, \left( (K_A \, W) \, \ov{W}_{\ov B} \, + W_A\, (K_{\ov {B}} \ov W) \right) \nonumber\\
& & \hskip-1cm V_3 := K^{{A} \ov {B}} \, W_A \, \ov{W}_{\ov B} \nonumber
\end{eqnarray}
Now, using the simplification results from {\bf Part: 1} and {\bf Part: 2}, we try to rewrite these three pieces $V_1, V_2$ and $V_3$ of the scalar potential in terms of new generalized flux combinations. The reason for such a collection will be clearer as we proceed in this section.

\subsection{Using new generalized flux orbits}
\subsubsection*{Rewriting $V_1$ using generalized flux orbits:}
After a detailed investigation of pieces within $V_1\equiv V_{cs1}=V_{cs1}^{(1)}+ V_{cs1}^{(2)} $  given in eqn. (\ref{eq:Vcs1aAndb}) and using new generalized flux orbits (\ref{eq:OddOrbitA})-(\ref{eq:OddOrbitC}), we find that first piece of simplified $V_{cs1}$ takes the form, 
\begin{eqnarray}
\label{eq:Vcs1a}
& & \hskip-1cm V_{cs1}^{(1)} = -\frac{1}{2} \, e^{-K_{cs}} \, \biggl[\left({\mathbb F}_\Lambda \, {\cal M}^{\Lambda \Delta} \,  {\mathbb F}_\Delta - {\mathbb F}_\Lambda \, {\cal M}^{\Lambda}_{\, \, \, \Delta} \, {\mathbb F}^\Delta+  {\mathbb F}^\Lambda \, {\cal M}_{\Lambda}^{\, \, \, \Delta} \, {\mathbb F}_\Delta- {\mathbb F}^\Lambda \, {\cal M}_{\Lambda \Delta} \, {\mathbb F}^\Delta\right) \\
& & \hskip2.0cm + s^2\, \, \left({\mathbb H}_\Lambda \, {\cal M}^{\Lambda \Delta} \,  {\mathbb H}_\Delta - {\mathbb H}_\Lambda \, {\cal M}^{\Lambda}_{\, \, \, \Delta} \, {\mathbb H}^\Delta+  {\mathbb H}^\Lambda \, {\cal M}_{\Lambda}^{\, \, \, \Delta} \, {\mathbb H}_\Delta- {\mathbb H}^\Lambda \, {\cal M}_{\Lambda \Delta} \, {\mathbb H}^\Delta\right) \nonumber\\
& & \hskip2.0cm + \left(\hat{\mathbb Q}_\Lambda \, {\cal M}^{\Lambda \Delta} \,  \hat{\mathbb Q}_\Delta - \hat{\mathbb Q}_\Lambda \, {\cal M}^{\Lambda}_{\, \, \, \Delta} \, \hat{\mathbb Q}^\Delta+  \hat{\mathbb Q}^\Lambda \, {\cal M}_{\Lambda}^{\, \, \, \Delta} \, \hat{\mathbb Q}_\Delta- \hat{\mathbb Q}^\Lambda \, {\cal M}_{\Lambda \Delta} \, \hat{\mathbb Q}^\Delta\right) \nonumber\\
& & \hskip2.0cm  -2 \, s \, \, \left({\mathbb H}_\Lambda \, {\cal M}^{\Lambda \Delta} \,  \hat{\mathbb Q}_\Delta - {\mathbb H}_\Lambda \, {\cal M}^{\Lambda}_{\, \, \, \Delta} \, \hat{\mathbb Q}^\Delta+  {\mathbb H}^\Lambda \, {\cal M}_{\Lambda}^{\, \, \, \Delta} \, \hat{\mathbb Q}_\Delta- {\mathbb H}^\Lambda \, {\cal M}_{\Lambda \Delta} \, \hat{\mathbb Q}^\Delta\right) \nonumber
\end{eqnarray}
and the second piece $V_{cs1}^{(2)}$ can be further rearranged as under,
\begin{eqnarray}
\label{eq:Vcs1b}
& & \hskip-1cm V_{cs1}^{(2)} = \frac{i}{2} \, e^{-K_{cs}} \,\left(\ov e_\Lambda  m^\Lambda - e_\Lambda  \, \ov m^\Lambda \right) \\
& & \hskip1cm = -\frac{1}{2} \, e^{-K_{cs}} \, \biggl[2\, s\, \,\left({\mathbb F}_\Lambda \, {\mathbb H}^\Lambda - {\mathbb F}^\Lambda \, {\mathbb H}_\Lambda \right) -2\, \left({\mathbb F}_\Lambda \, \hat{\mathbb Q}^{\Lambda} - {\mathbb F}^\Lambda \,\hat{\mathbb Q}_\Lambda \right) \biggr]\nonumber
\end{eqnarray}
where $\hat{\mathbb Q}_\Lambda:= \hat{\mathbb Q}^\alpha{}_\Lambda \sigma_\alpha$ and $\hat{\mathbb Q}^\Lambda:= \hat{\mathbb Q}^{\alpha{}\Lambda} \sigma_\alpha$ have been utilized in these expressions. The piece (\ref{eq:Vcs1b}) combines various NS-NS and RR-Bianchi identities and can be considered as generalized RR Tadpoles \cite{Gao:2015nra,Shukla:2015bca}, and these have to vanish by adding local sources. 

\subsubsection*{Rewriting $V_2$ using generalized flux orbits:}
As a next step, we consider the second piece $V_2$ which using the results of {\bf Part: 2} simplifies as under
\bea
\label{eq:V2xxx}
& & \hskip-1.1cm V_2: = V_{cs2} + \left(K^{{A} \ov {B}} \, K_A \, K_{\ov {B}} |W|^2 -\,3 |W|^2\right) + K^{{A} \ov {B}} \, \left( (K_A \, W) \, \ov{W}_{\ov B} \, + W_A\, (K_{\ov {B}} \ov W) \right)\nonumber\\
& & = (e_\Lambda - \ov e_\Lambda) Re({\cal X}^\Lambda \ov {\cal X}^\Delta) (e_\Delta - \ov e_\Delta) + (e_\Lambda - \ov e_\Lambda) Re({\cal X}^\Lambda \ov {\cal F}_\Delta) (m^\Delta - \ov m^\Delta) \nonumber\\
& & \hskip0.3cm (m^\Lambda - \ov m^\Lambda) Re({\cal F}_\Lambda \ov {\cal X}^\Delta) (e_\Delta - \ov e_\Delta) + (m^\Lambda - \ov m^\Lambda) Re({\cal F}_\Lambda \ov {\cal F}_\Delta) (m^\Delta - \ov m^\Delta)
\eea
Now, the eqn. (\ref{eq:V2xxx}) can be re-expressed using generalized flux orbits as, 
\begin{eqnarray}
\label{eq:V3aaa}
& & \hskip-0.65cm V_2 = 4\, \,Re({\cal X}^\Lambda \ov {\cal X}^\Delta) \biggl[-s^2 \, {\mathbb H}_\Lambda \, {\mathbb H}_\Delta +  \, s \, {\mathbb H}_\Lambda \, \hat{\mathbb Q}_\Delta +  \, s \, {\mathbb H}_\Delta \, \hat{\mathbb Q}_\Lambda - \hat{\mathbb Q}_\Lambda \, \hat{\mathbb Q}_\Delta \biggr] \nonumber\\
&& +4\, \,Re({\cal X}^\Lambda \ov {\cal F}_\Delta) \biggl[-s^2 \, {\mathbb H}_\Lambda \, {\mathbb H}^\Delta +  \, s \, {\mathbb H}_\Lambda \, \hat{\mathbb Q}^\Delta +  \, s \, {\mathbb H}^\Delta \, \hat{\mathbb Q}_\Lambda - \hat{\mathbb Q}_\Lambda \, \hat{\mathbb Q}^\Delta \biggr] \\
& & +4\, \,Re({\cal F}_\Lambda \ov {\cal X}^\Delta) \biggl[-s^2 \, {\mathbb H}^\Lambda \, {\mathbb H}_\Delta +  \, s \, {\mathbb H}^\Lambda \, \hat{\mathbb Q}_\Delta +  \, s \, {\mathbb H}_\Delta \, \hat{\mathbb Q}^\Lambda - \hat{\mathbb Q}^\Lambda \, \hat{\mathbb Q}_\Delta \biggr] \nonumber\\
& & + 4\, \,Re({\cal F}_\Lambda \ov {\cal F}_\Delta) \biggl[-s^2 \, {\mathbb H}^\Lambda \, {\mathbb H}^\Delta +  \, s \, {\mathbb H}^\Lambda \, \hat{\mathbb Q}^\Delta +  \, s \, {\mathbb H}^\Delta \, \hat{\mathbb Q}^\Lambda - \hat{\mathbb Q}^\Lambda \, \hat{\mathbb Q}^\Delta \biggr] \nonumber
\end{eqnarray}
\subsubsection*{Rewriting $V_3$ using generalized flux orbits:}
We consider the third collection of piece $V_3$ which is given as:
\begin{eqnarray}
& & \hskip-1.5cm V_3 := K^{{A} \ov {B}} \, W_A \, \ov{W}_{\ov B} = {\cal X}^\Lambda \ov {\cal X}^\Delta \biggl[ H_\Lambda \, K^{\tau\ov\tau} \, H_\Delta +  \omega_{a \Lambda} \, K^{G^a\ov G^b} \, \omega_{b \Delta} + \hat{Q}^\alpha{}_{\Lambda} \, K^{T_\alpha\ov T_\beta} \, \hat{Q}^\beta{}_{\Delta}  \nonumber\\
& & \hskip1cm +   H_\Lambda \, K^{\tau\ov G^a} \, \omega_{a \Delta} + \omega_{a \Lambda} \, K^{G^a \ov\tau} \, H_\Delta +  H_\Lambda \, K^{\tau\ov T_\alpha}\, \hat{Q}^\alpha{}_\Delta + \hat{Q}^\alpha{}_\Lambda \, K^{T_\alpha\ov \tau} \, H_\Delta  \nonumber\\
&& \hskip2cm +\hat{Q}^\alpha{}_\Lambda \, K^{T_\alpha\ov G^a} \, \omega_{a \Delta}  +  \omega_{a \Lambda} \, K^{G^a \ov T_\alpha} \, \hat{Q}^\alpha{}_\Delta \biggr] \\
& & \hskip-1cm + {\cal X}^\Lambda \ov {\cal F}_\Delta [---]+ {\cal F}_\Lambda \ov {\cal X}^\Delta [----] +{\cal F}_\Lambda \ov {\cal F}_\Delta [---]\nonumber
\end{eqnarray}
Using generalized flux combinations and K\"ahler metric components, we find the following rearrangement in $V_3$,
\begin{eqnarray}
\label{eq:V3a}
& & \hskip-0.65cm V_3 = 4\, \,Re({\cal X}^\Lambda \ov {\cal X}^\Delta) \biggl[s^2 \, {\mathbb H}_\Lambda \, {\mathbb H}_\Delta + \frac{s}{4}  \, {\mho}_{\Lambda a}\, {\cal G}^{ab}\, {\mho}_{b\,\Delta} + \frac{k_0^2}{9}\,\, \hat{\mathbb Q}_\Lambda^\alpha  \, \tilde{\cal G}_{\alpha \beta}\, \hat{\mathbb Q}^\beta{}_\Delta \biggr] \nonumber\\
&& +4\, \,Re({\cal X}^\Lambda \ov {\cal F}_\Delta) \biggl[s^2 \,{\mathbb H}_\Lambda \, {\mathbb H}^\Delta + \frac{s}{4}  \, {\mho}_{\Lambda a}\, {\cal G}^{ab}\,{\mho}_b{}^{\Delta} + \frac{k_0^2}{9}\,\hat{\mathbb Q}_\Lambda^\alpha  \, \tilde{\cal G}_{\alpha \beta}\, \hat{\mathbb Q}^{\beta\Delta} \biggr] \\
& & +4\, \,Re({\cal F}_\Lambda \ov {\cal X}^\Delta) \biggl[s^2 \,{\mathbb H}^\Lambda \, {\mathbb H}_\Delta + \frac{s}{4}\, {\mho}^{\Lambda}{}_{a} \, {\cal G}^{ab}\, {\mho}_{b\, \Delta} + \frac{k_0^2}{9}\,\hat{\mathbb Q}^{\Lambda\alpha}  \, \tilde{\cal G}_{\alpha \beta} \, \hat{\mathbb Q}^\beta{}_\Delta \biggr] \nonumber\\
& & + 4\, \,Re({\cal F}_\Lambda \ov {\cal F}_\Delta) \biggl[s^2 \,{\mathbb H}^\Lambda \, {\mathbb H}^\Delta + \frac{s}{4} \, {\mho}^{\Lambda}{}_{a}\, {\cal G}^{ab} \, {\mho}_b{}^{\Delta} + \frac{k_0^2}{9}\, \hat{\mathbb Q}^{\Lambda\alpha}  \, \tilde{\cal G}_{\alpha \beta} \, \hat{\mathbb Q}^{\beta\Delta} \biggr] \nonumber
\end{eqnarray}
One should observe that in the case when only $H_3$ and $F_3$ fluxes are present, the whole contribution of $|H|^2$ type is already embedded into $V_{cs1}$ \cite{Taylor:1999ii, Blumenhagen:2003vr}, and hence a cancellation of pure ${\mathbb H}$-flux pieces of $V_2$ and $V_3$ is anticipated, and we have
\begin{eqnarray}
\label{eq:V2+V3}
& & \hskip-0.90cm V_2+V_3 = 4\,Re({\cal X}^\Lambda \ov {\cal X}^\Delta) \biggl[2 \, s {\mathbb H}_\Lambda \, \hat{\mathbb Q}_\Delta  + \frac{s}{4}  \, {\mho}_{\Lambda a}\, {\cal G}^{ab}\, {\mho}_{b\,\Delta} + \frac{1 }{4}\, \, \, \hat{\mathbb Q}_\Lambda^\alpha \, \hat{\mathbb Q}^\beta{}_\Delta \left(\frac{4 \, k_0^2}{9} \, \tilde{\cal G}_{\alpha \beta}\,   - 4\, \sigma_\alpha \, \sigma_\beta \right)\biggr] \nonumber\\
& & \hskip-0.50cm+4\, \,Re({\cal X}^\Lambda \ov {\cal F}_\Delta) \biggl[2 \, s \, {\mathbb H}_\Lambda \, \hat{\mathbb Q}^\Delta + \frac{s}{4}  \, {\mho}_{\Lambda a}\, {\cal G}^{ab}\,{\mho}_b{}^{\Delta} + \frac{1}{4}\, \, \, \hat{\mathbb Q}_\Lambda^\alpha \, \hat{\mathbb Q}^{\beta{}\Delta} \left(\frac{4 \, k_0^2}{9} \, \tilde{\cal G}_{\alpha \beta}\,   - 4\, \sigma_\alpha \, \sigma_\beta \right)\biggr] \nonumber\\
& & \hskip-0.50cm +4\, \,Re({\cal F}_\Lambda \ov {\cal X}^\Delta) \biggl[2 \, s \, {\mathbb H}^\Lambda \, \hat{\mathbb Q}_\Delta  + \frac{s}{4}\, {\mho}^{\Lambda}{}_{a} \, {\cal G}^{ab}\, {\mho}_{b\, \Delta} + \frac{1}{4}\, \, \, \hat{\mathbb Q}^{\alpha{}{\Lambda}} \, \hat{\mathbb Q}^\beta{}_\Delta \left(\frac{4 \, k_0^2}{9} \, \tilde{\cal G}_{\alpha \beta}\,   - 4\, \sigma_\alpha \, \sigma_\beta \right) \biggr] \, \\
& &\hskip-0.50cm + 4\, \,Re({\cal F}_\Lambda \ov {\cal F}_\Delta) \biggl[2 \, s \, {\mathbb H}^\Lambda \, \hat{\mathbb Q}^\Delta  + \frac{s}{4} \, {\mho}^{\Lambda}{}_{a}\, {\cal G}^{ab} \, {\mho}_b{}^{\Delta} + \frac{1}{4}\, \, \, \hat{\mathbb Q}^{\alpha{}\Lambda} \, \hat{\mathbb Q}^{\beta{}\Delta} \left(\frac{4\, k_0^2}{9} \, \tilde{\cal G}_{\alpha \beta}\,   - 4\, \sigma_\alpha \, \sigma_\beta \right) \biggr] \nonumber
\end{eqnarray}
\subsection{Summary of first rearrangement} 
As a summary at this stage, we have the following scalar potential rearrangement, 
\bea
\label{eq:collection1}
& & \hskip-0.5cm V_{{\mathbb F} {\mathbb F}} = -\frac{1}{2} \, e^{K-K_{cs}} \,\biggl[{\mathbb F}_\Lambda \, {\cal M}^{\Lambda \Delta} \,  {\mathbb F}_\Delta - {\mathbb F}_\Lambda \, {\cal M}^{\Lambda}_{\, \, \, \Delta} \, {\mathbb F}^\Delta+  {\mathbb F}^\Lambda \, {\cal M}_{\Lambda}^{\, \, \, \Delta} \, {\mathbb F}_\Delta- {\mathbb F}^\Lambda \, {\cal M}_{\Lambda \Delta} \, {\mathbb F}^\Delta \biggr] \\
& & \hskip-0.5cm V_{{\mathbb H} {\mathbb H}} = -\frac{1}{2} \, e^{K-K_{cs}} \, \biggl[s^2 \, \, \left({\mathbb H}_\Lambda \, {\cal M}^{\Lambda \Delta} \,  {\mathbb H}_\Delta - {\mathbb H}_\Lambda \, {\cal M}^{\Lambda}_{\, \, \, \Delta} \, {\mathbb H}^\Delta+  {\mathbb H}^\Lambda \, {\cal M}_{\Lambda}^{\, \, \, \Delta} \, {\mathbb H}_\Delta- {\mathbb H}^\Lambda \, {\cal M}_{\Lambda \Delta} \, {\mathbb H}^\Delta\right)\biggr]\nonumber\\
& & \hskip-0.5cm V_{{\mathbb F} {\mathbb H}} = -\frac{1}{2} \, e^{K-K_{cs}} \, \biggl[(-2)\, s\, \,\left({\mathbb F}_\Lambda \, {\mathbb H}^\Lambda - {\mathbb F}^\Lambda \, {\mathbb H}_\Lambda \right) \biggr]\nonumber\\
& & \hskip-0.5cm V_{{\mathbb F} {\mathbb Q}} = -\frac{1}{2} \, e^{K-K_{cs}} \, \biggl[(+2)\, \left({\mathbb F}_\Lambda \, \hat{\mathbb Q}^{\Lambda} - {\mathbb F}^\Lambda \,\hat{\mathbb Q}_\Lambda \right) \biggr] \nonumber\\
& & \hskip-0.5cm V_{{\mathbb H} \hat{\mathbb Q}} = -\frac{1}{2} e^{K-K_{cs}} \,\bigg[ (-2) \, s \, \biggl\{ \left({\mathbb H}_\Lambda {\cal M}^{\Lambda \Delta} \,  \hat{\mathbb Q}_\Delta - {\mathbb H}_\Lambda \, {\cal M}^{\Lambda}_{\, \, \Delta} \hat{\mathbb Q}^\Delta+  {\mathbb H}^\Lambda  {\cal M}_{\Lambda}^{\, \, \Delta} \, \hat{\mathbb Q}_\Delta- {\mathbb H}^\Lambda \, {\cal M}_{\Lambda \Delta} \, \hat{\mathbb Q}^\Delta\right) \nonumber\\
& & \hskip2.5cm +\, 8 \, \,e^{K_{cs}} \, \biggl({\mathbb H}_\Lambda \, Re({\cal X}^\Lambda \ov {\cal X}^\Delta) \, \hat{\mathbb Q}_\Delta  + \, {\mathbb H}_\Lambda \, Re({\cal X}^\Lambda \ov {\cal F}_\Delta) \, \hat{\mathbb Q}^\Delta \nonumber\\
& & \hskip3.0cm +\, {\mathbb H}^\Lambda \, Re({\cal F}_\Lambda \ov {\cal X}^\Delta) \, \hat{\mathbb Q}_\Delta  + \, {\mathbb H}^\Lambda \, Re({\cal F}_\Lambda \ov {\cal F}_\Delta) \, \hat{\mathbb Q}^\Delta \biggr) \biggr\}  \biggr] \nonumber\\
& & \hskip-0.5cm V_{{\mathbb \mho} {\mathbb \mho}} = -\frac{1}{2} \, e^{K-K_{cs}} \,\bigg[ (-2)\, s\, e^{K_{cs}}\, {\cal G}^{ab}\,\, \biggl\{ {\mho}_{\Lambda a}\, Re({\cal X}^\Lambda \ov {\cal X}^\Delta) \, {\mho}_{b\,\Delta} +\, {\mho}_{\Lambda a}\, \,Re({\cal X}^\Lambda \ov {\cal F}_\Delta)\,{\mho}_b{}^{\Delta} \nonumber\\
& & \hskip2.5cm + \, {\mho}^{\Lambda}{}_{a} \,Re({\cal F}_\Lambda \ov {\cal X}^\Delta) \, {\mho}_{b\, \Delta}  +  \, {\mho}^{\Lambda}{}_{a}\, Re({\cal F}_\Lambda \ov {\cal F}_\Delta) \, {\mho}_b{}^{\Delta} \biggr\} \biggr]\nonumber\\
& & \hskip-0.5cm V_{\hat{\mathbb Q} \hat{\mathbb Q}}=  -\frac{1}{2} \, e^{K-K_{cs}} \, \biggl[\left(\hat{\mathbb Q}_\Lambda \, {\cal M}^{\Lambda \Delta} \,  \hat{\mathbb Q}_\Delta - \hat{\mathbb Q}_\Lambda \, {\cal M}^{\Lambda}_{\, \, \, \Delta} \, \hat{\mathbb Q}^\Delta+  \hat{\mathbb Q}^\Lambda \, {\cal M}_{\Lambda}^{\, \, \, \Delta} \, \hat{\mathbb Q}_\Delta- \hat{\mathbb Q}^\Lambda \, {\cal M}_{\Lambda \Delta} \, \hat{\mathbb Q}^\Delta\right) \nonumber\\
& & \hskip0.75cm -\, 2 \, e^{K_{cs}} \, \left(\frac{4\, k_0^2}{9} \, \tilde{\cal G}_{\alpha \beta}\,   - 4\, \sigma_\alpha \, \sigma_\beta \right) \biggl\{ \hat{\mathbb Q}_\Lambda^\alpha \, Re({\cal X}^\Lambda \ov {\cal X}^\Delta) \, \hat{\mathbb Q}^\beta{}_\Delta + \hat{\mathbb Q}_\Lambda^\alpha \, Re({\cal X}^\Lambda \ov {\cal F}_\Delta)\, \hat{\mathbb Q}^{\beta{}\Delta} \nonumber\\
& & \hskip2.5cm + \hat{\mathbb Q}^{\alpha{}{\Lambda}} \, Re({\cal F}_\Lambda \ov {\cal X}^\Delta)\, \hat{\mathbb Q}^\beta{}_\Delta   +  \hat{\mathbb Q}^{\alpha{}\Lambda} \, Re({\cal F}_\Lambda \ov {\cal F}_\Delta)\, \hat{\mathbb Q}^{\beta{}\Delta} \biggr\} \biggr] \,. \nonumber
\eea

\section{Rearrangement of scalar potential: Step 2}
\label{sec_using-symplecticIdentities}
In this section we will provide three sets of equivalent symplectic representations of the $F$-term scalar potential taking a next step to our first rearrangement in eqn. (\ref{eq:collection1}). For that purpose, let us first present a couple of very important symplectic identities.

\subsection{Invoking a  set of important symplectic identities}
We find that the following interesting and very analogous relation as compared to the definition of period matrix (\ref{eq:periodN}) holds, 
\bea
\label{eq:periodF}
& & {\cal F}_{\Lambda\Delta} = \ov{\cal N}_{\Lambda\Delta} + 2 \, i \, \frac{Im({\cal N}_{\Lambda\Gamma}) \, {\cal X}^\Gamma X^\Sigma \, (Im{\cal N}_{\Sigma \Delta}) }{Im({\cal N}_{\Gamma\Sigma}) {\cal X}^\Gamma X^\Sigma}
\eea
Moreover, similar to the definition of the period matrices (\ref{coff}), one can define another set of symplectic quantities as under,
\begin{eqnarray}
\label{coff2}
&& {\cal L}^{\Lambda \Delta} = {\rm Im{\cal F}}^{\Lambda \Delta} \nonumber\\
&& {\cal L}_{\Lambda}^{\, \, \, \, \, \Delta}  = {\rm Re{\cal F}}_{\Lambda \Gamma} \, \, {\rm Im{\cal F}}^{\Gamma \Delta} \\
&& {\cal L}^\Lambda_{\, \, \, \, \Delta} =- \left({\cal L}_{\Lambda}^{\, \, \, \, \, \Delta}\right)^{T} \nonumber\\
&& {\cal L}_{\Lambda \Delta}\, =  -{\rm Im{\cal F}}_{\Lambda \Delta} -{\rm Re{\cal F}}_{\Lambda \Sigma} \, \, {\rm Im{\cal F}}^{\Sigma \Gamma}\, \, {\rm Re{\cal F}}_{\Gamma \Delta} \nonumber
\end{eqnarray}
Now we will use these two sets of matrices ${\cal M}$ and ${\cal L}$ as building blocks and will define some new combinations of the same which will be useful for our scalar potential rearrangement purpose. In this context, we define three new sets of symplectic quantities ${\cal M}_1$, ${\cal M}_2$ and ${\cal M}_3$ as under,
\begin{eqnarray}
\label{coff3}
&& {{\cal M}_1}^{\Lambda \Delta} = {\cal M}^{\Lambda \Delta} + \, \, {\cal L}^{\Lambda \Delta} \nonumber\\
&& {{\cal M}_1}_{\Lambda}^{\, \, \, \, \, \Delta}  = {\cal M}_{\Lambda}^{\, \, \, \, \, \Delta} +  \, \, {\cal L}_{\Lambda}^{\, \, \, \, \, \Delta} \\
&& {{\cal M}_1}^\Lambda_{\, \, \, \, \Delta} = {\cal M}^\Lambda_{\, \, \, \, \Delta}  +  \, \,  {\cal L}^\Lambda_{\, \, \, \, \Delta} \nonumber\\
&& {{\cal M}_1}_{\Lambda \Delta}\, = {\cal M}_{\Lambda \Delta} + \, \, {\cal L}_{\Lambda \Delta} \nonumber
\end{eqnarray}
\begin{eqnarray}
\label{coff4}
&& {{\cal M}_2}^{\Lambda \Delta} =-\left({\cal M}^{\Lambda \Delta} + 2 \, \, {\cal L}^{\Lambda \Delta}\right) \nonumber\\
&& {{\cal M}_2}_{\Lambda}^{\, \, \, \, \, \Delta}  = -\left({\cal M}_{\Lambda}^{\, \, \, \, \, \Delta} + 2 \, \, {\cal L}_{\Lambda}^{\, \, \, \, \, \Delta}\right) \\
&& {{\cal M}_2}^\Lambda_{\, \, \, \, \Delta} =- \left( {\cal M}^\Lambda_{\, \, \, \, \Delta}  + 2 \, \,  {\cal L}^\Lambda_{\, \, \, \, \Delta} \right) \nonumber\\
&& {{\cal M}_2}_{\Lambda \Delta}\, =  -\left({\cal M}_{\Lambda \Delta} + 2 \, \, {\cal L}_{\Lambda \Delta} \right)\nonumber
\end{eqnarray}
and 
\begin{eqnarray}
\label{coff5}
&& {{\cal M}_3}^{\Lambda \Delta} = + \left({\cal M}^{\Lambda}{}_{ \Sigma} \, {\cal L}^{\Sigma \Delta} + {\cal M}^{\Lambda \Sigma} \, {\cal L}_{\Sigma}{}^{ \Delta} \right) \nonumber\\
&& {{\cal M}_3}_\Lambda^{\, \, \, \, \Delta} =+ \left({\cal M}_{\Lambda}{}_{ \Sigma} \, {\cal L}^{\Sigma \Delta} + {\cal M}_{\Lambda}{}^{ \Sigma} \, {\cal L}_{\Sigma}{}^{ \Delta} \right) -  {{\delta}}_\Lambda{}^{\Delta} \nonumber\\
&& {{\cal M}_3}^{\Lambda}{}_{ \Delta}  = -\left({\cal M}^{\Lambda}{}_{ \Sigma} \, {\cal L}^{\Sigma}{}_\Delta + {\cal M}^{\Lambda \Sigma} \, {\cal L}_{\Sigma \Delta}\right) + {{\delta}}^\Lambda{}_{\Delta}\\
&& {{\cal M}_3}_{\Lambda \Delta}\, =  -\left({\cal M}_{\Lambda \Sigma} \, {\cal L}^{\Sigma}{}_{ \Delta} + {\cal M}_{\Lambda}{}^{\Sigma} \, {\cal L}_{\Sigma \Delta} \right)\nonumber
\end{eqnarray}
Apart from these defining equations (\ref{coff3}-\ref{coff5}), there are some more relations among ${\cal M}_1, {\cal M}_2$ and ${\cal M}_3$ which we will present in the appendix \ref{sec_symplectic}. Now the most important relation which will serve as a bridging segment for the symplectic rearrangement of the scalar potential is given as under \footnote{The first equation of (\ref{eq:symp1}) can be also obtained by comparing eqns. (11) and (27) of \cite{Ceresole:1995ca}, and this has  motivated us to define what we call our ${\cal L}$ matrices and invoke for its three other components.},
\begin{eqnarray}
\label{eq:symp1}
& & 4 \, e^{K_{cs}} \, Re({\cal X}^\Lambda \ov {\cal X}^\Delta) = - \, {{\cal M}_1}^{\Lambda \Delta} \nonumber\\
& & 4 \, e^{K_{cs}} \, Re({\cal F}_\Lambda \ov {\cal X}^\Delta) = - \, {{\cal M}_1}_{\Lambda}{}^ \Delta \\
& & 4 \, e^{K_{cs}} \, Re({\cal X}^\Lambda \ov {\cal F}_\Delta) = + \, {{\cal M}_1}^{\Lambda}{}_{ \Delta} \nonumber\\
& & 4 \, e^{K_{cs}} \, Re({\cal F}_\Lambda \ov {\cal F}_\Delta) = + \, {{\cal M}_1}_{\Lambda \Delta} \nonumber
\end{eqnarray}
We have checked eqn. (\ref{eq:symp1}) for $h^{2,1}_-(CY)= 0, 1, 2$ and $3$ using pre-potential given in eqn. (\ref{eq:prepotential}). As the computations involve inverting complicated  matrices of order $(h^{2,1}_- +1)$, for $h^{2,1}_-(CY)\ge4$, it gets too huge to verify the identities, however  we expect the same to be generically true for an arbitrary number of complex structure moduli.

\subsection{Symplectic rearrangements}
As we have many symplectic identities with many quantities such as ${\cal M}, {\cal L}$ and ${\cal M}_i$'s, this will result in more than one equivalent rearrangements of the scalar potential. In order to prefer one over the other,  let us try to figure out some points as guidelines for our rearrangement,
\begin{itemize}
\item{Considering the moduli space metrices given in eqn.(\ref{eq:genMetrices}) we find that, 
\bea
\label{eq:metricrelationsNew}
& & {\cal G}^{ab} = -\frac{2}{3} \, k_0\, \hat{\kappa}^{ab} = - 4 \, {\cal V}_E \, \hat{k}^{ab} \\
& & \hskip-1.11cm \left(\frac{4\, k_0^2}{9} \, \tilde{\cal G}_{\alpha \beta}\,   - 4\, \sigma_\alpha \, \sigma_\beta \right) = -\frac{2}{3} \, k_0 \, ({\hat{d}}^{-1})_\alpha{}^{\alpha'} \, {k}_{\alpha'\beta'} \, ({\hat{d}}^{-1})_\beta{}^{\beta'}= - 4 \, {\cal V}_E \, ({\hat{d}}^{-1})_\alpha{}^{\alpha'} \, {k}_{\alpha'\beta'} \, ({\hat{d}}^{-1})_\beta{}^{\beta'} \nonumber
\eea
where $k_0 = 6 \, {\cal V}_E$ has been used. From eqn. (\ref{eq:collection1}), this shows that coefficient of the pieces with $Re({\cal X}^\Lambda \ov {\cal X}^\Delta)$, $Re({\cal X}^\Lambda \ov {\cal F}_\Delta)$  etc. in both $V_{\mathbb \mho \mathbb \mho}$ as well as $V_{\hat{\mathbb Q} \hat{\mathbb Q}}$ are similar, and so may be clubbed in a  similar manner in the rearrangement.}
\item{Apart from the first choice mention above, we also observe that  $(\sigma_\alpha \sigma_\beta)$ contributions in $\mathbb Q \mathbb Q$ piece via $\left(\frac{4\, k_0^2}{9} \, \tilde{\cal G}_{\alpha \beta}\,   - 4\, \sigma_\alpha \, \sigma_\beta \right)$ factor can be clubbed with the other one to look similar as $\left({\cal M}^{\Lambda \Delta} \, +  8 \, \,e^{K_{cs}} \, Re({\cal X}^\Lambda \ov {\cal X}^\Delta) \right)$ etc. which is similar to the only piece of ${\mathbb H}\hat{\mathbb Q}$ type. This appears to be a better one as then, one piece of both of $\mathbb Q \mathbb Q$ and $\mho\mho$ can be written with $h^{1,1}$ flux-indices being contracted by the even/odd sector metrics $\tilde{\cal G}_{\alpha\beta}$ and ${\cal G}^{ab}$. }
\end{itemize}
Here we note that the first four pieces of collection (\ref{eq:collection1}) are already in desired form as those are already written in the `suitable'  symplectic forms as we will see later. 
Now using relations (\ref{coff}), (\ref{coff3})-(\ref{coff5}), and considering the points above we can rearrange the $V_{{\mathbb H} \hat{\mathbb Q}}$, $V_{{\mathbb \mho} {\mathbb \mho}}$  and $V_{{\mathbb Q} \hat{\mathbb Q}} $ pieces of eqn. (\ref{eq:collection1}) in the following three representations,
\subsubsection*{Representation-I: Using ${\cal M}$ and ${\cal M}_1$ matrices}
\bea
\label{eq:rep1}
& & \hskip-0.5cm V_{{\mathbb H} \hat{\mathbb Q}} = -\frac{1}{2} \, e^{K-K_{cs}} \,\bigg[ (-2) \, s \, \, \biggl({\mathbb H}_\Lambda \, \left({{\cal M}}^{\Lambda \Delta}-2\, {{\cal M}_1}^{\Lambda \Delta} \right) \,  \hat{\mathbb Q}_\Delta - {\mathbb H}_\Lambda \, \left({{\cal M}}^{\Lambda}_{\, \, \, \Delta}-2\, {{\cal M}_1}^{\Lambda}_{\, \, \, \Delta}\right) \, \hat{\mathbb Q}^\Delta \nonumber\\
& & \hskip2.5cm +  {\mathbb H}^\Lambda \, \left({{\cal M}}_{\Lambda}^{\, \, \, \Delta}-2\, {{\cal M}_1}_{\Lambda}^{\, \, \, \Delta}\right) \, \hat{\mathbb Q}_\Delta- {\mathbb H}^\Lambda \, \left({{\cal M}}_{\Lambda \Delta}-2\, {{\cal M}_1}_{\Lambda \Delta}\right) \, \hat{\mathbb Q}^\Delta \biggr) \biggr]\nonumber\\
& & \hskip-0.5cm V_{{\mathbb \mho} {\mathbb \mho}} = -\frac{1}{2} \, e^{K-K_{cs}} \,\biggl[ (-2 \, {\cal V}_E \, s )\, \hat{k}^{ab}\, \biggl({\mathbb \mho}_{a\Lambda} \, {{\cal M}_1}^{\Lambda \Delta} \,  {\mathbb \mho}_{b \Delta} - {\mathbb \mho}_{a \Lambda} \, {{\cal M}_1}^{\Lambda}_{\, \, \, \Delta} \, {\mathbb \mho}_b{}^\Delta \, \\
& & \hskip5.0cm +  {\mathbb \mho}_a{}^\Lambda \, {{\cal M}_1}_{\Lambda}^{\, \, \, \Delta} \, {\mathbb \mho}_{b \Delta}- {\mathbb \mho}_a{}^\Lambda \, {{\cal M}_1}_{\Lambda \Delta} \,{\mathbb \mho}_b{}^\Delta \biggr) \biggr]\nonumber\\
& & \hskip-0.5cm V_{\hat{\mathbb Q} \hat{\mathbb Q}}=  -\frac{1}{2} \, e^{K-K_{cs}} \, \biggl[\left(-2 \, {\cal V}_E \, ({\hat{d}^{-1}})_{\alpha}{}^{\alpha'}\, k_{\alpha' \beta'}\,  ({\hat{d}^{-1}})_{\beta}{}^{\beta'}\right) \, \biggl(\hat{\mathbb Q}^\alpha{}_\Lambda \, {{\cal M}_1}^{\Lambda \Delta} \,  \hat{\mathbb Q}^\beta{}_\Delta - \hat{\mathbb Q}^\alpha{}_\Lambda \, {{\cal M}_1}^{\Lambda}_{\, \, \, \Delta} \, \hat{\mathbb Q}^{\beta\Delta}\nonumber\\
& & \hskip5.0cm +  \hat{\mathbb Q}^{\alpha\Lambda} \, {{\cal M}_1}_{\Lambda}^{\, \, \, \Delta} \, \hat{\mathbb Q}^\beta{}_\Delta- \hat{\mathbb Q}^{\alpha\Lambda} \, {{\cal M}_1}_{\Lambda \Delta} \, \hat{\mathbb Q}^{\beta\Delta}\biggr) \biggr]  \nonumber\\
& & \hskip0.75cm -\frac{1}{2} \, e^{K-K_{cs}} \, \biggl[\hat{\mathbb Q}_\Lambda \, {{\cal M}}^{\Lambda \Delta} \,  \hat{\mathbb Q}_\Delta - \hat{\mathbb Q}_\Lambda \, {{\cal M}}^{\Lambda}_{\, \, \, \Delta} \, \hat{\mathbb Q}^\Delta+  \hat{\mathbb Q}^\Lambda \, {{\cal M}}_{\Lambda}^{\, \, \, \Delta} \, \hat{\mathbb Q}_\Delta- \hat{\mathbb Q}^\Lambda \, {{\cal M}}_{\Lambda \Delta} \, \hat{\mathbb Q}^\Delta\biggr]\, . \nonumber
\eea
\subsubsection*{Representation-II: Using ${\cal M}$ and ${\cal M}_2$ matrices}
\bea
\label{eq:rep2}
& & \hskip-0.5cm V_{{\mathbb H} \hat{\mathbb Q}} = -\frac{1}{2} \, e^{K-K_{cs}} \,\bigg[ (-2) \, s \, \, \biggl({\mathbb H}_\Lambda \, {{\cal M}_2}^{\Lambda \Delta} \,  \hat{\mathbb Q}_\Delta - {\mathbb H}_\Lambda \, {{\cal M}_2}^{\Lambda}_{\, \, \, \Delta} \, \hat{\mathbb Q}^\Delta \nonumber\\
& & \hskip5.0cm +  {\mathbb H}^\Lambda \, {{\cal M}_2}_{\Lambda}^{\, \, \, \Delta} \, \hat{\mathbb Q}_\Delta- {\mathbb H}^\Lambda \, {{\cal M}_2}_{\Lambda \Delta} \, \hat{\mathbb Q}^\Delta \biggr) \biggr]\nonumber\\
& & \hskip-0.5cm V_{{\mathbb \mho} {\mathbb \mho}} = -\frac{1}{2} \, e^{K-K_{cs}} \,\biggl[ \frac{1}{4} \, s \times\, \, {\cal G}^{ab}\,\, \biggl({\mathbb \mho}_{a\Lambda} \, {{\cal M}}^{\Lambda \Delta} \,  {\mathbb \mho}_{b \Delta} - {\mathbb \mho}_{a \Lambda} \, {{\cal M}}^{\Lambda}_{\, \, \, \Delta} \, {\mathbb \mho}_b{}^\Delta \, \nonumber\\
& & \hskip5.0cm +  {\mathbb \mho}_a{}^\Lambda \, {{\cal M}}_{\Lambda}^{\, \, \, \Delta} \, {\mathbb \mho}_{b \Delta}- {\mathbb \mho}_a{}^\Lambda \, {{\cal M}}_{\Lambda \Delta} \,{\mathbb \mho}_b{}^\Delta \biggr) \biggr]\nonumber\\
& & \hskip0.5cm -\frac{1}{2} \, e^{K-K_{cs}} \,\biggl[ (s \,{\cal V}_E\, \hat{k}^{ab}) \times\, \biggl({\mathbb \mho}_{a\Lambda} \, {{\cal M}_2}^{\Lambda \Delta} \,  {\mathbb \mho}_{b \Delta} - {\mathbb \mho}_{a \Lambda} \, {{\cal M}_2}^{\Lambda}_{\, \, \, \Delta} \, {\mathbb \mho}_b{}^\Delta \, \\
& & \hskip5.0cm +  {\mathbb \mho}_a{}^\Lambda \, {{\cal M}_2}_{\Lambda}^{\, \, \, \Delta} \, {\mathbb \mho}_{b \Delta}- {\mathbb \mho}_a{}^\Lambda \, {{\cal M}_2}_{\Lambda \Delta} \,{\mathbb \mho}_b{}^\Delta \biggr) \biggr]\nonumber\\
& & \hskip-0.5cm V_{\hat{\mathbb Q} \hat{\mathbb Q}}=  -\frac{1}{2} \, e^{K-K_{cs}} \, \biggl[\frac{1}{4} \, \left(\frac{4\, k_0^2}{9} \, \tilde{\cal G}_{\alpha \beta}\,  \right) \times \biggl(\hat{\mathbb Q}^\alpha{}_\Lambda \, {{\cal M}}^{\Lambda \Delta} \,  \hat{\mathbb Q}^\beta{}_\Delta - \hat{\mathbb Q}^\alpha{}_\Lambda \, {{\cal M}}^{\Lambda}_{\, \, \, \Delta} \, \hat{\mathbb Q}^{\beta\Delta}\nonumber\\
& & \hskip5.0cm +  \hat{\mathbb Q}^{\alpha\Lambda} \, {{\cal M}}_{\Lambda}^{\, \, \, \Delta} \, \hat{\mathbb Q}^\beta{}_\Delta- \hat{\mathbb Q}^{\alpha\Lambda} \, {{\cal M}}_{\Lambda \Delta} \, \hat{\mathbb Q}^{\beta\Delta}\biggr) \biggr]  \nonumber\\
& & \hskip0.75cm -\frac{1}{2} \, e^{K-K_{cs}} \, \biggl[\left({\cal V}_E \, ({\hat{d}^{-1}})_{\alpha}{}^{\alpha'}\, k_{\alpha' \beta'}\,  ({\hat{d}^{-1}})_{\beta}{}^{\beta'} \right) \biggl(\hat{\mathbb Q}_\Lambda \, {{\cal M}_2}^{\Lambda \Delta} \,  \hat{\mathbb Q}_\Delta - \hat{\mathbb Q}_\Lambda \, {{\cal M}_2}^{\Lambda}_{\, \, \, \Delta} \, \hat{\mathbb Q}^\Delta\nonumber\\
& & \hskip3.5cm+  \hat{\mathbb Q}^\Lambda \, {{\cal M}_2}_{\Lambda}^{\, \, \, \Delta} \, \hat{\mathbb Q}_\Delta- \hat{\mathbb Q}^\Lambda \, {{\cal M}_2}_{\Lambda \Delta} \, \hat{\mathbb Q}^\Delta\biggr)\biggr]\,. \nonumber
\eea
\subsubsection*{Representation-III: Using ${\cal M}$ and ${\cal M}_3$ matrices}
\bea
\label{eq:rep3}
& & \hskip-0.50cm V_{{\mathbb H} \hat{\mathbb Q}} = -\frac{1}{2} \, e^{K-K_{cs}} \,\bigg[ (-2) \, s \, \, \biggl\{ \left({\mathbb H}_\Lambda \, {\cal M}^{\Lambda \Delta} \,  \hat{\mathbb Q}_\Delta - {\mathbb H}_\Lambda \, {\cal M}^{\Lambda}_{\, \, \, \Delta} \, \hat{\mathbb Q}^\Delta+  {\mathbb H}^\Lambda \, {\cal M}_{\Lambda}^{\, \, \, \Delta} \, \hat{\mathbb Q}_\Delta- {\mathbb H}^\Lambda \, {\cal M}_{\Lambda \Delta} \, \hat{\mathbb Q}^\Delta\right) \nonumber\\
& & \hskip2.5cm +\, 2\, \left({\mathbb H}_\Lambda \, {\cal M}^{\Lambda \Delta} \,  \tilde{{\cal Q}}_\Delta - {\mathbb H}_\Lambda \, {\cal M}^{\Lambda}_{\, \, \, \Delta} \, \tilde{{\cal Q}}^\Delta+  {\mathbb H}^\Lambda \, {\cal M}_{\Lambda}^{\, \, \, \Delta} \, \tilde{{\cal Q}}_\Delta- {\mathbb H}^\Lambda \, {\cal M}_{\Lambda \Delta} \, \tilde{{\cal Q}}^\Delta\right) \biggr\}  \biggr] \nonumber\\
& & \hskip-0.50cm V_{{\mathbb \mho} {\mathbb \mho}} = -\frac{1}{2} \, e^{K-K_{cs}} \,\biggl[\frac{s}{4}\,{\cal G}^{ab}\,\, \biggl(\tilde{\mathbb \mho}_{a\Lambda} \, {\cal M}^{\Lambda \Delta} \,  \tilde{\mathbb \mho}_{b \Delta} - \tilde{\mathbb \mho}_{a \Lambda} \, {\cal M}^{\Lambda}_{\, \, \, \Delta} \, \tilde{\mathbb \mho}_b{}^\Delta\nonumber\\
& & \hskip3.5cm +  \tilde{\mathbb \mho}_a{}^\Lambda \, {\cal M}_{\Lambda}^{\, \, \, \Delta} \, \tilde{\mathbb \mho}_{b \Delta}- \tilde{\mathbb \mho}_a{}^\Lambda \, {\cal M}_{\Lambda \Delta} \, \tilde{\mathbb \mho}_b{}^\Delta\biggr) \biggr]\nonumber\\
& & \hskip-0.50cm V_{\hat{\mathbb Q} \hat{\mathbb Q}}=  -\frac{1}{2} \, e^{K-K_{cs}} \, \biggl[\left(\hat{\mathbb Q}_\Lambda \, {\cal M}^{\Lambda \Delta} \,  \hat{\mathbb Q}_\Delta - \hat{\mathbb Q}_\Lambda \, {\cal M}^{\Lambda}_{\, \, \, \Delta} \, \hat{\mathbb Q}^\Delta+  \hat{\mathbb Q}^\Lambda \, {\cal M}_{\Lambda}^{\, \, \, \Delta} \, \hat{\mathbb Q}_\Delta- \hat{\mathbb Q}^\Lambda \, {\cal M}_{\Lambda \Delta} \, \hat{\mathbb Q}^\Delta\right) \nonumber\\
& & \hskip0.75cm +\frac{1}{4} \, \left(\frac{4\, k_0^2}{9} \, \tilde{\cal G}_{\alpha \beta}\,   - 4\, \sigma_\alpha \, \sigma_\beta \right) \biggl(\tilde{\cal Q}^\alpha{}_\Lambda \, {\cal M}^{\Lambda \Delta} \,  \tilde{\cal Q}^\beta{}_\Delta - \tilde{\cal Q}^\alpha{}_\Lambda \, {\cal M}^{\Lambda}_{\, \, \, \Delta} \, \tilde{\cal Q}^{\beta\Delta}\nonumber\\
& & \hskip2cm+  \tilde{\cal Q}^{\alpha\Lambda} \, {\cal M}_{\Lambda}^{\, \, \, \Delta} \, \tilde{\cal Q}^\beta{}_\Delta- \tilde{\cal Q}^{\alpha\Lambda} \, {\cal M}_{\Lambda \Delta} \, \tilde{\cal Q}^{\beta\Delta}\biggr) \biggr] \,, \nonumber
\eea
where eqn. (\ref{eq:mainSymp}) has been utilized for this representation via defining $\tilde{\cal \mho}_a$ and $\tilde{\cal Q}^\alpha$ as under,

\begin{eqnarray}
 & & \tilde{\mathbb \mho}_a = -\left({{\cal M}_3}^{\Sigma\Delta} {\mathbb \mho}_{a\Delta} + {{\cal M}_3}^\Sigma{}_\Delta {\mathbb \mho}_a{}^\Delta\right) \alpha_\Sigma + \left({{\cal M}_3}_{\Sigma}{}^{\Delta} {\mathbb \mho}_{a \Delta} + {{\cal M}_3}_{\Sigma\Delta} {\mathbb \mho}_a{}^\Delta\right) \beta^\Sigma \\
& & \tilde{\cal Q}^\alpha = -\left({{\cal M}_3}^{\Sigma\Delta} \hat{\mathbb Q}^\alpha{}_\Delta + {{\cal M}_3}^\Sigma{}_\Delta \hat{\mathbb Q}^{\alpha\Delta}\right) \alpha_\Sigma + \left({{\cal M}_3}_{\Sigma}{}^{\Delta} \hat{\mathbb Q}^\alpha{}_\Delta + {{\cal M}_3}_{\Sigma\Delta} \hat{\mathbb Q}^{\alpha\Delta}\right) \beta^\Sigma \nonumber
\end{eqnarray}
{\it There is a bit of abuse of notation as different quantities are denoted with similar (however not the same) notations; e.g. ${\mathbb \mho}, \, \hat{\mathbb \mho}, \,  \tilde{\mathbb \mho}$ as well as $\tilde{\cal Q}$ and $\hat{\mathbb Q}$ are different.}

\subsection{Adding D-term contributions}
As we have mentioned earlier, if the choice of homolorphic involution is such that one can have $h^{2,1}_+(CY) \ne 0$, then additional contributions to the effective scalar potential are introduced via D-terms written in new generalized flux orbits as under \cite{Shukla:2015bca, Blumenhagen:2015lta},
\bea
& & \hskip-1.5cm  D_K  = \frac{1}{2\, s\,{\cal V}_E}\, \biggl[ f^{-1} R_K \, {\cal V}_E - s\, t^\alpha \, \hat{\mho}_{\alpha K} \biggr]\, , \, \,  D^K  = -\frac{1}{2\, s\,{\cal V}_E}\, \biggl[ f^{-1} R^K \, {\cal V}_E - s\, t^\alpha \,\hat{\mho}_{\alpha}{}^{K} \biggr] \, . \nonumber
\eea
Now similar to the period matrices ${\cal M}$ of involutively  odd (2,1)-cohomology sector, following from the underlying $N=2$ symplectic structure, one can introduce similar matrices for the disjoint even sector as under,
\begin{eqnarray}
\label{coffeven}
&& \hat{\cal M}^{JK} = {\rm Im\hat{\cal N}}^{JK}, \quad \hat{\cal M}_{J}^{\, \, \, \, \, K}  = {\rm Re\hat{\cal N}}_{JI} \, \, {\rm Im\hat{\cal N}}^{IK} \\
&& \hat{\cal M}^J_{\, \, \, \, K} =- \left(\hat{\cal M}_{J}^{\, \, \, \, \, K}\right)^{T} \nonumber\\
&& \hat{\cal M}_{JK}\, =  -{\rm Im\hat{\cal N}}_{JK} -{\rm Re\hat{\cal N}}_{JI} \, \, {\rm Im\hat{\cal N}}^{IL}\, \, {\rm Re\hat{\cal N}}_{LK} \nonumber
\end{eqnarray}
Also, as the gauge kinetic function $\hat{\cal G}$ are given as \cite{Benmachiche:2006df},
\bea
& & \hat{\cal G}_{JK} = \left(- \frac{i}{2} \ov{\hat{\cal N}_{JK}}\right)_{{\rm at }\, \, ({z^K = 0 = \ov z^K})},  \, 
\eea 
where $\hat{\cal N}$ is the period matrix on the even (2,1)-cohomology sector similar to (\ref{eq:periodN}). Using these ingredients one finds the $D$-term contributions to the four dimensional scalar potential as under,
\bea
& & V_D^{(1)} = V_{\hat{\mathbb \mho} \hat{\mathbb \mho}} + V_{{\mathbb R} \hat{\mathbb \mho}} + V_{{\mathbb R} {\mathbb R}}
\eea
where
\bea
\label{eq:collection2}
& & \hskip-0.90cm V_{\hat{\mathbb \mho} \hat{\mathbb \mho}} = -\frac{1}{2} \, e^{K-K_{cs}} \,\biggl[ \hat{\mathbb \mho}_{J} \, \hat{\cal M}^{JK} \,  \hat{\mathbb \mho}_{K} - \hat{\mathbb \mho}_{J} \, \hat{\cal M}^{J}_{\, \, \, K} \, \hat{\mathbb \mho}^K+  \hat{\mathbb \mho}^J \, \hat{\cal M}_{J}^{\, \, \, K} \, \hat{\mathbb \mho}_{K}- \hat{\mathbb \mho}^J \, \hat{\cal M}_{JK} \, \hat{\mathbb \mho}^K \biggr] \\
& & \hskip-0.90cm V_{{\mathbb R} \hat{\mathbb \mho}} = -\frac{1}{2} \, e^{K-K_{cs}} \,\biggl[ (-2{\cal V}_E) \, \, \biggl({\mathbb R}_J \, {\hat{\cal M}}^{JK} \,  \hat{\mathbb \mho}_K - {\mathbb R}_J \, {\hat{\cal M}}^{J}_{\, \, \, K} \, \hat{\mathbb \mho}^K   +  {\mathbb R}^J \, {\hat{\cal M}}_{J}^{\, \, \, K} \, \hat{\mathbb \mho}_K- {\mathbb R}^J \, {\hat{\cal M}}_{JK} \, \hat{\mathbb \mho}^K \biggr)  \biggr] \nonumber\\
& & \hskip-0.90cm V_{{\mathbb R} {\mathbb R}} = -\frac{1}{2} \, e^{K-K_{cs}} \,  \biggl[\frac{{\cal V}_E^2}{s} \times \, \, \left({\mathbb R}_J \, \hat{\cal M}^{\Lambda \Delta} \,  {\mathbb R}_K - {\mathbb R}_J \, \hat{\cal M}^{J}_{\, \, \, K} \, {\mathbb R}^K+  {\mathbb R}^J \, \hat{\cal M}_{J}^{\, \, \, K} \, {\mathbb R}_K- {\mathbb R}^J \, \hat{\cal M}_{JK} \, {\mathbb R}^K \right)\biggr] \nonumber
\eea
Here $\hat{\mathbb \mho}_K = t^\alpha \, \hat{\mathbb \mho}_{\alpha K}$ and $\hat{\mathbb \mho}^K = t^\alpha \, \hat{\mathbb \mho}_{\alpha}{}^{K}$. We also mention that total D-term is positive definite and can be written as $-\frac{1}{4 {\cal V}_E^2} \int_{CY_3} D \wedge \hat{\ast} D$ for $D = D^K a_K + D_K b^K$. 

\subsection{Summary of final symplectic form}
The good thing about presenting several symplectic arrangements via introducing symplectic matrices ($\cal M, \hat{\cal M}$ and ${\cal M}_i$'s) is the fact that now one can express various pieces either as ${\cal O}_1\wedge \ast {\cal O}_2$ or ${\cal O}_1\wedge{\cal O}_2$ form. For example, considering the third representation, we can express the full scalar potential as,
\begin{eqnarray}
\label{eq:V2b}
& & \hskip-1.0cm V_{\rm eff} = {\bf  -}\frac{1}{2} \, e^{K-K_{cs}} \, \, \int_{CY_3}\,\biggl[{\mathbb F} \wedge \ast {\mathbb F} + s^2 \, {\mathbb H} \wedge \ast {\mathbb H} + \hat{\mathbb Q} \wedge \ast \hat{\mathbb Q} - 2 \, s \, {\mathbb H} \wedge \ast \hat{\mathbb Q}   \nonumber\\
& & - 4 \, s \, {\mathbb H} \wedge \ast \tilde{\cal Q} + \frac{s}{4}\,{\cal G}^{ab} \, \tilde{\mathbb \mho}_a \wedge \ast \tilde{\mathbb \mho}_b  + \frac{1}{4} \, \left(\frac{4\, k_0^2}{9} \, \tilde{\cal G}_{\alpha \beta}\,   - 4\, \sigma_\alpha \, \sigma_\beta \right)\tilde{\cal Q}^\alpha \wedge \ast \tilde{\cal Q}^\beta \\
& & + \frac{{\cal V}_E^2}{s}\, {\mathbb R} \wedge \hat{\ast} {\mathbb R} + s\, \hat{\mathbb \mho} \wedge \hat{\ast} \hat{\mathbb \mho} - 2 \, {\cal V}_E\, {\mathbb R} \wedge \hat{\ast} \hat{\mathbb \mho}   + 2 \, s \, {\mathbb F} \wedge {\mathbb H} - 2 \, \, {\mathbb F} \wedge \hat{\mathbb Q} \biggr]\, . \nonumber
\end{eqnarray}
where $\hat\ast$ and $\ast$ denote the Hodge star operations in the even/odd (2,1)-cohomology sector. {\it Here while introducing the integral sign, we have assumed that fluxes are constant parameters, and so ${\cal M}$ matrices have been simply replaced by their respective integral forms.} The last two pieces with ${\mathbb F} \wedge {\mathbb H}$ and ${\mathbb F} \wedge \hat{\mathbb Q}$ terms are nullified via a combination of NS-NS and RR Bianchi identities. In order words, the same can be nullified by adding contributions from local sources such as branes/orientifold planes. The same can be expressed as additional `generalized' $D3/D7$ contributions given as under
\bea
& & {V_{D}^{(2)}} = - {V_{{\mathbb F}{\mathbb H}}} - {V_{{\mathbb F}\hat{\mathbb Q}}}
\eea
Note that in addition to the actual RR tadpole constraints, setting the above ${V_{D}^{(2)}}$ to zero will need the following (subset of) NS-NS Bianchi identities obtained via demanding the nillpotency of the twisted differential operator as ${\cal D}^2 =0$ \cite{Robbins:2007yv}, 
\bea
\label{eq:BIs1}
& & \hskip1.0cm H^\Lambda \, \hat{Q}_\Lambda{}^\alpha - H_\Lambda \hat{Q}^{\alpha \Lambda} = 0, \quad \quad \quad H_\Lambda \, \omega_{a}{}^{\Lambda} - H^\Lambda \, \omega_{a k} = 0 \\
& & \hskip-1cm \hat{Q}^{\alpha \Lambda} \hat{Q}^\beta{}_{\Lambda} - \hat{Q}^{\beta \Lambda} \hat{Q}^\alpha{}_{\Lambda} = 0, \quad  \omega_{a}{}^{\Lambda} \omega_{b \Lambda} - \omega_{b}{}^{\Lambda} \omega_{a k} =0 , \quad \omega_{a \Lambda} \hat{Q}^{\alpha \Lambda} - \omega_{a}{}^{\Lambda} \hat{Q}^\alpha{}_{\Lambda} = 0\nonumber\\
& & \hskip1.0cm R^K \, \hat{\omega}_{\alpha K} - R_K \hat{\omega}_{\alpha}{}^{K} = 0, \quad \quad \quad R_K \, Q^{a K} - R^K \, Q^{a}{}_{K} = 0 \nonumber\\
& & \hskip-1cm \hat{\omega}_{\alpha}{}^{K} \hat{\omega}_{\beta K} - \hat{\omega}_{\beta}{}^{K} \hat{\omega}_{\alpha K} = 0, \quad  Q^{a K} Q^{b}{}_{K} -Q^{b K} Q^{a}{}_{K} =0 , \quad Q^{a K} \hat{\omega}_{\alpha K} - Q^{a}{}_{K} \hat{\omega}_{\alpha}^{K} = 0\nonumber
\eea
Finally, using generic tree level K\"ahler potential in eqn. (\ref{eq:K}), we get $e^{K-K_{cs}} = \frac{1}{2 \, s \, {\cal V}_E^2}$, and subsequently the total scalar potential takes a final form as under, 
\begin{eqnarray}
\label{eq:finalSymp}
& & \hskip-1.0cm V_{\rm eff} = {\bf  -}\frac{1}{4\, s\, {\cal V}_E^2} \, \, \int_{CY_3}\,\biggl[{\mathbb F} \wedge \ast {\mathbb F} + s^2 \, {\mathbb H} \wedge \ast {\mathbb H} + \hat{\mathbb Q} \wedge \ast \hat{\mathbb Q} - 2 \, s \, {\mathbb H} \wedge \ast \hat{\mathbb Q}   \nonumber\\
& & - 4 \, s \, {\mathbb H} \wedge \ast \tilde{\cal Q} + \frac{s}{4}\,{\cal G}^{ab} \, \tilde{\mathbb \mho}_a \wedge \ast \tilde{\mathbb \mho}_b  + \frac{1}{4} \, \left(\frac{4\, k_0^2}{9} \, \tilde{\cal G}_{\alpha \beta}\,   - 4\, \sigma_\alpha \, \sigma_\beta \right)\tilde{\cal Q}^\alpha \wedge \ast \tilde{\cal Q}^\beta \\
& & + \frac{{\cal V}_E^2}{s}\, {\mathbb R} \wedge \hat{\ast} {\mathbb R} + s\, \hat{\mathbb \mho} \wedge \hat{\ast} \hat{\mathbb \mho} - 2 \, {\cal V}_E\, {\mathbb R} \wedge \hat{\ast} \hat{\mathbb \mho}  \biggr]\, . \nonumber
\end{eqnarray}
It appears to be quite remarkable that the total F/D-term scalar potential of arbitrary number of complex structure moduli, K\"ahler moduli and odd-axions has been written out so compactly in terms of symplectic ingredients along with the moduli space metrices, and also without the need of knowing the Calabi Yau metric.

\subsection{Towards the ten-dimensional uplift of the symplectic rearrangement of the scalar potential}
So far our aim has been only to rearrange the 4D effective scalar potential which has been derived using a generalized version of the GVW flux superpotential. As an evidence that the collection of various 4D scalar potential pieces, written in terms of new generalized flux orbits, can indeed be derived from the dimensional reduction of a ten-dimensional theory, now we connect the various pieces of our symplectic formulation with those obtained from the reduction of Double Field Theory (DFT) on a Calabi Yau orientifold \footnote{We thank the referee for suggesting us to comment on the ten-dimensional origin of our symplectic rearrangement of the 4D scalar potential.} \cite{Blumenhagen:2015lta}. Although for details on the later, we refer the readers to \cite{Blumenhagen:2015lta}, we hereby collect the relevant results needed to establish the connection with our approach. The DFT Lagrangian on a Calabi Yau threefold can be given as the sum of following two pieces,
\bea
& & \hskip-2.5cm \ast L_{RR} = -\frac{1}{2} \, {\mathfrak G} \wedge \ast {\mathfrak G} \nonumber\\
& & \hskip-3cm \ast L_{NS \,\,NS} = -e^{-2 \, \phi} \biggl[\frac{1}{2} \, \chi \wedge \ast \ov\chi + \frac{1}{2}\, \Psi \wedge \ast \ov\Psi \\
& & -\frac{1}{4} \Big(\Omega\wedge \chi\Big)\wedge \ast\, \Big(\ov\Omega\wedge \ov \chi\Big)-{1\over 4}\, \Big(\Omega\wedge \ov\chi\Big)\wedge \ast \, \Big(\ov\Omega\wedge  \chi\Big) \biggr] \nonumber
\eea
where
\begin{itemize} 
\item{The $N=2$ string-frame definition of the flux combination $\chi$ (in our conventions) can be given as,
\bea
\label{eq:chi}
& & \hskip-1.5cm \chi \equiv  {\mathfrak D} \, e^{i J} ={\mathbb H} + {\mathbb \mho}\triangleleft (iJ)+{\mathbb Q}\triangleright
\left({(iJ)\wedge (iJ)\over 2}\right) + {\mathbb R}\bullet
     \left({(iJ)\wedge (iJ)\wedge (iJ)\over 6}\right)
\eea
where similar to the previously defined twisted differential operator ${\cal D}$, a new operator ${\mathfrak D} = d + \, {\mathbb H} \wedge . + \, {\mathbb \mho} \triangleleft . + \, {\mathbb Q} \triangleright .+ \, {\mathbb R} \bullet . $ have been introduced to incorporate the effects of $B_2$-field such that, 
\bea
& & {\mathbb H} = {H} + {\omega}\triangleleft B_2 + {Q}\triangleright \left({B_2\wedge B_2\over 2}\right) + {\mathbb R}\bullet
     \left({B_2\wedge B_2\wedge B_2\over 6}\right), \quad etc.
\eea}
\item{The second flux combination $\Psi$ is defined as,
\eq{
\label{eq:Psi}
 \Psi&\equiv \, {\mathfrak D}\, \Omega={\mathbb H}\wedge \Omega + {\mathbb \mho}\triangleleft \Omega+{\mathbb Q}\triangleright \Omega + {\mathbb R}\bullet  \Omega 
  }}
\item{The generalized RR three form field strength ${\mathfrak G}$ is given as,
\bea
\label{eq:mathfrak G}
& & {\mathfrak G}\equiv F + {\mathfrak D} \, {\cal C} = F + {\mathbb H} \wedge C^{(0)} \, + \mho \triangleleft C^{(2)} + {\mathbb Q}\triangleright C^{(4)} + {\mathbb R} \bullet C^{(6)}
\eea
where RR-form ${\cal C} = C^{(0)} + C^{(2)}+ C^{(4)} + C^{(6)} + ... $ }
\end{itemize}
Now, we will investigate the four types of terms in $L_{NS \,NS}$ and $L_{RR}$ to connect with those of ours. Note that, as we have already converted the total scalar potential into real moduli/axions (and the final collection does not use the chiral superfields), so we can directly check the connection by simply considering the orientifold projected version of various terms in $L_{NS \,NS}$ and $L_{RR}$.
\subsubsection*{Matching the RR sector:} As a very first observation, let us recall that in our approach the generalized RR field strength is as given in eqn. (\ref{eq:OddOrbitC}),
\bea
& &  \hskip-0.9cm {\mathbb F}_\Lambda = F_\Lambda + \omega_{a\Lambda} \, {c}^a + \hat{Q}^\alpha{}_\Lambda \, \left({\rho}_\alpha + \hat{\kappa}_{\alpha a b} c^a b^b\right)   + c_0 \, {\mathbb H}_\Lambda, \nonumber\\
& & \hskip2cm {\mathbb F}^\Lambda = F^\Lambda + \omega_a{}^\Lambda \, {c}^a + \hat{Q}^{\alpha \Lambda} \, \left({\rho}_\alpha + \hat{\kappa}_{\alpha a b} c^a b^b\right)\, + c_0 \, {\mathbb H}^\Lambda . \, \nonumber
\eea
 Thus noting from orbit collection in eqn. (\ref{eq:OddOrbitA}) that $\mho_{a\Lambda}  = \omega_{a\Lambda} + \hat{Q}^\alpha{}_\Lambda \, \hat{\kappa}_{\alpha a b} c^a b^b$ and $\mho_a{}^\Lambda =\omega_a{}^\Lambda + \hat{Q}^{\alpha \Lambda} \,  \hat{\kappa}_{\alpha a b} b^b$, we have the first identification from eqn. (\ref{eq:mathfrak G}) and eqn. (\ref{eq:OddOrbitC}),
 \bea
 & & {\mathfrak G} \equiv {\mathbb F}\, .
 \eea
This identifies the RR sectors of the 4D scalar potentials in the two approaches as,
\bea
\label{eq:Vk00000}
 & & \hskip-2cm (I):= \quad -\frac{1}{4\, s \, {\cal V}_E^2} \int_{CY_3/{\cal O}} {\mathfrak G} \wedge \ast \ov{\mathfrak G} = -\frac{1}{4\, s \, {\cal V}_E^2} \int_{CY_3/{\cal O}} {\mathbb F} \wedge \ast {\mathbb F} 
\eea
\subsubsection*{Matching the NS-NS sector:}
Under the orientifold involution, for the flux combination $\chi$ being defined in eqn. (\ref{eq:chi}), we get the following splitting into the even/odd (2,1)-cohomology bases,
\bea
& & \hskip-2cm \chi \equiv \chi^\Lambda \, {\cal A}_\Lambda + \chi_\Lambda \, {\cal B}^\Lambda \\
& & \hskip-1cm = \biggl[\left({\mathbb H}^\Lambda \, -\frac{1}{s} \, \hat{\mathbb Q}^{\alpha\Lambda} \, \sigma_\alpha \right) {\cal A}_\Lambda + \left({\mathbb H}_\Lambda \, -\frac{1}{s} \, \hat{\mathbb Q}^{\alpha}{}_{\Lambda} \, \sigma_\alpha \right) \, {\cal B}^\Lambda \biggr] \nonumber\\
& & + i \, \frac{1}{\sqrt s} \biggl[ \left(\hat{\mho}_\alpha{}^K \, t^\alpha - \frac{1}{s} \, {\cal V}_E\, {\mathbb R}^K \right) a_K + \left(\hat{\mho}_{\alpha{}K} \, t^\alpha - \frac{1}{s} \, {\cal V}_E\, {\mathbb R}_K \right) b^K\biggr] \nonumber
\eea
Note that some factors of ``s'' are introduced as we have changed various orientifold projected pieces of $\chi$ into Einstein-frame. Using the collection of $\chi$ in terms of our generalized flux combinations, we find the following identification of pieces in Einstein-frame,
\begin{eqnarray}
\label{eq:V11111}
 & & \hskip-1.5cm (II):= \quad {\bf  -}\frac{1}{4\, s\, {\cal V}_E^2} \int_{CY_3/{\cal O}} s^2\,{\chi} \wedge \ast \ov{\chi} \\
 & & = {\bf  -}\frac{1}{4\, s\, {\cal V}_E^2} \, \, \int_{CY_3/{\cal O}} \,\biggl[s^2 \, {\mathbb H} \wedge \ast {\mathbb H} + \hat{\mathbb Q} \wedge \ast \hat{\mathbb Q} - 2 \, s \, {\mathbb H} \wedge \ast \hat{\mathbb Q}   \nonumber\\
& & \hskip4.5cm + \frac{{\cal V}_E^2}{s}\, {\mathbb R} \wedge \hat{\ast} {\mathbb R} + s\, \hat{\mathbb \mho} \wedge \hat{\ast} \hat{\mathbb \mho} - 2 \, {\cal V}_E\, {\mathbb R} \wedge \hat{\ast} \hat{\mathbb \mho}  \biggr] \nonumber
\end{eqnarray}
Comparing above with our symplectic collection given in eqn. (\ref{eq:finalSymp}), we find that the pieces in the first line are from $F$-term superpotential contribution while those in the last line are induced via $D$-terms.

Further, as the holomorphic three-form $\Omega$ is odd under orientifold involution, the multi-degree form $\Psi$, as defined in eqn. (\ref{eq:Psi}), will have three components appearing as $6$-form, $4$-form and $2$-form respectively. Therefore Einstein-frame expression of $(\Psi\wedge\ast \ov\Psi)$ piece can be expanded as,
\begin{eqnarray}
\label{eq:Vk22222}
 & & \hskip-1.5cm (III):= \quad {\bf  -}\frac{1}{4\, s\, {\cal V}_E^2} \int_{CY_3/{\cal O}} s^2\,{\Psi} \wedge \ast \ov{\Psi} = {\bf  -}\frac{1}{4\, s\, {\cal V}_E^2} \int_{CY_3/{\cal O}} \bigg[s^2\,{({\mathbb H}\wedge \Omega)} \wedge \ast ({\mathbb H}\wedge \ov\Omega) \nonumber\\
 & & \hskip1.5cm  + s\, {({\mathbb \mho}\triangleleft \Omega)} \wedge \ast ({\mathbb \mho}\triangleleft \ov\Omega) +({\mathbb Q}\triangleright \Omega) \wedge \ast ({\mathbb Q}\triangleright \ov\Omega) \biggr] 
\end{eqnarray}
Now let us consider the two cross-pieces of $L_{NS\, NS}$ which are given as under,
\begin{eqnarray}
\label{eq:Vk33333}
& & \hskip-2.9cm  (IV):= \quad \frac{1}{8\, s\, {\cal V}_E^2}\, \int_{CY_3/{\cal O}} \bigg[ \Big(\Omega\wedge \chi\Big)\wedge
   *\, \Big(\ov\Omega\wedge \ov \chi\Big)+ \Big(\Omega\wedge \ov\chi\Big)\wedge
   *\, \Big(\ov\Omega\wedge  \chi\Big) \biggr] \nonumber\\
& & \hskip1.3cm \equiv \frac{1}{4\, s\, {\cal V}_E^2}\, \int_{CY_3/{\cal O}} \Big(\Omega\wedge Re({\chi})\Big)\wedge
   *\, \Big(\ov\Omega\wedge  Re({\chi})\Big)\nonumber\\
& & \hskip-1.9cm = \frac{1}{4\, s\, {\cal V}_E^2} \int_{CY_3/{\cal O}} \biggl[{s^2\, ({\mathbb H}\wedge \Omega)} \wedge \ast ({\mathbb H}\wedge \ov\Omega) + {(\hat{\mathbb Q}\wedge \Omega)} \wedge \ast (\hat{\mathbb Q}\wedge \ov\Omega) \nonumber\\
& & \hskip1.5cm -s\, {({\mathbb H}\wedge \Omega)} \wedge \ast (\hat{\mathbb Q}\wedge \ov\Omega)-s\,  {(\hat{\mathbb Q}\wedge \Omega)} \wedge \ast ({\mathbb H}\wedge \ov\Omega) \biggr]
\end{eqnarray}
where in the last equality, we have used $Re(\chi) = {\mathbb H} - \frac{1}{s}\, \hat{\mathbb Q}$ with appropriate indices. Now, notice that the first piece in eqns. (\ref{eq:Vk22222}) and (\ref{eq:Vk33333}) cancel each other. Recall that this is the same cancellation which we have observed in eqns. (\ref{eq:V3aaa}) and (\ref{eq:V3a}) while considering $V_2 + V_3$ in the analysis of previous section. Now using the symplectic relations (\ref{eq:symp10})-(\ref{eq:symp30}) we find  that $(III)+(IV)$ results in the remaining following pieces of our collection,
\begin{eqnarray}
& & \hskip-1.5cm (III)+(IV) \equiv {\bf  -}\frac{1}{4\, s\, {\cal V}_E^2} \, \, \int_{CY_3/{\cal O}}\,\biggl[\frac{s}{4}\,{\cal G}^{ab} \, \tilde{\mathbb \mho}_a \wedge \ast \tilde{\mathbb \mho}_b \\
& & \hskip2cm - 4 \, s \, {\mathbb H} \wedge \ast \tilde{\cal Q}   + \frac{1}{4} \, \left(\frac{4\, k_0^2}{9} \, \tilde{\cal G}_{\alpha \beta}\,   - 4\, \sigma_\alpha \, \sigma_\beta \right)\tilde{\cal Q}^\alpha \wedge \ast \tilde{\cal Q}^\beta \biggr]\, . \nonumber
\end{eqnarray}
In this way, we are able to ensure that the various pieces of the scalar potential rearrangement collected in our symplectic formalism can indeed be derived from a ten-dimensional theory (namely DFT) after compactifying the same on a Calabi Yau orientifold. Now we will examine the proposal in two concrete examples.

\section{Explicit examples for checking the proposal}
\label{sec_2examples}
Here we will present two explicit examples to illustrate the insights of our symplectic formulation of the four dimensional scalar potential.
 
\subsection{Example A: Type IIB $\hookrightarrow$ ${\mathbb T}^6/({\mathbb Z}_2\times{\mathbb Z}_2$)-orientifold}
Let us briefly revisit the relevant features of a setup within type IIB superstring theory compactified on  $ {\mathbb T}^6 / \left(\mathbb Z_2\times \mathbb Z_2\right)$ orientifold. The complex coordinates $z_i$'s on each of the tori in ${\mathbb T}^6={\mathbb T}^2\times {\mathbb T}^2\times {\mathbb T}^2$ are defined as
\bea
z^1=x^1+ U_1  \, x^2, ~ z^2=x^3+ U_2 \, x^4,~ z^3=x^5+ U_3 \, x^6 ,
\eea
where the three complex structure moduli $U_i$'s can be written as
$U_i= v_i + i\, u_i,\,\,i=1,2,3$. Further, the two $\mathbb Z_2$ orbifold actions are being defined as
\bea
\label{thetaactions}
& & \theta:(z^1,z^2,z^3)\to (-z^1,-z^2,z^3)\\
 & &   \ov\theta:(z^1,z^2,z^3)\to (z^1,-z^2,-z^3)\, . \nonumber
 \eea
Moreover, the full orientifold action is: ${\cal O} \equiv  (\Omega_p\,(-1)^{F_L}\,  \sigma)$ has the holomorphic involution $\sigma$ being defined as
\bea
\label{eq:orientifold}
& & \sigma : (z^1,z^2,z^3)\rightarrow (-z^1,-z^2,-z^3)\,,
\eea
resulting in a setup with the presence of $O3/O7$-plane. The complex structure moduli dependent pre-potential is given as,
\bea
\label{eq:prepotentialA}
& & {\cal F} = \frac{{\cal X}^1 \, {\cal X}^2 \, {\cal X}^3}{{\cal X}^0} = U_1 \, U_2 \, U_3
\eea 
which results in the following period-vectors,
\bea
& & {\cal X}^0=1\,, \quad   {\cal X}^1=U_1\, , \quad {\cal X}^2=U_2\, , \quad  {\cal X}^3=U_3\, , \\
& & {\cal F}_0=\, -\, \, U_1 \, U_2 \, U_3\, , \quad  {\cal F}_1= U_2 \, U_3\, , \quad  {\cal F}_2=U_3 \, U_1\, , \quad  {\cal F}_3=U_1 \, U_2\,  \nonumber
\eea
Now, the holomorphic three-form $\Omega_3=dz^1\wedge dz^2\wedge dz^3$ can be expanded as,
\bea
& & \hskip-1.5cm  \Omega_3\,  = \alpha_0 +  \, U_1 \, \alpha_1 + U_2 \, \alpha_2 + U_3 \alpha_3 \nonumber\\
& & \hskip1.0cm +\, U_1 \, U_2 \,  U_3 \, \beta^0 -U_2 \, U_3 \, \beta^1- U_1 \, U_3 \, \beta^2 - U_1\, U_2 \, \beta^3 \nonumber
\eea
where we have choosen the following basis of closed three-forms
\bea
\label{formbasis}
      \alpha_0&=dx^1\wedge dx^3\wedge dx^5\,, \quad
      \beta^0= dx^2\wedge dx^4\wedge dx^6\, ,\nonumber\\
     \alpha_1&=dx^2\wedge dx^3\wedge dx^5\, , \quad
    \beta^1= -\, dx^1\wedge dx^4\wedge dx^6\, ,\\
       \alpha_2&=dx^1\wedge dx^4\wedge dx^5\, , \quad
      \beta^2=-\, dx^2\wedge dx^3\wedge dx^6\, ,\nonumber\\
      \alpha_3&=dx^1\wedge dx^3\wedge dx^6\, , \quad
      \beta^3=-\, dx^2\wedge dx^4\wedge dx^5\,  \nonumber
\eea
Subsequently we find that 
\bea
& & K_{cs} = -\ln\biggl[i \, \left( \ov{\cal X}^\Lambda {\cal F}_\Lambda - {\cal X}^\Lambda \ov{\cal F}_\Lambda \right)\biggr] = -\sum_{j = 1}^{3} \ln\left(i (U_j - \ov U_j)\right).
\eea
This also demands that $Im(U_i) <0$ which is rooted from the condition of physical domain to be defined via period matrix (\ref{eq:periodN}) condition $Im({\cal N}_{\Lambda\Delta})<0$ \cite{Ceresole:1995ca,Blumenhagen:2003vr}. This condition $Im(U_i) <0$ is equally important as to demand $Im(\tau) > 0$ and $Im(T_\alpha) < 0$ which are related to string coupling and volume moduli to take positive values, or in other words positive definiteness of moduli space metrices. Now, the basis of orientifold even two-forms and four-forms are as under,
\bea
& & \hskip3cm \mu_1 = dx^1 \wedge dx^2,  \quad \mu_2 = dx^3 \wedge dx^4, \quad  \mu_3 = dx^5 \wedge dx^6 \\
& & \hskip-0.7cm \tilde{\mu}^1 = dx^3 \wedge dx^4 \wedge dx^5 \wedge dx^6,  \quad \tilde{\mu}^2 = dx^1 \wedge dx^2 \wedge dx^5 \wedge dx^6, \quad  \tilde{\mu}^3 = dx^1 \wedge dx^2 \wedge dx^3 \wedge dx^4 \nonumber
\eea
implying that $\hat{d}_\alpha{}^\beta = \delta_\alpha{}^\beta$. The only non-trivial triple intersection number ($k_{\alpha\beta\gamma}$) is given as $\kappa_{123}=k_{123} = 1$ which implies the volume form of the sixfold to be ${\cal V}_E = t_1 \, t_2 t_3$ and so the four cycle volume moduli are given as,
$
\tau_1 =  \, t_2 \, t_3, \,\,\tau_2 =  \, t_3 \, t_1, \, \, \,\,\tau_3 =   \, t_1 \, t_2.
$
implying that
\bea
\label{twoinfour}
& & t_1=\sqrt{\tau_2\,  \tau_3\over \tau_1}\,, ~~  t_2=\sqrt{\tau_1\,  \tau_3\over \tau_2}\,, ~~ t_3=\sqrt{\tau_1\,  \tau_2\over \tau_3}\, \Longrightarrow {\cal V}_E = \sqrt{\tau_1 \tau_2 \tau_3}
\eea
Let us mention that for this example there are no two-forms
anti-invariant under the orientifold projection, i.e. $h^{1,1}_-(X_6) = 0$, and  therefore no ${B}_2$ and $C_2$ moduli as well as no geometric flux components such as $\omega_{a \Lambda}, \omega_{a}{}^\Lambda$ are present. Moreover, as $h^{2,1}_+(X_6) = 0$, so no geometric as well as non-geometric flux components with index $K\in h^{2,1}_+(X_6)$ are present, and this implies that respective $D$-terms will not be induced. The only D-term can arise from the local sources such as branes and orientifold planes to cancel the RR tadpoles. Now, the expressions for K\"ahler potential and the generalized flux-induced superpotential take the following forms,
\bea
\label{eq:Kexample1}
& & \hskip-1.6cm K = -\ln\left(-i(\tau-\ov\tau)\right) -\sum_{j=1}^{3} \ln\left(i(U_j - \ov U_j)\right) - \sum_{\alpha=1}^{3} \ln\left(\frac{i\,(T_\alpha - \ov T_\alpha)}{2}\right)  \\
\label{eq:Wexample1}
& & \hskip-1.6cm W = \biggl[\left(F_\Lambda + \tau\, H_\Lambda  + \hat{Q}^{\alpha}{}_{\Lambda} \, T_\alpha\right)\, {\cal X}^\Lambda + \left(F^\Lambda + \tau\, H^\Lambda  + \hat{Q}^{\alpha \Lambda} \, T_\alpha\right) \, {\cal F}_\Lambda \biggr]\, ,
\eea
where $\Lambda = 0, 1, 2, 3$ and $\alpha = 1,2,3$ implying the presence of 8 components for each of three form fluxes $H_3$ and $F_3$ given as,
\bea
\label{eq:fluxconversionA1}
& & H_0 , \, \quad H_1, \, \quad H_2, \, \quad H_3, \, \, \quad H^0 , \, \quad H^1 , \, \quad H^2 , \, \quad H^3 \nonumber\\
& & F_0 , \,  \quad F_1 , \, \quad F_2 , \, \quad F_3 , \, \, \quad F^0 , \, \quad F^1 , \, \quad F^2 , \, \quad F^3 \nonumber
\eea
and similarly 24 $Q$-flux components can be written for $\hat{Q}_\Lambda^\alpha$ and $\hat{Q}^{\alpha\Lambda}$. Now to analyze and express the total F-term scalar potential in our symplectic formulation, we do the followings, 
\begin{itemize}
\item{First, we utilize the K\"ahler potential (\ref{eq:Kexample1}) and superpotential (\ref{eq:Wexample1}) which results in 2422 terms in total. }
\item{Subsequently, using new generalized flux orbits and the relevant symplectic relations given in appendix \ref{sec_symplectic}, we enumerate terms in each of the three rearrangements, and find that the counting of terms can be distributed into the various pieces of our symplectic formulation given in eqn. (\ref{eq:finalSymp}) as under,
\bea
& & \hskip-1.0cm V_{\mathbb F \mathbb F} = {\bf  -}\frac{1}{4\, s\, {\cal V}_E^2} \, \,\int_{CY_3} \,{\mathbb F} \wedge \ast {\mathbb F}, \hskip5.0cm \#(V_{\mathbb F \mathbb F})= 1630 \nonumber\\
& & \hskip-1.0cm V_{\mathbb H \mathbb H} ={\bf  -}\frac{s}{4\, {\cal V}_E^2} \, \,\int_{CY_3} \, {\mathbb H} \wedge \ast {\mathbb H}, \hskip5.0cm \#(V_{\mathbb H \mathbb H})= 76  \nonumber\\
& & \hskip-1.0cm V_{\mathbb Q \mathbb Q} ={\bf  -}\frac{1}{4\, s\, {\cal V}_E^2} \, \,\int_{CY_3} \left( \hat{\mathbb Q} \wedge \ast \hat{\mathbb Q} -({\cal V}_E\, \hat{\kappa}_{\alpha\beta})\, \, \tilde{\cal Q}^\alpha \wedge \ast \tilde{\cal Q}^\beta\right) , \, \, \, \, \#(V_{\mathbb Q \mathbb Q})= 408 \nonumber\\
& & \hskip-1.0cm V_{\mathbb H \mathbb Q} ={\bf  -}\frac{1}{4\,\, {\cal V}_E^2} \, \,\int_{CY_3}\left(-2 \, {\mathbb H} \wedge \ast \hat{\mathbb Q}  - 4 \, {\mathbb H} \wedge \ast \tilde{\cal Q} \right) , \hskip1.7cm  \#(V_{\mathbb H \mathbb Q})=  180 \nonumber\\
& & \hskip-1.0cm V_{\mathbb H \mathbb F} ={\bf  -}\frac{1}{4\, {\cal V}_E^2} \, \,\int_{CY_3} \left(2 \, \mathbb F \wedge \mathbb H \right), \hskip4.8cm \#(V_{\mathbb H \mathbb F})=  32 \nonumber\\
& & \hskip-1.0cm V_{\mathbb F \mathbb Q} ={\bf  -}\frac{1}{4\, s\, {\cal V}_E^2} \, \,\int_{CY_3} \left(-2 \, \mathbb F \wedge \mathbb Q \right) , \hskip4.2cm \#(V_{\mathbb F \mathbb Q})=  96 \nonumber
\eea}
\item{Here, for checking the symplectic formulation, we have used the following relations, 
\bea
& & \hskip-1.5cm \left(\frac{4}{9}\, k_0^2 \tilde{\cal G}_{\alpha \beta}\right) \equiv 4 \, \sigma_\alpha \sigma_\beta - 4 {\cal V}_E\, k_{\alpha\beta}= \left(
 \begin{array}{ccc}
4\, \sigma_1^2 & 0 & 0\\
0 & 4\, \sigma_2^2  & 0 \\
0 &0 & 4\, \sigma_3^2 \\
\end{array}
\right)
\eea
and 
\bea
& & \hskip-1.5cm  {\cal V}_E \, k_{\alpha\beta} = \left(
 \begin{array}{ccc}
0&  \sigma_1 \, \sigma_2 &  \sigma_1 \, \sigma_3\\
 \sigma_1 \, \sigma_2 &0 &  \sigma_2 \, \sigma_3 \\
 \sigma_1 \, \sigma_3 &  \sigma_2 \, \sigma_3 & 0 \\
\end{array}
\right)
\eea
}
\end{itemize}
Thus we are able to rewrite the total F-term scalar potential in terms of symplectic ingredients and without using internal background metric. As mentioned earlier, the last two pieces $V_{\mathbb H \mathbb F}$ and $V_{\mathbb F \mathbb Q}$ correspond to {\it generalized tadpole contributions} and these have to be canceled by local sources plus satisfying a subset of NS-NS Bianchi identities given in (\ref{eq:BIs1}).

Note that our example A is too simple to illustrate all the features of proposed symplectic formalism in eqn. (\ref{eq:finalSymp}) basically in two sense; first it does not have odd axions due to $h^{1,1}_-(X_6)=0$ and so use of generalized flux orbits corrected via odd-axions ${B}_2/C_2$ have not been demonstrated. Second, this example could not introduce non-geometric $R$-flux due to a trivial sector of even (2,1)-cohomology as $h^{2,1}_+(X_6) = 0$.  For that purpose, we now come to our example B in the next subsection.

\subsection{Example B: Type IIB $\hookrightarrow$ ${\mathbb T}^6/{\mathbb Z}_4$-orientifold}
In this case, we consider a type IIB compactification setup on the orientifold of ${\mathbb T}^6/{\mathbb Z}_4$ orbifold and analyze the scalar potential for the untwisted sector moduli/axions. This setup has $h^{2,1}(X) = 1_+ +0_-$, and $h^{1,1}(X) = 3_+ +2_-$, i.e. there are three complexified K\"ahler moduli ($T_\alpha$), two complexified odd axions ($G^a$) and no complex structure moduli. The only non-zero intersection numbers are: $k_{311} = 1/2, k_{322} = -1$ which results in overall volume form ${\cal V}_E = \frac{1}{4} (t_1^2-2\, t_2^2)\, t_3$. In addition, one has odd intersection numbers as $\hat{k}_{311} = -1, \hat{k}_{322} = -1/2$ along with $\hat{d}_\alpha{}^\beta = diag\{1/2, -1, 1/4\}$ and $d^a{}_b=diag\{-1,-1/2\}$. Also, given that $h^{2,1}_-(X) =0$, complex structure moduli dependent piece of the K\"ahler potential is just a constant piece. Here we fix our conventions by considering ${\cal X}^0 = 1, {\cal F}_0 = -i$, which results in $e^{Kcs} = 1/2$. While we leave additional orientifold construction related details to be directly referred from \cite{Robbins:2007yv,Shukla:2015bca}, here we simply provide the explicit expressions of K\"ahler- and super-potentials for analyzing F-term scalar potential. The K\"ahler potential is given as under,
\bea
\label{typeIIBK}
K &= -\ln2 -\ln\left(-i(\tau-\ov \tau)\right) - 2 \, \ln{\cal V}_E(T_\alpha, \tau, G^a; \ov T_\alpha, \ov \tau, \ov G^a)
\eea
where the Einstein frame volume is given as,
\bea
\label{eq:Volume}
& & {\cal V}_E\equiv {\cal V}_E(T_\alpha, S, G^a) = \frac{1}{4}\, \left(\frac{i(T_3 -\ov T_3)}{2} - \frac{i}{4 (\tau -\ov \tau)} \, \hat{\kappa}_{3 a b} \, (G^a -\ov G^a) (G^b -\ov G^b)\right)^{1/2} \nonumber\\
& & \hskip2cm \times  \biggl[\biggl(\frac{i(T_1 -\ov T_1)}{2}\biggr)^2 -2\, \biggl(\frac{i(T_2 -\ov T_2)}{2}\biggr)^2\biggr]^{1/2} \, \, 
\eea
Further, the generic form of the tree level flux superpotential with all allowed fluxes being included is given as,
\bea
\label{eq:Wsimp1}
& & \hskip-1.5cm W = \biggl[\left(F_0 + \tau\, H_0 + \omega_{0a} \, G^a + \hat{Q}^{\alpha}{}_{0} \, T_\alpha \right) -i\,  \left(F^0 + \tau\, H^0 + \omega_a{}^0 \, G^a + \hat{Q}^{\alpha 0} \, T_\alpha \right)\biggr]\,,
\eea
where $a = \{1,2\}$ and $\alpha=\{1,2,3\}$. Now, one can compute the full F-term scalar potential from these explicit expressions of $K$ and $W$. For this toroidal setup, the new generalized flux orbits given in eqns. (\ref{eq:OddOrbitA}), (\ref{eq:OddOrbitB}) and (\ref{eq:OddOrbitC}) are simplified. The ones with odd-indexed fluxes are given as under,
\bea
\label{eq:orbitsB2Ex2}
& & \hskip-0.5cm {\mathbb H}_0 = H_0 + (\omega_{01} \, {b}^1+\omega_{02} \, {b}^2) + \hat{Q}^3{}_0 \, \left(\frac{1}{2}\, \hat{\kappa}_{3 11} (b^1)^2 + \frac{1}{2}\, \hat{\kappa}_{3 22} (b^2)^2\right)  \nonumber\\
& & \hskip-0.5cm  {\mathbb F}_0 = c_0 \, {\mathbb H}_0 + \biggl[F_0 + (\omega_{01} \, {c}^1 + \omega_{02} \, {c}^2) + \hat{Q}^1{}_0 \,{\rho}_1 \nonumber\\
& & \hskip2cm + \hat{Q}^2{}_0 \,{\rho}_2+ \hat{Q}^3{}_0\, \left({\rho}_3 + \hat{\kappa}_{3 11} c^1 b^1 + \hat{\kappa}_{3 22} c^2 b^2\right) \biggr] \, \\
& & \hskip-0.5cm {\mathbb \mho}_{01} = \biggl[\omega_{01} + \hat{Q}^3{}_0 \, \left(\hat{\kappa}_{311}\, b^1\right)  \biggr] , \quad \quad  {\mathbb \mho}_{02} = \biggl[\omega_{02} + \hat{Q}^3{}_0 \, \left(\hat{\kappa}_{322}\, b^2\right)  \biggr]\nonumber\\
& & \hskip-0.5cm {\hat{\mathbb Q}^{1}{}_{0}} = \hat{Q}^{1}{}_{0} , \quad {\hat{\mathbb Q}^{2}{}_{0}} = \hat{Q}^{2}{}_{0}, \quad {\hat{\mathbb Q}^{3}{}_{0}} = \hat{Q}^{3}{}_{0}\nonumber
\eea
while the ones with even-indexed fluxes are given as,
\bea
& & \hskip-1.9cm {\mathbb R}_1= f^{-1}R_1, \, \, \hat{\mho}_{11} = \hat{\omega}_{11},\, \, \hat{\mho}_{21} = \hat{\omega}_{21}, \,  \hat{\mho}_{31} = \hat{\omega}_{31} - \left( Q_1{}^1 b_1 + Q_1{}^2 b_2 \right) - R_1 (2 b_1^2 + b_2^2 ) 
\eea
and similarly flux components with upper index `$\Lambda= 0$ and $K=1$' can be analogously written. Using these flux orbits one gets a total of 382 terms in F-term contribution while 72 terms in $V_{D}^{(1)}$.  The F-term pieces can be rearranged as,
\bea
\label{eq:VEx2a}
& & \hskip-1cm V_{{\mathbb F}{\mathbb F}} =  \frac{1}{4 \, s\,{\cal V}_E^2}\, \biggl[{\mathbb F}_0^2+{({\mathbb F}^0)}^2\biggr] \hskip7.3cm \#(V_{{\mathbb F}{\mathbb F}})=178\nonumber\\
& & \hskip-1cm V_{{\mathbb H}{\mathbb H}} =  \frac{s }{4 \, {\cal V}_E^2}\, \biggl[{\mathbb H}_0^2 +\, {({\mathbb H}^0)}^2 \biggr] \hskip7.2cm \#(V_{{\cal H}{\mathbb H}})=30 \, \nonumber\\
& & \hskip-1cm V_{{\mathbb F}{\mathbb H}} =  \frac{1}{4 \, {\cal V}_E^2} \times 2 \biggl[{\mathbb H}_0 {\mathbb F}^0-{\mathbb F}_0 {\mathbb H}^0\biggr] \hskip6.4cm \#(V_{{\mathbb F}{\mathbb H}})=60 \nonumber\\
& & \hskip-1cm V_{{\mathbb F}\hat{\mathbb Q}} =  \frac{1}{4 \, s \, {\cal V}_E^2} \times 2\biggl[{\mathbb F}_0 \left(\hat{\mathbb Q}^{0\alpha}\sigma_\alpha \right) - \left(\hat{\mathbb Q}_0{}^\alpha \, \sigma_\alpha \right) {\mathbb F}^0 \biggr] \hskip3.6cm \#(V_{{\mathbb F}{\mathbb Q}})= 64 \nonumber\\
& & \hskip-1cm V_{{\mathbb \mho}{\mathbb \mho}} = \frac{1}{4 \, {\cal V}_E^2} \biggl[ \sigma_3 \left(\mho_{01} \mho_{01} + \mho_{1}{}^{0} \mho_{1}{}^{0}\right) +2\, \sigma_3 \left(\mho_{02} \mho_{02} + \mho_{2}{}^{0} \mho_{2}{}^{0}\right)\biggr] \nonumber\\
& & \hskip-1cm V_{{\mathbb H}\hat{\mathbb Q}} =  \frac{1}{4  \, {\cal V}_E^2} \times (+2)\biggl[3 \,{\mathbb H}^0 \left(\hat{\mathbb Q}^{0\alpha}\sigma_\alpha \right) + 3\, \left(\hat{\mathbb Q}_0{}^\alpha \, \sigma_\alpha \right) {\mathbb H}_0 \biggr]  \hskip1.3cm \#(V_{{\mho}{\mho}} + V_{{\mathbb H}{\mathbb Q}})=34 \nonumber\\
& & \hskip-1.0cm V_{\hat{\mathbb Q}\hat{\mathbb Q}} =\frac{1}{4 \,s\, {\cal V}_E^2} \biggl[\left(4 \sigma_2^2 - \sigma_1^2\right) \hat{\mathbb Q}_0{}^1 \hat{\mathbb Q}_0{}^1 + \left(\sigma_1^2 - \sigma_2^2\right) \hat{\mathbb Q}_0{}^2 \hat{\mathbb Q}_0{}^2  + \sigma_3^2 \hat{\mathbb Q}_0{}^3 \hat{\mathbb Q}_0{}^3 \nonumber\\
& & + 2 \, \sigma_1 \sigma_2 \hat{\mathbb Q}_0{}^1 \hat{\mathbb Q}_0{}^2 - 6 \, \sigma_2 \sigma_3 \hat{\mathbb Q}_0{}^2 \hat{\mathbb Q}_0{}^3- 6 \, \sigma_1 \sigma_3 \hat{\mathbb Q}_0{}^1 \hat{\mathbb Q}_0{}^3 \biggr]  \\
& & \hskip-0.0cm + \frac{1}{4 \,s\, {\cal V}_E^2} \biggl[\left(4 \sigma_2^2 - \sigma_1^2\right) \hat{\mathbb Q}^{01} \hat{\mathbb Q}^{01} + \left(\sigma_1^2 - \sigma_2^2\right) \hat{\mathbb Q}^{02} \hat{\mathbb Q}^{02}  + \sigma_3^2 \hat{\mathbb Q}^{03} \hat{\mathbb Q}^{03} \nonumber\\
& & + 2 \, \sigma_1 \sigma_2 \hat{\mathbb Q}^{01} \hat{\mathbb Q}^{02} - 6 \, \sigma_2 \sigma_3 \hat{\mathbb Q}^{02} \hat{\mathbb Q}^{03}- 6 \, \sigma_1 \sigma_3 \hat{\mathbb Q}^{01} \hat{\mathbb Q}^{03} \biggr] \hskip2.3cm \#(V_{\hat{\mathbb Q}\hat{\mathbb Q}})=16 \,.\nonumber
\eea
while the pieces coming from D-term $V_{D}^{(1)}$ are as under,
\bea
\label{eq:VEx2c}
& & \hskip-1.0cm V_{{\mathbb R}{\mathbb R}} = \frac{{\mathbb R}_1^2+({\mathbb R}^1)^2}{4 \,s^2}  \hskip6.9cm \#(V_{{\mathbb R}{\mathbb R}}) = 2, \nonumber\\
& &  \hskip-1.0cm V_{{\hat{\cal \mho}}{\hat{\cal \mho}}} =  \frac{\left(t^\alpha \, \,\hat{{\cal \mho}}_{\alpha 1}\right)^2+{(t^\alpha\,\, {\hat{\cal \mho}_\alpha{}^1})}^2 }{4\, \,{\cal V}_E^{2}}\, \hskip4.80cm \#( V_{{\hat{\cal \mho}}{\hat{\cal \mho}}}) =56, \\
& & \hskip-1.0cm V_{{\mathbb R}{\hat{\cal \mho}}}=-2 \times \, \frac{{\mathbb R}_1 \left(t^\alpha \,\, \hat{{\cal \mho}}_{\alpha 1}\right)+ {\mathbb R}^1 \, \left(t^\alpha\,\, {\hat{\cal \mho}_\alpha{}^1}\right)}{4 \,s\, {\cal V}_E^{2}} \hskip2.8cm \#(V_{{\cal R}{\hat{\cal \mho}}}) = 14\, , \nonumber
\eea
This example also illustrates how a huge scalar potential can be so compactly rewritten using the new generalized flux orbits. Moreover, this rearrangement of the total scalar potential can be easily seen from our eqn. (\ref{eq:finalSymp}) and any of the three symplectic representations in eqns. (\ref{eq:rep1})-(\ref{eq:rep3}) after supplementing the following symplectic ingredients,
\bea
& & {\cal M}^{00} = -1, \quad {\cal M}^{0}{}_{0} = 0, \quad  {\cal M}_0{}^0 = 0, \quad {\cal M}_{00} = 1 \nonumber\\ 
& & {\cal L}^{00} = -1,  \quad {\cal L}^{0}{}_{0} = 0, \quad  {\cal L}_0{}^0 = 0, \quad {\cal L}_{00} = 1 \nonumber\\ 
& & {({\cal M}_1)}^{00} = -2, \quad  {({\cal M}_1)}^{0}{}_{0} = 0, \quad {({\cal M}_1)}_0{}^0 = 0, \quad {({\cal M}_1)}_{00} = 2 \\ 
& & {({\cal M}_2)}^{00} = 3, \quad {({\cal M}_2)}^{0}{}_{0} = 0, \quad  {({\cal M}_2)}_0{}^0 =0, \quad {({\cal M}_2)}_{00} = -3 \nonumber\\ 
& & {({\cal M}_3)}^{00} = 0, \quad {({\cal M}_3)}^{0}{}_{0} = 2, \quad  {({\cal M}_3)}_0{}^0 =-2, \quad {({\cal M}_3)}_{00} = 0 \nonumber\\ 
& & \hat{\cal M}^{11} = -1, \quad \hat{\cal M}^{1}{}_{1} = 0, \quad  \hat{\cal M}_1{}^1=0, \quad \hat{\cal M}_{11} = 1 \nonumber
\eea
For example, $V_{{\mathbb \mho}{\mathbb \mho}}$ can be known simply by considering ${\cal G}^{ab} = - 4 \, {\cal V}_E \, \hat{k}^{ab}$. Now $\hat{k}_{ab} = \hat{k}_{\alpha a b} t^\alpha$, so one has the only non-zero components given as ${\cal G}^{11} = \sigma_3$ and ${\cal G}^{22} = 2\, \sigma_3$ as can be seen from collection in eqn. (\ref{eq:VEx2a}). Similarly, for the largest piece $V_{\hat{\mathbb Q}\hat{\mathbb Q}}$, let us consider the followings, 
\bea
& & \hskip-2.5cm \left(\frac{4}{9}\, k_0^2 \tilde{\cal G}_{\alpha \beta} - 4 \sigma_\alpha \, \sigma_\beta\right):=- 4\, {\cal V}_E\, ({\hat{d}^{-1}})_{\alpha}{}^{\alpha'}\, k_{\alpha' \beta'}\,  ({\hat{d}^{-1}})_{\beta}{}^{\beta'}\\
& &  = \left(
 \begin{array}{ccc}
4 \sigma_2^2 - 2 \sigma_1^2& 0 & -4 \sigma_1 \, \sigma_3\\
0 & \sigma_1^2 - 2 \sigma_2^2 & -4 \sigma_2 \, \sigma_3 \\
-4 \sigma_1 \, \sigma_3 & -4 \sigma_2 \, \sigma_3 & 0 \\
\end{array}
\right)\nonumber
\eea
Now one can immediately read off the precise sum of two $\hat{\mathbb Q}\hat{\mathbb Q}$ pieces as given in collection (\ref{eq:VEx2a}) from the following coefficient matrix,
\bea
& & \hskip-1.5cm \left(\frac{4}{9}\, k_0^2 \tilde{\cal G}_{\alpha \beta} - 4 \sigma_\alpha \, \sigma_\beta\right) + \sigma_\alpha \, \sigma_\beta = \left(
 \begin{array}{ccc}
4 \sigma_2^2 -  \sigma_1^2& \sigma_1 \, \sigma_2 & -3 \sigma_1 \, \sigma_3\\
\sigma_1 \, \sigma_2 & \sigma_1^2 -  \sigma_2^2 & -3 \sigma_2 \, \sigma_3 \\
-3 \sigma_1 \, \sigma_3 & -3 \sigma_2 \, \sigma_3 & \sigma_3^2 \\
\end{array}
\right)
\eea
Here we recall that $\sigma_\alpha = \frac{1}{2}\, \kappa_{\alpha\beta\gamma} t^\alpha \, t^\beta \, t^\gamma$ and results in $\sigma_1 = t_1\, t_3, \sigma_2 = t_2 \, t_3$ and $\sigma_3=(t_1^2-2\, t_2^2)$. Thus we have illustrated our generic proposal of symplectic rearrangement in two Toroidal examples.

\section{Conclusions and future directions}
\label{sec_conclusion}

In \cite{Taylor:1999ii}, the four dimensional effective potentials obtained in the context of type IIB superstring compactificaion with superpotentials induced by the standard NS-NS and RR three form fluxes ($H_3$ and $F_3$) have been expressed in terms of symplectic ingredients using \cite{Ceresole:1995ca}. In this article, we have extended that symplectic formulation for a superpotential induced by generalized fluxes turned-on on generic Calabi Yau orientifold backgrounds. This has been done in a two-step strategy. First we have rewritten the total scalar potential into suitable pieces using a set of {\it new} generalized flux orbits, and subsequently after invoking some non-trivial symplectic relations we have further rearranged various pieces into a symplectic formulation.

As a check of our proposal, we have considered two concrete examples of type IIB superstring compactification on the orientifolds of ${\mathbb T}^6/({\mathbb Z}_2 \times {\mathbb Z}_2)$ and ${\mathbb T}^6/{\mathbb Z}_4$. Both of these simple examples have their own advantages and limitations. For example, the first example with ${\mathbb T}^6/({\mathbb Z}_2 \times {\mathbb Z}_2)$-orientifold illustrates the utility of period matrix part in the symplectic rearrangement as it has 3 complex structure moduli, however this example does neither support involutively odd-axions nor has involutively even $(2,1)$-cohomology sector to illustrate the appearance of D-term involving $R$-flux. On the other hand, the second example with ${\mathbb T}^6/{\mathbb Z}_4$-orientifold has two odd axions as $h^{1,1}_-(CY) =2$, and moreover $h^{2,1}_+(CY) =1$ which help in demonstrating the crucial use of the {\it new} generalized flux combinations we have, and also in the embedding of D-terms. However, the second example does not have any complex structure moduli and so the information within period matrix sector has been indeed trivial. Thus we can say that the two examples considered in this article compliment each other quite well, and at the same time remain simple enough to perform explicit analytic computations needed to check the proposal.

The symplectic rearrangement of the 4D scalar potential proposed in this article has many possible advantages and applications; for example,
\begin{itemize}
 \item{The total symplectic rearrangement is very compact, and helps in rewriting the scalar potential consisting of thousands of terms into a few lines. Moreover, we do not need to know the Calabi Yau metric as the desired relevant pieces of information for rewriting the total scalar potential can be extracted via the moduli space matrices and the period matrices.}
 \item{The symplectic rearrangement is what we call `suitable' for dimensional oxidation purpose (on the lines of \cite{Blumenhagen:2013hva, Gao:2015nra, Shukla:2015rua, Shukla:2015bca}), and at least for the scenarios when the fluxes are treated as constant parameters, one could naively guess the ten-dimensional uplift of the four dimensional scalar potential. In fact, we have connected the various pieces of our rearrangement with those of a scalar potential obtained by dimensional reduction of Double Field Theory on a CY orientifold \cite{Blumenhagen:2015lta}.}
 \item{Moreover, the scalar potential under consideration is valid for an arbitrary Calabi Yau orientifold compactification, and so is equipped with arbitrary numbers of complex structure moduli, K\"ahler moduli and odd-axions. In addition, the symplectic rearrangement generically consists of all kinds of (non-)geometric fluxes along with the standard $H_3$ and $F_3$ fluxes, however which of those can be consistently turned-on on a given background still needs an answer.}
 \item{In the light of the aforesaid points, the present analysis should be helpful in the model independent studies of phenomenological aspects, e.g. moduli stabilization, searching de-Sitter solutions etc. }
\end{itemize}
For example, to elaborate on the point of moduli stabilization, let us consider an orientifold setup with $h^{2,1}_+ (CY_3) =0$ which has been very common in the setups of previous moduli stabilization studies and let us say that we want to focus on the stabilization of universal axion $(c_0)$ and dilaton ($s$), then using eqn.(\ref{eq:finalSymp}) the total effective potential can be rewritten as,
\bea
\label{eq:Vsc0}
& & V(c_0, s ; .....) = \left(\frac{l_1}{s}  \, + {l_2} + {s}\, \, l_3 \right) +  \frac{l_4}{s} \, c_0 + \frac{l_5}{s} \, c_0^2
\eea
where $l_i$'s depend on all the moduli/axions except the dilaton ($s$) and RR axion $c_0$. Note that it has been possible to extract the dilaton dependence from all the pieces as we have already expressed the symplectic collection into Einstein-frame. Further, the explicit expressions of $l_i$'s can be collected as under,
\bea
& & l_1 = {\bf  -}\frac{1}{4\, {\cal V}_E^2} \, \, \int_{CY_3}\, \biggl[{\mathbb G} \wedge \ast {\mathbb G}  + \hat{\mathbb Q} \wedge \ast \hat{\mathbb Q} + \frac{1}{4} \, \left(\frac{4\, k_0^2}{9} \, \tilde{\cal G}_{\alpha \beta}\,   - 4\, \sigma_\alpha \, \sigma_\beta \right)\tilde{\cal Q}^\alpha \wedge \ast \tilde{\cal Q}^\beta \biggr] \nonumber\\
& & l_2 = {\bf  -}\frac{1}{4\,  {\cal V}_E^2} \, \, \int_{CY_3}\, \biggl[- 2 \, {\mathbb H} \wedge \ast \hat{\mathbb Q}  - 4 \, {\mathbb H} \wedge \ast \tilde{\cal Q} + \frac{1}{4}\,{\cal G}^{ab} \, \tilde{\mathbb \mho}_a \wedge \ast \tilde{\mathbb \mho}_b \biggr], \quad \nonumber\\
& & l_3 = {\bf  -}\frac{1}{4\,  {\cal V}_E^2} \, \, \int_{CY_3}\, {\mathbb H} \wedge \ast {\mathbb H} \equiv  l_5, \quad \\
& & l_4 = {\bf  -}\frac{1}{4\,  {\cal V}_E^2} \, \, \int_{CY_3}\, \left({\mathbb G} \wedge \ast {\mathbb H} + {\mathbb H} \wedge \ast {\mathbb G} \right), \quad \nonumber
\eea
where ${\mathbb G} = F + \mho_a \, c^a + \hat{\mathbb Q}^\alpha \, \rho_\alpha$ while other flux combinations are as defined in eqn. (\ref{eq:OddOrbitA}). Now extremizing the potential (\ref{eq:Vsc0}) w.r.t. universal axion and dilaton, one finds that
\bea
& & \ov{c_0} = - \frac{l_4}{2\, l_5}\,, \quad \quad \quad \ov{s} = \frac{\sqrt{4\, l_1 l_5 - l_4^2}}{2 \, \sqrt{l_3 \, l_5}}
\eea
Moreover, the two-field analysis shows that Hessian at the above critical point leads to,
\bea
& & V_{c_0 c_0} = \frac{4 \, \sqrt{l_3} \, l_5^{3/2}}{\sqrt{4\, l_1 l_5 - l_4^2}} \, \quad , V_{c_0 s} = 0 = V_{s c_0} \,, \quad  V_{ss} = \frac{4\, \sqrt{l_5} \, l_3^{3/2}}{\sqrt{4\, l_1 l_5 - l_4^2}}
\eea
By this two-field analysis we have shown some indications how the symplectic rearrangement could be useful for performing a model independent moduli stabilization. Finally we may agree that many things work quite nicely, however there are several issues to be settled in order to have a complete understanding of the setups with non-geometric fluxes. Moreover, which and how many fluxes can be truly and consistently turned on simultaneously remains an open issue which is essential for studying the moduli stabilization and subsequent phenomenology, and we hope to get back on some of these issues in future.

\section*{Acknowledgments}
I am very grateful to Ralph Blumenhagen for useful discussions and encouragements throughout. Moreover, I am thankful to Ralph Blumenhagen, Anamaria Font, Xin Gao, Daniela Herschmann, Oscar Loaiza-Brito and Erik Plauschinn for useful discussions during earlier collaboration. This work was supported by the Compagnia di San Paolo contract ``Modern Application of String Theory'' (MAST) TO-Call3-2012-0088.

\clearpage
\appendix

\section{Useful symplectic relations}
\label{sec_symplectic}
Using the symplectic matrices ${\cal M}, {\cal M}_1,{\cal M}_2$ and ${\cal M}_3$ defined in eqns. (\ref{coff}), (\ref{coff3}), (\ref{coff4}) and  (\ref{coff5}) respectively, one finds that,
\bea
\label{eq:symp10}
& & Re({\cal X}^\Lambda \, \ov{\cal X}^\Delta) = -\frac{1}{4} \, e^{-K_{cs}} \, \left( {\cal M}^{\Lambda \Delta} + \, \, {\cal L}^{\Lambda \Delta} \right) = -\frac{1}{4} \, e^{-K_{cs}} \, {{\cal M}_1}^{\Lambda \Delta} \\
& & Re({\cal F}_\Lambda \, \ov{\cal X}^\Delta) = - \frac{1}{4} \, e^{-K_{cs}} \, \left(  {\cal M}_{\Lambda}^{\, \, \, \, \, \Delta} +  \, \, {\cal L}_{\Lambda}^{\, \, \, \, \, \Delta}\right) = - \frac{1}{4} \, e^{-K_{cs}} \, {{\cal M}_1}_{\Lambda}^{\, \, \, \, \, \Delta}\nonumber\\
& & Re({\cal X}^\Lambda \, \ov{\cal F}_\Delta) = +\frac{1}{4} \, e^{-K_{cs}} \, \left({\cal M}^\Lambda_{\, \, \, \, \Delta}  +  \, \,  {\cal L}^\Lambda_{\, \, \, \, \Delta} \right) = +\frac{1}{4} \, e^{-K_{cs}} \,{{\cal M}_1}^\Lambda_{\, \, \, \, \Delta} \nonumber\\
& & Re({\cal F}_\Lambda \, \ov{\cal F}_\Delta) = + \frac{1}{4} \, e^{-K_{cs}} \, \left({\cal M}_{\Lambda \Delta} + \, \, {\cal L}_{\Lambda \Delta}\right) = + \frac{1}{4} \, e^{-K_{cs}} \, {{\cal M}_1}_{\Lambda \Delta}\nonumber
\eea
and 
\bea
\label{eq:symp20}
& & \left({\cal M}^{\Lambda \Delta} + 8 \, e^{K_{cs}}\, Re({\cal X}^\Lambda \, \ov{\cal X}^\Delta) \right)= {{\cal M}_2}^{\Lambda \Delta} = \left({{\cal M}}^{\Lambda \Delta}  - 2 {{\cal M}_1}^{\Lambda \Delta}\right) \\
& & \left({\cal M}_{\Lambda}^{\, \, \, \, \, \Delta} + 8 \, e^{K_{cs}}\,  Re({\cal F}_\Lambda \, \ov{\cal X}^\Delta) \right)= {{\cal M}_2}_{\Lambda}^{\, \, \, \, \, \Delta} =  \left({{\cal M}}_{\Lambda}^{\, \, \, \, \, \Delta} - 2  {{\cal M}_1}_{\Lambda}^{\, \, \, \, \, \Delta} \right)\nonumber\\
& & \left(-{\cal M}^\Lambda_{\, \, \, \, \Delta} + 8 \, e^{K_{cs}}\, Re({\cal X}^\Lambda \, \ov{\cal F}_\Delta)\right) = -{{\cal M}_2}^\Lambda_{\, \, \, \, \Delta} = - \left({{\cal M}}^\Lambda_{\, \, \, \, \Delta} - 2 {{\cal M}_1}^\Lambda_{\, \, \, \, \Delta}  \right)\nonumber\\
& & \left(-{\cal M}_{\Lambda \Delta} + 8 \, e^{K_{cs}} \,Re({\cal F}_\Lambda \, \ov{\cal F}_\Delta) \right) = -{{\cal M}_2}_{\Lambda \Delta} = -\left({{\cal M}}_{\Lambda \Delta}  - 2 \, {{\cal M}_1}_{\Lambda \Delta} \right)\nonumber
\eea
and so equivalently we have another set of relations as under,
\bea
\label{eq:symp30}
& & {{\cal M}_1}^{\Lambda \Delta} = \frac{1}{2} \left({{\cal M}}^{\Lambda \Delta}  - {{\cal M}_2}^{\Lambda \Delta}\right)\, , \,\\
& &  {{\cal M}_1}_{\Lambda}^{\, \, \, \, \, \Delta}  = \frac{1}{2} \left( {{\cal M}}_{\Lambda}^{\, \, \, \, \, \Delta} - {{\cal M}_2}_{\Lambda}^{\, \, \, \, \, \Delta} \right) \,,\nonumber\\
& & {{\cal M}_1}^\Lambda_{\, \, \, \, \Delta}  = \frac{1}{2} \left({{\cal M}}^\Lambda_{\, \, \, \, \Delta}  - {{\cal M}_2}^\Lambda_{\, \, \, \, \Delta} \right)\, ,\nonumber\\
& & {{\cal M}_1}_{\Lambda \Delta} = \frac{1}{2} \left({{\cal M}}_{\Lambda \Delta}  - {{\cal M}_2}_{\Lambda \Delta} \right) \nonumber
\eea
\begin{eqnarray}
\label{eq:mainSymp}
 & & \hskip-0.7cm \, {{\cal M}_1}^{\Gamma \Delta} = \frac{1}{2}\biggl[{{\cal M}_3}_{\Lambda}^{\,\,\, \Gamma} \left({\cal M}^{\Lambda \Sigma} \, {{\cal M}_3}_{\Sigma}^{\, \, \, \, \, \Delta} + {\cal M}^\Lambda_{\, \, \, \, \Sigma} \, {{\cal M}_3}^{\Sigma \Delta}\right) + {{\cal M}_3}^{\Lambda \Gamma} \left({\cal M}_{\Lambda}^{\,\,\, \Sigma} \, {{\cal M}_3}_{\Sigma}^{\, \, \, \, \, \Delta} + {\cal M}_{\Lambda \, \Sigma} \, {{\cal M}_3}^{\Sigma \Delta}\right)\biggr] \nonumber\\
 & & \hskip-0.7cm \, {{\cal M}_1}_{\Gamma}^{\, \, \, \, \, \Delta}  = \frac{1}{2}\biggl[{{\cal M}_3}_{\Lambda \Gamma} \left({\cal M}^{\Lambda \Sigma} \, {{{\cal M}_3}}_{\Sigma}^{\, \, \, \, \, \Delta} + {\cal M}^\Lambda_{\, \, \, \, \Sigma} \, {{\cal M}_3}^{\Sigma \Delta}\right) + {{{\cal M}_3}}^{\Lambda}_{\,\,\,\,\, \Gamma} \left({\cal M}_{\Lambda}^{\,\,\, \Sigma} \, {{{\cal M}_3}}_{\Sigma}^{\, \, \, \, \, \Delta} + {\cal M}_{\Lambda \, \Sigma} \, {{\cal M}_3}^{\Sigma \Delta}\right) \biggr]\nonumber\\
  & & \hskip-0.7cm \, {{\cal M}_1}^\Gamma_{\, \, \, \, \Delta}= -\frac{1}{2}\biggl[{{{\cal M}_3}}_{\Lambda}^{\, \, \, \, \, \Gamma} \left({\cal M}^{\Lambda \Sigma} \, {{\cal M}_3}_{\Sigma \Delta} + {\cal M}^\Lambda_{\, \, \, \, \Sigma} \, {{{\cal M}_3}}^{\Sigma}_{\,\,\,\, \,\,\Delta}\right) + {{{\cal M}_3}}^{\Lambda \Gamma} \left({\cal M}_{\Lambda}^{\,\,\, \Sigma} \, {{\cal M}_3}_{\Sigma \, \Delta} + {\cal M}_{\Lambda \, \Sigma} \, {{{\cal M}_3}}^{\Sigma}_{\,\,\,\,\, \Delta}\right) \biggr]\nonumber\\
 & & \hskip-0.7cm \, {{\cal M}_1}_{\Gamma \Delta}=- \frac{1}{2}\biggl[{{\cal M}_3}_{\Lambda \Gamma} \left({\cal M}^{\Lambda \Sigma} \, {{\cal M}_3}_{\Sigma \Delta} + {\cal M}^\Lambda_{\, \, \, \, \Sigma} \, {{{\cal M}_3}}^{\Sigma}_{\,\,\,\, \,\,\Delta}\right) + {{{\cal M}_3}}^{\Lambda}_{\,\,\,\,\, \Gamma} \left({\cal M}_{\Lambda}^{\,\,\, \Sigma} \, {{\cal M}_3}_{\Sigma \, \Delta} + {\cal M}_{\Lambda \, \Sigma} \, {{{\cal M}_3}}^{\Sigma}_{\,\,\,\,\, \Delta}\right) \biggr]\nonumber\\
\end{eqnarray}
Being directly related to produce Hodge star of three-forms as in eqn. (\ref{stardef}), we consider that ${\cal M}$ should be present in all the rearrangement of the scalar potential pieces, and so we choose either of ${\cal M}_1, {\cal M}_2$ and ${\cal M}_3$ along with ${\cal M}$ for rewriting the various pieces. This leads to three rearrangements of the scalar potential.
\subsection*{Verifying the non-trivial symplectic identities}
Though verifying these symplectic identities is quite non-trivial for generic $h^{2,1}_-(CY)$ case, let us present some verification of the same by considering particular cases in the limit of not presenting too huge expressions.
\subsubsection*{Case 1: $h^{2,1}_-(CY)= 0$} 
For the case of frozen complex structure moduli (e.g. models studied in \cite{Blumenhagen:2015qda, Blumenhagen:2015kja, Shukla:2015rua, Shukla:2015bca}), we can have all the $l_{ijk}, l_{ij}$ and $l_i$ to be zero while choosing the pure imaginary number $l_0$ as $l_0=-\, i$, and so we have
\bea
& & {\cal X}^0 = 1, \quad {\cal F}_0 = - i
\eea
implying that $K_{cs} := - \ln\left(i\, (\ov{\cal X}^\Lambda {\cal F}_\Lambda -{\cal X}^\Lambda \ov{\cal F}_\Lambda) \right) = -\ln 2$, and subsequently from the respective definitions, one has
\bea
& & {\cal M}^{00} = -1, \quad {\cal M}^{0}{}_{0} = 0, \quad  {\cal M}_0{}^0 = 0, \quad {\cal M}_{00} = 1 \nonumber\\ 
& & {\cal L}^{00} = -1,  \quad {\cal L}^{0}{}_{0} = 0, \quad  {\cal L}_0{}^0 = 0, \quad {\cal L}_{00} = 1 \nonumber\\ 
& & {({\cal M}_1)}^{00} = -2, \quad  {({\cal M}_1)}^{0}{}_{0} = 0, \quad {({\cal M}_1)}_0{}^0 = 0, \quad {({\cal M}_1)}_{00} = 2 \nonumber\\ 
& & {({\cal M}_2)}^{00} = 3, \quad {({\cal M}_2)}^{0}{}_{0} = 0, \quad  {({\cal M}_2)}_0{}^0 =0, \quad {({\cal M}_2)}_{00} = -3 \nonumber\\ 
& & {({\cal M}_3)}^{00} = 0, \quad {({\cal M}_3)}^{0}{}_{0} = 2, \quad  {({\cal M}_3)}_0{}^0 =-2, \quad {({\cal M}_3)}_{00} = 0 \nonumber\\ 
& & \hat{\cal M}^{11} = -1, \quad \hat{\cal M}^{1}{}_{1} = 0, \quad  \hat{\cal M}_1{}^1=0, \quad \hat{\cal M}_{11} = 1 \nonumber
\eea
Using these ingredients, we find that identities (\ref{eq:symp10}), (\ref{eq:symp20}), (\ref{eq:symp30}) and (\ref{eq:mainSymp}) follow quite immediately.
\subsubsection*{Case 1: $h^{2,1}_-(CY)= 1$} 
The pre-potential for this case can be written as,
\bea
& & \hskip-1cm {\cal F}({\cal X}^0, {\cal X}^1)=\frac{1}{6\, {\cal X}^0}\biggl[3\, l_{0} \, ({\cal X}^0)^3 + 6\, l_{1} \, ({\cal X}^0)^2 \, {\cal X}^1 +   3 \,l_{11} \, ({\cal X}^1)^2 \, {\cal X}^0 + l_{111} \, ({\cal X}^1)^3 \biggr] 
\eea
Even for this simple pre-potential, the period matrix ${\cal N}$ as well as other symplectic matrices are quite huge to represent, so just for the sake of simple illustration, let us assume that $l \equiv l_{111} \ne 0$ and other triple intersection numbers to be zero. Subsequently, setting ${\cal X}^0 = 1$ and ${\cal X}^1 = v + i \, u$,  the various symplectic matrices are simplified as under,

\bea
& & \hskip-2cm {\cal M}^{\Lambda \Sigma} = \left(
\begin{array}{cc}
 \frac{6}{l u^3} & \frac{6 v}{l u^3} \\
 \frac{6 v}{l u^3} & \frac{2 \left(u^2+3 v^2\right)}{l u^3} \\
\end{array}
\right), \quad {\cal M}^\Lambda{}_\Sigma = \left(
\begin{array}{cc}
 \frac{v^3}{u^3} & -\frac{3 v^2}{u^3} \\
 \frac{v^2 \left(u^2+v^2\right)}{u^3} & -\frac{2 u^2 v+3 v^3}{u^3} \\
\end{array}
\right)\\
& & \hskip-2cm {\cal M}_\Lambda{}^\Sigma = \left(
\begin{array}{cc}
 -\frac{v^3}{u^3} & -\frac{v^2 \left(u^2+v^2\right)}{u^3} \\
 \frac{3 v^2}{u^3} & \frac{2 u^2 v+3 v^3}{u^3} \\
\end{array}
\right), \quad {\cal M}_{\Lambda \Sigma} = \left(
\begin{array}{cc}
 -\frac{l \left(u^2+v^2\right)^3}{6 u^3} & \frac{l v \left(u^2+v^2\right)^2}{2 u^3} \\
 \frac{l v \left(u^2+v^2\right)^2}{2 u^3} & -\frac{l \left(u^4+4 u^2 v^2+3 v^4\right)}{2 u^3} \\
\end{array}
\right) \nonumber
\eea

\bea
& & \hskip-1.2cm {\cal L}^{\Lambda \Sigma} = \left(
\begin{array}{cc}
 -\frac{3}{l u^3} & -\frac{3 v}{l u^3} \\
 -\frac{3 v}{l u^3} & \frac{u^2-3 v^2}{l u^3} \\
\end{array}
\right), \quad {\cal L}^\Lambda{}_\Sigma = \left(
\begin{array}{cc}
 -\frac{3 u^2 v+v^3}{2 u^3} & \frac{3 \left(u^2+v^2\right)}{2 u^3} \\
 -\frac{\left(u^2+v^2\right)^2}{2 u^3} & \frac{v \left(u^2+3 v^2\right)}{2 u^3} \\
\end{array}
\right)\\
& & \hskip-1.2cm {\cal L}_\Lambda{}^\Sigma = \left(
\begin{array}{cc}
 \frac{3 u^2 v+v^3}{2 u^3} & \frac{(u^2+v^2)^2}{2 u^3} \\
 -\frac{3 (u^2+v^2)}{2 u^3} & -\frac{v \left(u^2+3 v^2\right)}{2 u^3} \\
\end{array}
\right), \quad  {\cal L}_{\Lambda \Sigma} = \left(
\begin{array}{cc}
 \frac{l \left(u^2+v^2\right)^3}{12 u^3} & -\frac{l v \left(u^2+v^2\right)^2}{4 u^3} \\
 -\frac{l v \left(u^2+v^2\right)^2}{4 u^3} & -\frac{l \left(u^4-2 u^2 v^2-3 v^4\right)}{4 u^3} \\
\end{array}
\right) \nonumber
\eea

\bea
& & \hskip-1.9cm {{\cal M}_1}^{\Lambda \Sigma} = \left(
\begin{array}{cc}
 \frac{3}{l u^3} & \frac{3 v}{l u^3} \\
 \frac{3 v}{l u^3} & \frac{3 \left(u^2+v^2\right)}{l u^3} \\
\end{array}
\right), \quad {{\cal M}_1}^\Lambda{}_\Sigma = \left(
\begin{array}{cc}
 -\frac{v^3-3 u^2 v}{2 u^3} & \frac{u^4-v^4}{2 u^3} \\
 -\frac{3 \left(u^2-v^2\right)}{2 u^3} & \frac{3 v \left(u^2+v^2\right)}{2 u^3} \\
\end{array}
\right) \\
& & \hskip-1.9cm {{\cal M}_1}_\Lambda{}^\Sigma = \left(
\begin{array}{cc}
 \frac{v^3-3 u^2 v}{2 u^3} & \frac{3 \left(u^2-v^2\right)}{2 u^3} \\
 \frac{v^4-u^4}{2 u^3} & -\frac{3 v \left(u^2+v^2\right)}{2 u^3} \\
\end{array}
\right), \quad {{\cal M}_1}_{\Lambda \Sigma} = \left(
\begin{array}{cc}
 -\frac{l \left(u^2+v^2\right)^3}{12 u^3} & \frac{l v \left(u^2+v^2\right)^2}{4 u^3} \\
 \frac{l v \left(u^2+v^2\right)^2}{4 u^3} & -\frac{3 l \left(u^2+v^2\right)^2}{4 u^3} \\
\end{array}
\right) \nonumber
\eea

\bea
& & \hskip-6.0cm {{\cal M}_2}^{\Lambda \Sigma} = \left(
\begin{array}{cc}
 0 & 0 \\
 0 & -\frac{4}{l u} \\
\end{array}
\right), \quad {{\cal M}_2}^\Lambda{}_\Sigma = \left(
\begin{array}{cc}
 -\frac{3 v}{u} & -\frac{u^2+v^2}{u} \\
 \frac{3}{u} & -\frac{v}{u} \\
\end{array}
\right) \\
& & \hskip-6.0cm  {{\cal M}_2}_\Lambda{}^\Sigma = \left(
\begin{array}{cc}
 \frac{3 v}{u} & -\frac{3}{u} \\
 \frac{v^2}{u}+u & \frac{v}{u} \\
\end{array}
\right), \, {{\cal M}_2}_{\Lambda \Sigma} = \left(
\begin{array}{cc}
 0 & 0 \\
 0 & \frac{l \left(u^2+v^2\right)}{u} \\
\end{array}
\right) \nonumber
\eea

\bea
& & \hskip-2.5cm {{\cal M}_3}^{\Lambda \Sigma} = \left(
\begin{array}{cc}
 0 & \frac{3}{l u^2} \\
 -\frac{3}{l u^2} & 0 \\
\end{array}
\right), \quad 
{{\cal M}_3}^\Lambda{}_\Sigma = \left(
\begin{array}{cc}
 \frac{3 v^2}{2 u^2}+\frac{1}{2} & \frac{v^3}{u^2}+v \\
 -\frac{3 v}{u^2} & -\frac{3 v^2}{2 u^2}-\frac{1}{2} \\
\end{array}
\right), \\
& & \hskip-2.5cm {{\cal M}_3}_\Lambda{}^\Sigma =\left(
\begin{array}{cc}
 -\frac{3 v^2}{2 u^2}-\frac{1}{2} & \frac{3 v}{u^2} \\
 -\frac{v \left(u^2+v^2\right)}{u^2} & \frac{3 v^2}{2 u^2}+\frac{1}{2} \\
\end{array}
\right), \quad {{\cal M}_3}_{\Lambda \Sigma} = \left(
\begin{array}{cc}
 0 & \frac{l \left(u^2+v^2\right)^2}{4 u^2} \\
 -\frac{l \left(u^2+v^2\right)^2}{4 u^2} & 0 \\
\end{array}
\right) \nonumber
\eea
Now using $e^{Kcs} = \frac{3}{4 l u^3}$ and ${\cal F}_0 =-\frac{1}{6} l (v+i u)^3 , {\cal F}_1 = \frac{1}{2} l (v+i u)^2$ for the simplified ansatz, one can verify that 
\bea
& & e^{K_{cs}}  \, Re({\cal X}^\Lambda \, \ov{\cal X}^\Delta) = \left(
\begin{array}{cc}
 -\frac{3}{4 l u^3} & -\frac{3 v}{4 l u^3} \\
 -\frac{3 v}{4 l u^3} & -\frac{3 \left(u^2+v^2\right)}{4 l u^3} \\
\end{array}
\right) \\
& & e^{K_{cs}}  \, Re({\cal F}_\Lambda \, \ov{\cal X}^\Delta) = \left(
\begin{array}{cc}
 \frac{v^3-3 u^2 v}{8 u^3} & \frac{v^4-u^4}{8 u^3} \\
 \frac{3 \left(u^2-v^2\right)}{8 u^3} & -\frac{3 v \left(u^2+v^2\right)}{8 u^3} \\
\end{array}
\right)\nonumber\\
& & e^{K_{cs}}  \, Re({\cal X}^\Lambda \, \ov{\cal F}_\Delta) = \left(
\begin{array}{cc}
 \frac{v^3-3 u^2 v}{8 u^3} & \frac{3 \left(u^2-v^2\right)}{8 u^3} \\
 \frac{v^4-u^4}{8 u^3} & -\frac{3 v \left(u^2+v^2\right)}{8 u^3} \\
\end{array}
\right)\nonumber\\
& & e^{K_{cs}}  \, Re({\cal F}_\Lambda \, \ov{\cal F}_\Delta) = \left(
\begin{array}{cc}
 -\frac{l \left(u^2+v^2\right)^3}{48 u^3} & \frac{l v \left(u^2+v^2\right)^2}{16 u^3} \\
 \frac{l v \left(u^2+v^2\right)^2}{16 u^3} & -\frac{3 l \left(u^2+v^2\right)^2}{16 u^3} \\
\end{array}
\right)\nonumber
\eea
which are precisely $(-\frac{1}{4}{\cal M}_1)$ matrices, and hence we verified identities in eqn. (\ref{eq:symp10}) though for a simplified ansatz to show analytic form of intermediate matrices involved. Following similar procedure, and using generic pre-potential (\ref{eq:prepotential}), we can verify these identities for $h^{2,1}_-(CY) = 0, 1, 2, 3$ and we conjecture the same to be generically true.

\subsection*{Useful symplectic expressions for Example A}
Considering the pre-potential (\ref{eq:prepotentialA}), we get,
\bea
\label{eq:Fij}
& & {\cal F}_{\Lambda \Sigma} = \left(
\begin{array}{cccc}
 2 U_1 U_2 U_3 & -U_2 U_3 & -U_1 U_3 & -U_1 U_2 \\
 -U_2 U_3 & 0 & U_3 & U_2 \\
 -U_1 U_3 & U_3 & 0 & U_1 \\
 -U_1 U_2 & U_2 & U_1 & 0 \\
\end{array}
\right)
\eea
Using which one can compute the real and imaginary parts of period matrix ${\cal N}$ which are given as,
\bea
\label{eq:ReN}
& & Re \, {\cal N} = \left(
\begin{array}{cccc}
 2 v_1\,  v_2\, v_3 & -v_2 \, v_3 & -v_1 \, v_3 & -v_1 \, v_2 \\
 -v_2 \, v_3 & 0 & v_3 & v_2 \\
 -v_1 \, v_3 & v_3 & 0 & v_1 \\
 -v_1 \, v_2 & v_2 & v_1 & 0 \\
\end{array}
\right)
\eea
and
\bea
\label{eq:ImN}
& & \hskip-1.5cm Im \, {\cal N} = \left(
 \begin{array}{cccc}
 \frac{u_2 \, u_3 \, v_1^2}{u_1}+\frac{u_1 \, u_3 \, v_2^2}{u_2}+\frac{u_1 \, u_2 \, v_3^2}{u_3}+u_1 \, u_2 \, u_3 &
   -\frac{u_2 \, u_3 \, v_1}{u_1} & -\frac{u_1 \, u_3 \, v_2}{u_2} & -\frac{u_1 \, u_2 \, v_3}{u_3} \\
 -\frac{u_2 \, u_3 \, v_1}{u_1} & \frac{u_2 \, u_3}{u_1} & 0 & 0 \\
 -\frac{u_1 \, u_3 \, v_2}{u_2} & 0 & \frac{u_1 \, u_3}{u_2} & 0 \\
 -\frac{u_1\,  u_2 \, v_3}{u_3} & 0 & 0 & \frac{u_1 \, u_2}{u_3} \\
\end{array}
\right)
\eea
Recall that condition for physical domain is $Im{\cal N} <0$ which is ensured by $(u_1 \, u_2 \, u_3) <0$. Using these ingredients, we get the four sets of period matrices ${\cal M}$ defined in eqn. (\ref{eq:periodN}) to get expressions given as under,
\bea
\label{eq:Amatrix}
& & {\cal M}_{\Lambda}{}^{\Sigma} = \left(
\begin{array}{cccc}
 -\frac{v_1 v_2 v_3}{u_1 u_2 u_3} & -\frac{v_2 v_3 \left(u_1^2+v_1^2\right)}{u_1 u_2 u_3} &
   -\frac{v_1 v_3 \left(u_2^2+v_2^2\right)}{u_1 u_2 u_3} & -\frac{v_1 v_2
   \left(u_3^2+v_3^2\right)}{u_1 u_2 u_3} \\
 \frac{v_2 v_3}{u_1 u_2 u_3} & \frac{v_1 v_2 v_3}{u_1 u_2 u_3} & \frac{v_3
   \left(u_2^2+v_2^2\right)}{u_1 u_2 u_3} & \frac{v_2 \left(u_3^2+v_3^2\right)}{u_1 u_2 u_3}
   \\
 \frac{v_1 v_3}{u_1 u_2 u_3} & \frac{v_3 \left(u_1^2+v_1^2\right)}{u_1 u_2 u_3} & \frac{v_1
   v_2 v_3}{u_1 u_2 u_3} & \frac{v_1 \left(u_3^2+v_3^2\right)}{u_1 u_2 u_3} \\
 \frac{v_1 v_2}{u_1 u_2 u_3} & \frac{v_2 \left(u_1^2+v_1^2\right)}{u_1 u_2 u_3} & \frac{v_1
   \left(u_2^2+v_2^2\right)}{u_1 u_2 u_3} & \frac{v_1 v_2 v_3}{u_1 u_2 u_3} \\
\end{array}
\right)
\eea
\bea
\label{eq:Bmatrix}
& & \hskip-1cm {\cal M}_{\Lambda \Sigma} = \left(
\begin{array}{cccc}
 -\frac{\left(u_1^2+v_1^2\right) \left(u_2^2+v_2^2\right) \left(u_3^2+v_3^2\right)}{u_1 u_2
   u_3} & \frac{v_1 \left(u_2^2+v_2^2\right) \left(u_3^2+v_3^2\right)}{u_1 u_2 u_3} &
   \frac{v_2 \left(u_1^2+v_1^2\right) \left(u_3^2+v_3^2\right)}{u_1 u_2 u_3} & \frac{v_3
   \left(u_1^2+v_1^2\right) \left(u_2^2+v_2^2\right)}{u_1 u_2 u_3} \\
 \frac{v_1 \left(u_2^2+v_2^2\right) \left(u_3^2+v_3^2\right)}{u_1 u_2 u_3} &
   -\frac{\left(u_2^2+v_2^2\right) \left(u_3^2+v_3^2\right)}{u_1 u_2 u_3} & -\frac{v_1 v_2
   \left(u_3^2+v_3^2\right)}{u_1 u_2 u_3} & -\frac{v_1 v_3 \left(u_2^2+v_2^2\right)}{u_1 u_2
   u_3} \\
 \frac{v_2 \left(u_1^2+v_1^2\right) \left(u_3^2+v_3^2\right)}{u_1 u_2 u_3} & -\frac{v_1 v_2
   \left(u_3^2+v_3^2\right)}{u_1 u_2 u_3} & -\frac{\left(u_1^2+v_1^2\right)
   \left(u_3^2+v_3^2\right)}{u_1 u_2 u_3} & -\frac{v_2 v_3 \left(u_1^2+v_1^2\right)}{u_1 u_2
   u_3} \\
 \frac{v_3 \left(u_1^2+v_1^2\right) \left(u_2^2+v_2^2\right)}{u_1 u_2 u_3} & -\frac{v_1 v_3
   \left(u_2^2+v_2^2\right)}{u_1 u_2 u_3} & -\frac{v_2 v_3 \left(u_1^2+v_1^2\right)}{u_1 u_2
   u_3} & -\frac{\left(u_1^2+v_1^2\right) \left(u_2^2+v_2^2\right)}{u_1 u_2 u_3} 
\end{array}
\right)\nonumber\\
& & 
\eea
\bea
\label{eq:Cmatrix}
& & {\cal M}^{\Lambda \Sigma} = \left(
\begin{array}{cccc}
 \frac{1}{u_1 \, u_2 \, u_3} & \frac{v_1}{u_1 u_2 u_3} & \frac{v_2}{u_1 \, u_2 \, u_3} & \frac{v_3}{u_1 \, u_2 \, u_3} \\
 \frac{v_1}{u_1 \,  u_2 u_3} & \frac{u_1^2+v_1^2}{u_1 u_2 u_3} & \frac{v_1 v_2}{u_1 u_2 u_3} &
   \frac{v_1 v_3}{u_1 u_2 u_3} \\
 \frac{v_2}{u_1 u_2 u_3} & \frac{v_1 v_2}{u_1 u_2 u_3} & \frac{u_2^2+v_2^2}{u_1 u_2 u_3} &
   \frac{v_2 v_3}{u_1 u_2 u_3} \\
 \frac{v_3}{u_1 u_2 u_3} & \frac{v_1 v_3}{u_1 u_2 u_3} & \frac{v_2 v_3}{u_1 u_2 u_3} &
   \frac{u_3^2+v_3^2}{u_1 u_2 u_3} \\
\end{array}
\right) 
\eea
\bea
\label{eq:Dmatrix}
& & {\cal M}^{\Lambda}{}_{ \Sigma} = \left(
\begin{array}{cccc}
 \frac{v_1 v_2 v_3}{u_1 u_2 u_3} & -\frac{v_2 v_3}{u_1 u_2 u_3} & -\frac{v_1 v_3}{u_1 u_2 u_3}
   & -\frac{v_1 v_2}{u_1 u_2 u_3} \\
 \frac{v_2 v_3 \left(u_1^2+v_1^2\right)}{u_1 u_2 u_3} & -\frac{v_1 v_2 v_3}{u_1 u_2 u_3} &
   -\frac{v_3 \left(u_1^2+v_1^2\right)}{u_1 u_2 u_3} & -\frac{v_2
   \left(u_1^2+v_1^2\right)}{u_1 u_2 u_3} \\
 \frac{v_1 v_3 \left(u_2^2+v_2^2\right)}{u_1 u_2 u_3} & -\frac{v_3
   \left(u_2^2+v_2^2\right)}{u_1 u_2 u_3} & -\frac{v_1 v_2 v_3}{u_1 u_2 u_3} & -\frac{v_1
   \left(u_2^2+v_2^2\right)}{u_1 u_2 u_3} \\
 \frac{v_1 v_2 \left(u_3^2+v_3^2\right)}{u_1 u_2 u_3} & -\frac{v_2
   \left(u_3^2+v_3^2\right)}{u_1 u_2 u_3} & -\frac{v_1 \left(u_3^2+v_3^2\right)}{u_1 u_2 u_3}
   & -\frac{v_1 v_2 v_3}{u_1 u_2 u_3} \\
\end{array} 
\right)
\eea
Now, we provide the ${\cal M}_2$-matrices are given as under,
\bea
& & {{\cal M}_2}^{\Lambda\Sigma} = \left(
\begin{array}{cccc}
 0 & 0 & 0 & 0 \\
 0 & 0 & -\frac{1}{u_3} & -\frac{1}{u_2} \\
 0 & -\frac{1}{u_3} & 0 & -\frac{1}{u_1} \\
 0 & -\frac{1}{u_2} & -\frac{1}{u_1} & 0 \\
\end{array}
\right) 
\eea
\bea
\label{eq:Mmatrix}
& & \hskip-1.0cm {{\cal M}_2}_\Lambda{}^\Sigma = \left(
\begin{array}{cccc}
 -\frac{v_1}{u_1}-\frac{v_2}{u_2}-\frac{v_3}{u_3} & -\frac{u_1^2+v_1^2}{u_1} & -\frac{u_2^2+v_2^2}{u_2} & -\frac{u_3^2+v_3^2}{u_3}
   \\
 \frac{1}{u_1} & \frac{v_1}{u_1}-\frac{v_2}{u_2}-\frac{v_3}{u_3} & 0 & 0 \\
 \frac{1}{u_2} & 0 & -\frac{v_1}{u_1}+\frac{v_2}{u_2}-\frac{v_3}{u_3} & 0 \\
 \frac{1}{u_3} & 0 & 0 & -\frac{v_1}{u_1}-\frac{v_2}{u_2}+\frac{v_3}{u_3} \\
\end{array}
\right) 
\eea
\bea
& & \hskip-1.0cm {{\cal M}_2}^\Lambda{}_\Sigma= \left(
\begin{array}{cccc}
 \frac{v_1}{u_1}+\frac{v_2}{u_2}+\frac{v_3}{u_3} & -\frac{1}{u_1} & -\frac{1}{u_2} & -\frac{1}{u_3} \\
 \frac{v_1^2}{u_1}+u_1 & -\frac{v_1}{u_1}+\frac{v_2}{u_2}+\frac{v_3}{u_3} & 0 & 0 \\
 \frac{v_2^2}{u_2}+u_2 & 0 & \frac{v_1}{u_1}-\frac{v_2}{u_2}+\frac{v_3}{u_3} & 0 \\
 \frac{v_3^2}{u_3}+u_3 & 0 & 0 & \frac{v_1}{u_1}+\frac{v_2}{u_2}-\frac{v_3}{u_3} \\
\end{array}
\right) 
\eea
\bea
& & {{\cal M}_2}_{\Lambda \Delta}= \left(
\begin{array}{cccc}
 0 & 0 & 0 & 0 \\
 0 & 0 & \frac{v_3^2}{u_3}+u_3 & \frac{v_2^2}{u_2}+u_2 \\
 0 & \frac{v_3^2}{u_3}+u_3 & 0 & \frac{v_1^2}{u_1}+u_1 \\
 0 & \frac{v_2^2}{u_2}+u_2 & \frac{v_1^2}{u_1}+u_1 & 0 \\
\end{array}
\right) 
\eea
Other matrices are quite large to present here, however using ${\cal M}$ and ${\cal M}_2$ all of those can be determined.

\newpage
\bibliographystyle{utphys}
\bibliography{CYReference}

\end{document}